\documentclass[11pt]{article}
\usepackage{fancyheadings}
\usepackage{amssymb,amsmath}
\usepackage{ascmac}
\usepackage{hyperref}
\usepackage{graphicx}
\usepackage{epsfig}
\usepackage{array}
\usepackage{lscape}
\usepackage{accents}
\usepackage{color}
\usepackage{braket}
\usepackage{bm}
\usepackage{multirow}

\numberwithin{equation}{section}

\makeatletter
\def\alt{\mathrel{\mathpalette\gl@align<}}
\def\agt{\mathrel{\mathpalette\gl@align>}}
\def\gl@align#1#2{\lower.6ex\vbox{\baselineskip\z@skip\lineskip\z@
\ialign{$\m@th#1\hfil##\hfil$\crcr#2\crcr\sim\crcr}}}
\makeatother

\newcommand{\doi}[1]{\href{https://doi.org/#1}{doi:#1}}

\begin{document}

\begin{flushright}
{\large
October, 2026
}
\end{flushright}
\vspace*{3mm}

\begin{center}
\baselineskip 20pt

{\bf
\LARGE
Lepton number and flavor violation mediated by \\
doubly charged scalars at $\mu$TRISTAN $\mu^+ \mu^+$ collider} 

\vspace{1.0cm}

{\Large 
Takeshi Fukuyama$\,{}^{a}$,
Yukihiro Mimura$\,{}^{b}$
and
Yuichi Uesaka$\,{}^{c}$
}

\vspace{5mm}

{\large
${}^a${\it 
Research Center for Nuclear Physics (RCNP),
Osaka University, \\Ibaraki, Osaka, 567-0047, Japan
}\\
\vspace{3mm}
${}^{b}${\it
Department of Physical Sciences, College of Science and Engineering, \\
Ritsumeikan University, Shiga 525-8577, Japan
}\\
\vspace{3mm}
${}^{c}${\it
Department of Fundamental Education,
Dokkyo Medical University, \\
880 Kitakobayashi,
Mibu, Shimotsuga, Tochigi 321-0293, Japan
}\\
}

\vspace{8mm}

{\Large
{\bf Abstract}}\end{center}
\baselineskip 18pt
{\large
Charged lepton flavor violation is an important probe of
physics beyond the Standard Model.
Although $\Delta L_\mu = -\Delta L_e = \pm 1$ lepton flavor violating (LFV) processes,
such as $\mu \to e\gamma$ and $\mu \to 3e$,
are severely constrained experimentally,
there is still plenty of room for LFV processes involving lepton number violation (LNV).
In this paper,
we investigate such processes at a $\mu^+\mu^+$ collider
that remain even if processes with $\Delta L_\mu = -\Delta L_e = \pm 1$
are forbidden by discrete symmetries.
We consider processes induced by doubly charged scalars,
which can be either an $SU(2)_L$ singlet or a component of an $SU(2)_L$ triplet.
The cross sections of these processes at the collider depend on the representation of the scalar
unless the muon beam is unpolarized.
In the case of $\Delta L_\mu = -\Delta L_e = 2$,
we discuss the combination of the muonium-to-antimuonium transition
and the $\mu^+\mu^+ \to e^+e^+\gamma/Z$ processes at the collider,
and how the model parameters, such as the scalar couplings and mass,
can be determined from experimental measurements.
Doubly charged scalars are often utilized in models for generating neutrino masses,
and they can also be used to construct models with LNV processes at a muon collider.
If neutrino masses are generated at tree level,
the LNV processes at the collider are suppressed,
and therefore, we consider models
that do not induce neutrino masses at tree level.
LNV processes, such as $\mu^+\mu^+ \to t\bar b t\bar b$
and $\mu^+\mu^+ \to \mu^+ e^- t\bar b t\bar b$, are possible in these models.
Even if LNV is indicated by searches at hadron colliders, 
it is worthwhile to pursue these processes as clean LNV signals at a lepton collider.
}

\thispagestyle{empty}
\newpage
\addtocounter{page}{-1}

\section{Introduction}
\label{sec1}

\fontsize{13pt}{18pt}\selectfont

Lepton flavor violation (LFV) is one of the keys
to discussing new physics (NP) in the lepton sector,
since it can directly indicate the existence of a new particle or interaction
beyond the Standard Model (SM).
The LFV processes that serve as indicators of NP are often referred to
as charged lepton flavor violation (cLFV)
to distinguish them from LFV caused by neutrino oscillations.
In the SM with neutrino masses,
the cLFV processes are strongly suppressed due to the tiny differences
in the squared neutrino masses.
The electron number, $L_e$, muon number, $L_\mu$, and tau number, $L_\tau$,
are good conserved quantum numbers perturbatively, up to neutrino oscillations.
The cLFV processes, such as $\mu \to e\gamma$, $\mu \to 3e$ decays, and $\mu$-$e$ conversion in nuclei,
have not yet been observed \cite{MEGII:2025gzr,Bellgardt:1987du,SINDRUMII:2006dvw},
and their non-observation only gives severe bounds on the model parameters at present.
We refer to them as $\Delta L_\mu = -\Delta L_e = \pm 1$ processes.
Even if the $\Delta L_\mu = -\Delta L_e = \pm 1$ processes do not exist at all,
it does not necessarily follow that cLFV does not exist.
There is still plenty of room for NP in the lepton sector
because the lepton sector may be closely related to discrete symmetries.
Though it is surely important to search for the $\Delta L_\mu = -\Delta L_e = \pm 1$ processes,
we need to carefully examine models beyond the SM
so as not to have a preconception that those $\Delta L_\mu = -\Delta L_e = \pm 1$ processes play a leading role
in NP with cLFV.
Indeed, flavor number violation can be observed at lepton colliders,
even if the $\Delta L_\mu = -\Delta L_e = \pm 1$ processes are forbidden by a discrete symmetry.

If we limit our consideration to the first and second generations,
the following three types of cLFV processes are possible at the lowest order
even when $\Delta L_\mu = -\Delta L_e = \pm 1$ is prohibited:
\begin{enumerate}
\item
$(\Delta L_e, \Delta L_\mu, \Delta L_\tau) = (\pm 2, \mp 2, 0)$

Electron and muon numbers, $L_e$ and $L_\mu$, are violated, while
lepton number $L = L_e + L_\mu + L_\tau$ is conserved.
An example process is
$\mu^+ \mu^+ \to e^+ e^+$.

\item
$(\Delta L_e, \Delta L_\mu, \Delta L_\tau) = (\pm 2, 0,0),(0,\pm 2,0)$

Lepton number is violated, while one of $L_e$ and $L_\mu$ is conserved.
An example process is
$\mu^+ \mu^+ \to X$, where $X$ stands for multiple particles
and $X$ does not have a lepton number.

\item
$(\Delta L_e, \Delta L_\mu, \Delta L_\tau) = (\pm 1, \pm 1, 0)$

Lepton number is violated, while $L_e - L_\mu$ is conserved.
An example process is
$\mu^+ \mu^+ \to \mu^+ e^- X$,
where $X$ does not have a lepton number.

\end{enumerate}
Examples of discrete symmetries that forbid $\Delta L_\mu = -\Delta L_e = \pm 1$ processes
are a $Z_2$ discrete symmetry in each generation in cases 1 and 2,
and a $Z_3$ symmetry in case 3 (e.g., the $Z_3$ charges of $e^+$, $\mu^+$, and $\tau^+$
are 1, 2, and 0, respectively. The $Z_3$ charge of $e^-$ is 2 to allow the electron mass).
These three types of processes can be induced by various mediators.
Doubly charged scalars, which can couple to a pair of charged leptons,
can induce all three types.
In cases 1 and 2 with the $Z_2$ discrete symmetry,
the coupling matrix becomes flavor diagonal,
while in case 3, $e\mu$ and $\tau\tau$ couplings are allowed by the $Z_3$ symmetry.
Models of loop-induced neutrino masses employ
doubly charged scalars \cite{Babu:1988ki,Gustafsson:2012vj,Liu:2016mpf},
and these scalars are compatible with models involving lepton number violation (LNV).
The purpose of this paper is to investigate these processes induced by
the doubly charged scalar at a lepton collider.

Ultra-cold muon technology, originally developed
for measurements of muon $g-2$ and electric dipole moment at J-PARC, provides
muons ($\mu^+$) with very small spatial and momentum spreads,
and a realistic muon collider called $\mu$TRISTAN is being considered
by accelerating the low-emittance $\mu^+$ beam up to 1 TeV \cite{Hamada:2022mua}.
In addition to a $\mu^+ e^-$ collider,
a $\mu^+ \mu^+$ collider is possible with a center-of-mass energy
$\sqrt{s} = 2$ TeV.
The physics potential of $\mu$TRISTAN has been explored through sensitivity analyses of NP within the SM effective field theory~\cite{Hamada:2022uyn,Mukherjee:2026kmp} and studies of Higgs production~\cite{Hamada:2024ojj}.
The initial state of the $\mu^+ \mu^+$ collider has
a muon number $L_\mu = -2$,
and it is suitable for investigating LFV and LNV,
as examples are given in the list.
Furthermore,
it is also ideal for studying doubly charged scalars at the TeV scale.
According to Ref.~\cite{Hamada:2022mua},
the instantaneous luminosity of the $\mu^+\mu^+$ collider
is $5.7 \times 10^{32}\, {\rm cm}^{-2}\, {\rm s}^{-1}$,
which corresponds to 12 ${\rm fb}^{-1}/{\rm year}$
with about 240 days of running in a year.
This luminosity enables us to investigate $\Delta L_\mu = - \Delta L_e = 2$
LFV processes below the current experimental bound.
There are various phenomenological studies of the $\mu^+\mu^+$ option of $\mu$TRISTAN, addressing doubly charged scalars and other components of the same scalar multiplets~\cite{Lichtenstein:2023iut,Yang:2023ojm} and LFV or LNV in other NP scenarios~\cite{Jiang:2023mte}.

Transitions of muonium (Mu: $\mu^+ e^-$) to
antimuonium ($\overline{\rm Mu}$: $\mu^- e^+$) \cite{Pontecorvo:1957cp,Feinberg:1961zza,Lee:1977tib,Halprin:1982wm}
are important phenomena for $\Delta L_\mu = - \Delta L_e = 2$ processes.
Although the transitions can be induced through one-loop box diagrams
using $\Delta L_\mu = - \Delta L_e = \pm 1$ interactions or Majorana masses
of SM singlet neutrinos,
the transition probability is much smaller than the current experimental bound 
due to the constraints on the $\Delta L_\mu = - \Delta L_e = \pm 1$ processes
and the requirement of avoiding excessively large radiatively induced neutrino masses \cite{Fukuyama:2021iyw}.
The doubly charged scalars can induce the transition at tree level \cite{Chang:1989uk,Swartz:1989qz}
while satisfying the constraints from the $\Delta L_\mu = - \Delta L_e = \pm 1$ processes,
and
it is possible to generate a transition probability just below the current experimental bound.
The experimental bound on Mu-to-$\overline{\rm Mu}$ transitions obtained decades ago \cite{Willmann:1998gd}
will be updated in the near future \cite{Han:2021nod,Bai:2022sxq,Bai:2024skk,Kawamura:2021lqk,Kawamura:2025lhk}.
The transition may already have been observed
by the time the lepton collider is built.

There are two choices of the doubly charged scalar that couples to a pair of charged leptons:
an $SU(2)_L$ triplet and an $SU(2)_L$ singlet with appropriate hypercharges.
The $SU(2)_L$ triplet can be utilized in the so-called type-II seesaw model \cite{Schechter:1980gr,Cheng:1980qt,Lazarides:1980nt,Mohapatra:1980yp}
to induce neutrino masses at tree level.
The $SU(2)_L$ singlet doubly charged scalar is
also utilized in models that generate neutrino masses through multi-loop diagrams, as mentioned above.
Therefore,
we can expect that the doubly charged scalars also generate LNV processes
at the lepton collider.
In the case of loop-induced neutrino masses,
the scalar couplings to lepton pairs
are not necessarily small,
and they can induce sizable $\Delta L_\mu = - \Delta L_e = 2$ processes.
On the other hand,
in the minimal type-II seesaw model,
where only an $SU(2)_L$ triplet scalar is added to the SM,
LNV processes at the collider are related to neutrino masses,
and thus, their cross sections are suppressed.

In this paper,
we investigate the three types of processes listed above
at the $\mu$TRISTAN $\mu^+\mu^+$ collider.
For $\Delta L_\mu = - \Delta L_e = 2$ processes mediated by doubly charged scalars,
we suppose that
the probability of the Mu-to-$\overline{\rm Mu}$ transition is experimentally measured,
and study what can be obtained
from the combination of the transition amplitudes
and the measurements at the $\mu$TRISTAN.
We find that $\mu^+ \mu^+ \to e^+ e^+ \gamma$
is one of the important processes
to determine the scalar couplings to leptons, scalar mass,
and scalar decay width,
especially if
the scalar mass is smaller than the center-of-mass energy of the collider.
The branching fraction of the scalar decays to charged leptons
can be determined in principle
by additionally measuring the scalar coupling to muons
from $\mu^+\mu^+ \to \mu^+\mu^+$ scattering at the collider.

For LNV processes in cases 2 and 3,
we employ a model for the singlet doubly charged scalar,
where neutrino masses can be induced by a two-loop diagram.
The model has two Higgs doublets,
and an $SU(2)_L$ singlet singly charged scalar.
The singly charged scalar can mix with the charged scalar in the Higgs doublets,
which can couple to quarks.
As a result,
the doubly charged scalar can decay into quark pairs.
If the scalar coupling to muon pairs is not zero,
$\mu^+ \mu^+ \to t\bar b t\bar b$ (case 2 in the list) is possible at the $\mu^+\mu^+$ collider.
If the scalar coupling to muon and electron is not zero,
$\mu^+ \mu^+ \to \mu^+ e^- t\bar b t\bar b$ (case 3 in the list) is possible.
As mentioned,
in the minimal type-II seesaw model with an $SU(2)_L$ triplet,
LNV processes are suppressed due to the tiny neutrino masses.
Introducing an inert $SU(2)_L$ doublet
allows us to adopt a similar structure to that in the singlet scenario described above,
making it possible to induce the LNV process in the $\mu^+\mu^+$ collider
without acquiring large neutrino masses.
In this case, the simplest non-leptonic decay mode of the doubly charged scalar is $t\bar b W^+(\to jj)$, instead of $t\bar b t\bar b$.
Observing the case 3 process is more challenging than case 2.
We study the
$\mu^+ \mu^+ \to \mu^+ e^- X$ process for $\Delta L_\mu = \Delta L_e = 1$ LFV and LNV,
whose detection at the $\mu^+\mu^+$ collider
seems more likely than observing the $pp \mu^- \to nn e^+$ process in nuclei.

This paper is organized as follows:
In Sec.~\ref{sec2},
we introduce an LNV model with an $SU(2)_L$ singlet doubly charged scalar,
and its $SU(2)_L$ triplet version.
The coupling matrices of the scalar with the charged leptons
that do not generate $\Delta L_\mu = -\Delta L_e = \pm 1$ processes are also introduced.
In Sec.~\ref{sec3},
we describe experimental constraints:
Mu-to-$\overline{\rm Mu}$ transitions,
LEP bounds,
constraints on the triplet scalar coupling to left-handed electron and muon,
LNV processes in nuclei (neutrinoless double beta decay and $pp \mu^- \to nn e^+$),
and LHC bounds.
In Sec.~\ref{sec4},
the setups for colliders and detectors considered in this paper are summarized:
muon polarization and cuts for event generation.
We also comment on the missing energy veto
for the selection of cLFV processes at the lepton collider.
In Sec.~\ref{sec5},
we explore the $\mu^+\mu^+ \to e^+ e^+$ and $\mu^+\mu^+ \to e^+ e^+\gamma/Z$
processes at the $\mu^+\mu^+$ collider,
and study how the model parameters, such as coupling constants, scalar mass, and decay width,
can be obtained.
From experimental measurements, the branching ratio of the scalar decay
to lepton pairs can be determined if
the scalar mass is smaller than the center-of-mass energy.
In Sec.~\ref{sec6},
the non-leptonic decays of the scalars for the LNV processes at the collider are studied.
In Sec.~\ref{sec7},
we investigate the $\Delta L_e = \Delta L_\mu = 1$ LFV and LNV process at the $\mu^+\mu^+$ collider.
Sec.~\ref{sec8} is devoted to conclusions and discussions.

\section{Doubly charged scalars}
\label{sec2}

We consider doubly charged scalars that can couple to two charged leptons.
In the SM, the representation of the right-handed charged leptons $l_{aR} = e_R, \mu_R, \tau_R$ is
$({\bf 1}, -1)$ under $SU(2)_L \times U(1)_Y$,
and the representation of the left-handed lepton doublets $\ell_a = (\nu_{aL}, l_{aL})$ is $({\bf 2}, -\frac12)$.
The possible representations that contain doubly charged scalars are
$({\bf 1}, 2)$ and $({\bf 3},1)$.
We denote an $SU(2)_L$ singlet scalar as $k^{++}$, and
an $SU(2)_L$ triplet scalar as $\Delta$.

\subsection{{\boldmath$SU(2)_L$} singlet scalar { \boldmath$k^{++}$} }

We consider the following Lagrangian for the $k^{++}$ coupling to right-handed charged leptons:
\begin{equation}
-{\cal L} = g_{ab} \overline{l_{aR}^c} l_{bR} k^{++} + m_{hhk} h^- h^- k^{++} + m_{\Phi\Phi h} \Phi_1 \Phi_2 h^+ + {\rm H.c.},
\label{lagrangian-1}
\end{equation}
where $``c"$ denotes charge conjugation (safely distinguishable from a flavor index).
We also employ an $SU(2)_L$ singlet scalar $h^+$: $({\bf 1}, 1)$,
and two $SU(2)_L$ Higgs doublets $\Phi_1,\Phi_2$: $({\bf 2}, -\frac12)$, which couple to quarks and leptons.
The coupling matrix $g_{ab}$ is symmetric under the flavor indices $a,b$: $g_{ab} = g_{ba}$.
Because the lepton number of $\Phi_{1,2}$ is zero,
lepton number is violated if the couplings in Eq.~(\ref{lagrangian-1}) are non-zero.
In fact, these interactions can generate neutrino masses via a two-loop diagram \cite{Liu:2016mpf,Fukuyama:2022dhe}.
We note that neutrino masses can be generated via a one-loop diagram for $m_{\Phi\Phi h} \neq 0$
if $\ell\ell h^+$ terms,
\begin{equation}
f_{ab} \overline{\ell_a^c} \ell_{b} h^+ = f_{ab} (\overline{\nu_{aL}^c} l_{bL} - \overline{\nu_{bL}^c} l_{aL}) h^+ ,
\label{fab}
\end{equation}
are allowed \cite{Zee:1980ai,Zee:1985id}.
We suppose that the coupling constants $f_{ab}$ are small and are not important for the collider studies.

In this paper, we do not assume that the observed neutrino masses and mixings are necessarily generated
only from Eq.~(\ref{lagrangian-1}).
We concentrate on the possibility that
the first- and second-generation couplings of $g_{ab}$ ($g_{ee}, g_{e\mu}, g_{\mu\mu}$) can be observed at lepton colliders.
Those couplings can generate severely constrained cLFV processes,
$\mu^+ \to e^+ e^- e^+$ at tree level and $\mu^+ \to e^+ \gamma$ via a one-loop diagram.
We consider the following two cases for the coupling matrix that do not generate these processes:
\begin{enumerate}
\item $Z_2$ flavor symmetry in each generation,
\begin{equation}
g = \left(
\begin{array}{ccc}
g_{ee} & 0 & 0 \\
0 & g_{\mu\mu} & 0 \\
0 & 0 & g_{\tau\tau}
\end{array}
\right).
\label{g-1}
\end{equation}
\item $Z_3$ flavor symmetry ($L_e - L_\mu$ symmetry),
\begin{equation}
g = \left(
\begin{array}{ccc}
0 & g_{e\mu} & 0 \\
g_{e\mu} & 0 & 0 \\
0 & 0 & g_{\tau\tau}
\end{array}
\right).
\label{g-2}
\end{equation}
\end{enumerate}
In these cases,
the $\mu^+ \to e^+ e^- e^+$ and $\mu^+ \to e^+ \gamma$ decays (which are $\Delta L_\mu = -\Delta L_e = 1$ processes)
are forbidden, up to tiny contributions from neutrino masses.
However, the following cLFV processes are still allowed:
$\Delta L_e = -\Delta L_\mu = \pm 2$ processes are allowed
in case 1 ($g_{ee},g_{\mu\mu} \neq 0$),
and $\Delta L_e = \Delta L_\mu = \pm 1$ processes
($\Delta L_e - \Delta L_\mu$ is conserved) are allowed in case 2 ($g_{e\mu} \neq 0$).

In the case of $g_{ee},g_{\mu\mu} \neq 0$,
the following processes can occur at $\mu$TRISTAN ($\mu^+ e^-$ and $\mu^+ \mu^+$ colliders):
\begin{align}
\mu^+ e^- \to \mu^- e^+ ,\\
\mu^+ \mu^+ \to e^+ e^+.
\end{align}
These are related to the muonium-to-antimuonium transition (a bound state of $\mu^+ e^-$ is called muonium,
and $\mu^- e^+$ is called antimuonium).
For $g_{\mu\mu} \neq 0$,
\begin{equation}
\mu^+ \mu^+ \to h^+ h^+
\label{mumuhh}
\end{equation}
can occur at the same-sign muon collider.
Due to the $\Phi_1 \Phi_2 h^+$ term in the Lagrangian,
$h^+$ and a charged scalar in the Higgs doublets $\Phi_1$ and $\Phi_2$ are mixed.
Therefore, $h^+$ can decay into $t\bar b$,
and we expect to observe an LNV process ($\Delta L = \Delta L_\mu = 2$) at the collider.
Similarly,
\begin{equation}
\mu^+ \mu^+ \to k^{++} Z/\gamma
\label{mumukZgamma}
\end{equation}
can occur,
and it can also be an LNV process through the decay chain
$k^{++} \to h^+ h^+ \to t\bar b t\bar b$.
Of course, $k^{++}$ can also decay into $e^+ e^+$ or $\tau^+ \tau^+$,
in which case the process is not LNV but cLFV.

In the case of $g_{e\mu} \neq 0$,
we can expect to observe the process
\begin{equation}
\mu^+ \mu^+ \to \mu^+ e^- k^{++},
\label{mumumuek}
\end{equation}
at the same-sign muon collider.
The doubly charged scalar $k^{++}$ can decay into $h^+ h^+$,
and $h^+$ can decay into quarks as above.
Therefore, this is a cLFV process with
$\Delta L_\mu = \Delta L_e = 1$
and an LNV process with $\Delta L = 2$.

\subsection{{\boldmath$SU(2)_L$} triplet scalar $\Delta$}

As a left-handed version of Eq.~(\ref{lagrangian-1}),
the Lagrangian for the triplet scalar $\Delta$ coupling to left-handed leptons is
\begin{equation}
-{\cal L} = \kappa_{ab} \overline{\ell_{a}^c} \ell_{b} \Delta + m_{\Delta\Phi\eta} \Delta \Phi \eta + {\rm H.c.},
\label{lagrangian-2}
\end{equation}
where $\Phi$ and $\eta$ are $SU(2)_L$ doublets with hypercharge $-1/2$,
and $\Phi$ is the SM Higgs doublet.
We omit the $SU(2)_L$ indices for the $\ell\ell\Delta$ term, which is expanded in components as
\begin{equation}
\overline{\ell_{a}^c} \ell_{b} \Delta
= \overline{\nu_{aL}^c} \nu_{bL} \Delta^0 - \frac1{\sqrt2} (\overline{\nu_{aL}^c} l_{bL} + \overline{\nu_{bL}^c} l_{aL} )\Delta^+
+ \overline{l^c_{aL}} l_{bL} \Delta^{++} .
\end{equation}
Neutrino masses are generated from the $\ell\ell\Delta$ term when $\Delta^0$ acquires a vacuum expectation value (vev).
Our purpose is to observe the $\kappa$ coupling for a TeV-scale mass of $\Delta$,
and thus the vev of $\Delta^0$ should be tiny.
Therefore, we suppose that a possible $\Delta\Phi\Phi$ scalar coupling is tiny.
The vev of $\Delta^0$ is obtained as
$m_{\Delta\Phi\eta} \langle \Phi^0\rangle \langle \eta^0\rangle/M_\Delta^2$,
and thus, the vev of $\eta^0$ should be tiny.
The ``inert'' doublet scalar $\eta$ can also couple to quarks and leptons,
but those Yukawa couplings are not necessarily related to their masses.
If $\eta$ couples to leptons with $m_{\Delta\Phi\eta} \neq 0$,
neutrino masses can be generated at one loop.
We, therefore, assume that the Yukawa couplings of $\overline{l_R} \ell \eta$ are small.
The $\Delta^+$ and $\eta^+$ can mix through the $m_{\Delta\Phi\eta}$ coupling,
and thus $\Delta^+$ can decay into $t\bar b$.
The doubly charged scalar $\Delta^{++}$ can decay into $\Delta^+ W^+$ (if it is kinematically allowed).
As a result, there can be LNV processes at the lepton collider, as in the case of $k^{++}$.

We assume the $\kappa$ coupling matrix to have a structure similar to the two cases in Eqs.~(\ref{g-1}) and (\ref{g-2}).
Similar processes to those discussed above can be considered.
In addition to processes similar to those involving $k^{++}$ couplings to right-handed charged leptons,
we can consider
\begin{equation}
\mu^+ \mu^+ \to W^+ \Delta^+.
\label{mumu-WD}
\end{equation}
As noted above,
$\Delta^+$ can decay into quarks,
and this can result in an LNV process.
However,
we do not consider $\mu^+ \mu^+ \to \Delta^+ \Delta^+$, a process similar to Eq.~(\ref{mumuhh}).
A quartic term of the form $\Delta \Delta \Delta^\dagger \Delta^\dagger$
can contain $\Delta^{++} \Delta^- \Delta^- \Delta^0$.
The scattering amplitude for $\mu^+ \mu^+ \to \Delta^+ \Delta^+$ is proportional to
$\kappa_{\mu\mu} \langle \Delta^0 \rangle$,
which is the $(\mu,\mu)$ element of the neutrino mass matrix,
and therefore the cross section of this process is negligibly small.
The amplitude for $\mu^+ \mu^+ \to W^+ W^+$ is also proportional to
$\kappa_{\mu\mu} \langle \Delta^0 \rangle$.
Therefore, in the minimal case of the type-II seesaw model with only the Higgs doublet $\Phi$,
the amplitudes of the LNV processes at the collider are small.
However, there is a loophole, and the LNV process in Eq.~(\ref{mumu-WD}) can be observed by employing the inert doublet as described above.

The $\kappa_{e\mu}$ coupling (as well as $f_{e\mu}$ in Eq.~(\ref{fab}))
can induce muon decays,
and therefore the size of $\kappa_{e\mu}$ is constrained by precision experimental measurements,
which we will describe in the next section.

\subsection{Notation convention}

For notational simplicity, we denote the doubly charged scalar collectively by $S^{++}$
and its coupling to charged leptons by $y_{ab}$.
In the singlet case, $S^{++} = k^{++}$ and $y_{ab} = g_{ab}$,
whereas in the triplet case, $S^{++} = \Delta^{++}$ and $y_{ab} = \kappa_{ab}$.

\section{Experimental constraints}
\label{sec3}

In this section,
we describe the experimental constraints on the
doubly charged scalar couplings to leptons, $y_{ab}$,
and the scalar mass $m_S$,
which are related to LFV and LNV.

One of the important building blocks for the
$\Delta L_\mu = -\Delta L_e = 2$ LFV process
is the muonium-to-antimuonium (Mu-to-$\overline{\rm Mu}$) transition,
whose experimental bound will be updated in the near future.
The coupling $y_{ee}$ has been constrained by LEP experiments.
In addition to direct constraints from collider experiments,
the $\kappa_{e\mu}$ coupling is also constrained by precision measurements.
We also study the constraints from LNV processes
such as $nn\to pp e^- e^-$ (neutrinoless double beta decay)
and $pp \mu^- \to nn e^+$ in atomic nuclei.
We also describe the bounds on the doubly charged scalar mass at the LHC,
which depend on the scalar decay modes,
while the decay modes are not fixed in this paper.
We also mention the prospect of an $ep$ collider as another major collider.

\subsection{Muonium-to-antimuonium transition}

The doubly charged scalars
can generate the Mu-to-$\overline{\rm Mu}$ transition
\cite{Chang:1989uk,Swartz:1989qz}.
The transition operators induced by the doubly charged scalars are
\begin{equation}
-{\cal L} \supset \frac{4G_1}{\sqrt2} (\bar\mu \gamma_\alpha P_L e)(\bar\mu \gamma^\alpha P_L e)
+ \frac{4G_2}{\sqrt2} (\bar\mu \gamma_\alpha P_R e)(\bar\mu \gamma^\alpha P_R e)
+ {\rm H.c.},
\label{G1-G2-operators}
\end{equation}
where $P_{L}$ and $P_{R}$ are chiral projection operators.
The $\kappa_{ab}$ and $g_{ab}$ couplings in
Eqs.~(\ref{lagrangian-1}) and (\ref{lagrangian-2})
induce
\begin{equation}
G_1 = -\frac{\kappa_{ee}\kappa_{\mu\mu}^*}{4\sqrt2 m_\Delta^2},
\quad
G_2 = -\frac{g_{ee}g_{\mu\mu}^*}{4\sqrt2 m_k^2}.
\label{G1andG2}
\end{equation}
The experimental bounds on the Mu-to-$\overline{\rm Mu}$ transition
can be interpreted as \cite{Willmann:1998gd}
\begin{equation}
\frac{|G_1 + G_2|}{G_F} < 3.0 \times 10^{-3},
\end{equation}
if there are no other operators that generate the transition
(see Ref.~\cite{Fukuyama:2021iyw} for more detail).

\subsection{LEP bounds}
\label{sec3.2}

The LEP experiments can constrain the doubly charged scalar couplings
to electrons.
The $y_{ee}$ couplings are constrained by
the $e^+ e^- \to e^+ e^-$ Bhabha scattering data,
and the $y_{e\mu}$ couplings are constrained
by the $e^+ e^- \to \mu^+\mu^-$ scattering data.
Ref.~\cite{ALEPH:2013dgf}
analyzes the bounds on NP contributions from the LEP experimental data.
The bounds are shown
by classifying whether the NP contributions are constructive or destructive with respect to the SM ones.

The $e^+ e^- \to e^+ e^-$ process
has $s$- and $t$-channel exchange diagrams of $\gamma$ and $Z$ bosons in the SM,
and the $u$-channel exchange diagram of the doubly charged scalars using $y_{ee}$ couplings.
The main contribution in the SM comes from the $t$-channel exchange
(which has the opposite sign to the $s$-channel exchange).
The contribution from the doubly charged scalars
is destructive with respect to the SM contribution.
Therefore, we use the NP bounds for the destructive case given in Ref.~\cite{ALEPH:2013dgf}.
We note that it seems that
Ref.~\cite{Nomura:2017abh} uses the NP bounds for the constructive case
to obtain the bounds on the doubly charged scalar couplings.
We find that the LEP bounds imply
\begin{equation}
\frac{|G_1|}{G_F} < 2.98 \times 10^{-3} \times \left|\frac{\kappa_{\mu\mu}}{\kappa_{ee}}\right|,
\qquad
\frac{|G_2|}{G_F} < 3.05 \times 10^{-3} \times \left|\frac{g_{\mu\mu}}{g_{ee}}\right|.
\end{equation}
It is interesting to note that
the LEP bounds are as strong as the experimental bounds on the Mu-$\overline{\rm Mu}$ transition.

The $e^+ e^- \to \mu^+ \mu^-$ process has only $s$-channel exchange diagrams of $\gamma$ and $Z$ bosons in the SM,
and the $u$-channel exchange of the doubly charged scalar using $y_{e\mu}$ couplings.
The contribution from the doubly charged scalars in this case is constructive with respect to the SM contribution.
We obtain the following bounds on the couplings:
\begin{equation}
|\kappa_{e\mu}| < 0.205 \times \frac{m_\Delta}{1\ {\rm TeV}},
\qquad
|g_{e\mu}| < 0.216 \times \frac{m_k}{1\ {\rm TeV}}.
\label{LEPbound:emu}
\end{equation}

\subsection{Relation between {\boldmath$\kappa_{e\mu}$} coupling and {\boldmath $W$} boson mass}

Since the $\kappa_{e\mu} (= \kappa_{\mu e})$ coupling can modify the Fermi constant,
it influences the electroweak precision observables.

The Lagrangian we consider is
\begin{align}
-{\cal L} &=
(2 \kappa_{e\mu} \overline{\ell_e^c} \ell_\mu \Delta + {\rm H.c.}) + m_\Delta^2 {\rm tr}\,\Delta \Delta^\dagger \nonumber \\
&\supset
(-\sqrt2 (\kappa_{e\mu} \overline{\nu_e^c} \mu_L \Delta^+
+ \kappa_{e\mu}^* \overline{e_L} \nu_\mu^c \Delta^-) + {\rm H.c.}) + m_\Delta^2 \Delta^+ \Delta^-.
\end{align}
Integrating out $\Delta^+$,
we obtain
\begin{equation}
-{\cal L} \supset - 2 \frac{|\kappa_{e\mu}|^2}{m_\Delta^2} (\overline{e_L} \nu_\mu^c)  (\overline{\nu_e^c} \mu_L) + {\rm H.c.}
=
- \frac{|\kappa_{e\mu}|^2}{m_\Delta^2} (\overline{e_L} \gamma \nu_e)  (\overline{\nu_\mu} \gamma \mu_L) + {\rm H.c.},
\end{equation}
where the indices of the gamma matrices being contracted are omitted.
Using Fierz transformation, one obtains
\begin{equation}
(\overline{e_L} \nu_\mu^c)  (\overline{\nu_e^c} \mu_L)
= -\frac12 (\overline{e_L} \gamma \mu_L)  (\overline{\nu_e^c} \gamma \nu_\mu^c)
= \frac12 (\overline{e_L} \gamma \mu_L)  (\overline{\nu_\mu} \gamma \nu_e)
= \frac12 (\overline{e_L} \gamma \nu_e)  (\overline{\nu_\mu} \gamma \mu_L).
\end{equation}
The muon decay four-fermion coupling,
\begin{equation}
-{\cal L} \supset \frac{4G_F}{\sqrt2} (\overline{e_L} \gamma \nu_e)  (\overline{\nu_\mu} \gamma \mu_L),
\end{equation}
is modified at tree level,
\begin{equation}
\frac{4G_F}{\sqrt2} = \frac{4G_F^0}{\sqrt2} - \frac{|\kappa_{e\mu}|^2}{m_\Delta^2}.
\end{equation}
The SM tree-level $W$ boson mass using the accurately measured quantities $\alpha$, $m_Z$, and $G_F$
is
\begin{equation}
m_W^2 = \frac12 m_Z^2 \left(1+\sqrt{1- \frac{4\pi \alpha}{\sqrt2 G_F M_Z^2}}\right),
\end{equation}
which comes from
\begin{equation}
\sin^2 2 \theta_W = \frac{4\pi \alpha}{\sqrt2 G_F M_Z^2}.
\end{equation}
Due to the negative contribution to $G_F$,
the $W$ boson mass can shift to a larger value compared to
the SM prediction using the measured value of $G_F$.
According to the analysis of the CDF $W$ boson mass anomaly in Ref.~\cite{Endo:2022kiw},
the 1$\sigma$ region is
\begin{equation}
0.20 < \frac{|\kappa_{e\mu}|}{m_\Delta} < 0.24 \ ({\rm TeV}^{-1}).
\label{CDF-W-anomaly-emu}
\end{equation}

In the case of the $f_{e\mu}$ coupling in Eq.~(\ref{fab}),
the Lagrangian is
\begin{align}
-{\cal L} &=
(2 f_{e\mu} \overline{\ell_e^c} \ell_\mu h^+ + {\rm H.c.}) + m_h^2 h^+ h^- \nonumber \\
&\supset
(2f_{e\mu} \overline{\nu_e^c} \mu_L h^+
- 2f_{e\mu}^* \overline{e_L} \nu_\mu^c h^- + {\rm H.c.}) + m_h^2 h^+ h^-.
\end{align}
Integrating out $h^+$,
we obtain
\begin{equation}
-{\cal L} \supset \frac{4|f_{e\mu}|^2}{m_h^2} (\overline{e_L} \nu_\mu^c)  (\overline{\nu_e^c} \mu_L) + {\rm H.c.}
=
\frac{2|f_{e\mu}|^2}{m_h^2} (\overline{e_L} \gamma \nu_e)  (\overline{\nu_\mu} \gamma \mu_L) + {\rm H.c.}
\end{equation}
The modification of the $W$ boson mass has the opposite sign to that in the case of $\kappa_{e\mu}$.

If the $W$ boson mass anomaly fades out or is explained by other effects,
$\kappa_{e\mu}$ should be small
or the modification of the Fermi constant should be canceled between the
$\kappa_{e\mu}$ and $f_{e\mu}$ contributions.

\subsection{{\boldmath$nn \to pp e^- e^-$} and {\boldmath$pp \mu^- \to nn e^+$}}
 
We suppose that neutrinos are Majorana.
Long-range neutrino exchanges can induce
$nn \to pp e^- e^-$ (neutrinoless double beta decay, $0\nu 2\beta$) and $pp \mu^- \to nn e^+$
from
\begin{equation}
{\cal L} = \frac{4G_F}{\sqrt2} (\overline{u_L} d_R) ( \epsilon_{ee} \overline{e_L} \nu_e + \epsilon_{e\mu}
\overline{e_L} \nu_\mu + \epsilon_{\mu e} \overline{\mu_L} \nu_e)
+ \frac{4G_F}{\sqrt2} (\overline{u_L} \gamma d_L) (\overline{e_L} \gamma \nu_e + \overline{\mu_L} \gamma \nu_\mu)
+ {\rm H.c.}
\end{equation}
One can also consider $(\overline{u_R} d_L)$ pieces instead of $(\overline{u_L} d_R)$ in the operators.

We consider the following pieces of the Lagrangian for the triplet scalar model:
\begin{equation}
-{\cal L} \supset \lambda_{11} \overline{u_L} d_R \eta^+ - \sqrt2 \kappa_{ee} \overline{\nu_e} e_L \Delta^+
- \sqrt2 \kappa_{e\mu} (\overline{\nu_e} \mu_L + \overline{\nu_\mu} e_L) \Delta^+ + {\rm H.c.}
\end{equation}
The scalars $\eta^+$ and $\Delta^+$ are mixed by a mixing angle $\phi$,
and the mass eigenstates are $h_1^+$ and $h_2^+$.
Integrating out the scalars with masses $m_{h_{1,2}^+}$, we obtain
\begin{equation}
\frac{4G_F}{\sqrt2} \epsilon_{ab} = - \sqrt2 \kappa_{ab}^* \lambda_{11} c_\phi s_\phi \left(\frac{1}{m_{h_1^+}^2} - \frac1{m_{h_2^+}^2}\right).
\end{equation}

With a nuclear matrix element (NME) in Ref.~\cite{Deppisch:2012nb},
the $0\nu 2\beta$ bound $T_{1/2} > 3.8 \times 10^{26}$ yrs for Xe at KamLAND-Zen \cite{KamLAND-Zen:2024eml}
gives
\begin{equation}
|\epsilon_{ee}| < 1.6 \times 10^{-9}.
\end{equation}
Consequently, if the Mu-to-$\overline{\rm Mu}$ transition is large (namely, $\kappa_{ee}$ is sizable) and
the mixing $\sin 2\phi$ is large enough to observe the LNV at the $\mu^+\mu^+$ collider,
the inert doublet coupling $\lambda_{11}$ to the $u,d$ quarks
must be tiny.

We next study $pp\mu^- \to nn e^+$ for $\kappa_{e\mu} \neq 0$.
We suppose $\kappa_{ee} = 0$ in this case to eliminate the $\mu^- \to e^-$ LFV processes.
The rate of $pp\mu^- \to nn e^+$ in Refs.~\cite{Domin:2004tk,Lee:2021hnx} is
given as
\begin{equation}
R^{\mu e^+} = 2.6 \times 10^{-22} \times
\left\{\begin{array}{l}
\left|\frac{(m_\nu)_{\mu e}}{m_e}\right|^2 |{\cal M}_\nu|^2, \\
\left| (M_N^{-1})_{\mu e}  m_p \right|^2 |{\cal M}_N|^2.
\end{array}
\right.
\end{equation}
We write the rate induced by the $\epsilon_{e\mu}$, $\epsilon_{\mu e}$
as
\begin{equation}
R^{\mu e^+} = 2.6 \times 10^{-22} \times |\epsilon_{e\mu} {\cal M}_{\epsilon 1} +\epsilon_{\mu e} {\cal M}_{\epsilon 2}|^2.
\end{equation}
Because the $\kappa$ coupling matrix is symmetric, we can write
\begin{equation}
\epsilon_{e\mu} {\cal M}_{\epsilon 1} + \epsilon_{\mu e} {\cal M}_{\epsilon 2} = 2 \epsilon_{e\mu} {\cal M}_{\epsilon},
\end{equation}
where ${\cal M}_{\epsilon}$ is an average of the NMEs,
${\cal M}_{\epsilon} = ({\cal M}_{\epsilon 1} + {\cal M}_{\epsilon 2})/2$.
Using a momentum transfer $q$,
one can roughly estimate
\begin{align}
\frac{{\cal M}_\nu}{m_e} :  {\cal M}_\epsilon  :  m_p {\cal M}_N
\sim \frac{1}{q^2} : \frac{1}{q} : 1.
\end{align}
The NME can be roughly written as
\begin{equation}
{\cal M}_\epsilon \sim \frac{m_p}{q} {\cal M}_N, \qquad
q^2 = m_e m_p \frac{{\cal M}_N}{{\cal M}_\nu},
\end{equation}
and
\begin{equation}
{\cal M}_{\epsilon} \sim \sqrt{\frac{m_p}{m_e} {\cal M}_\nu {\cal M}_N}.
\end{equation}
The NME ${\cal M}_\epsilon$ is estimated to be $O(10)$--$O(100)$
using NMEs in Refs.~\cite{Domin:2004tk,Lee:2021hnx}.
Then, the rate is obtained as
\begin{equation}
R^{\mu e^+} = 1.0 \times 10^{-17} \times |\epsilon_{e\mu}|^2 \left|\frac{{\cal M}_\epsilon}{100} \right|^2,
\label{Rmue+}
\end{equation}
and
\begin{equation}
R^{\mu e^+} \simeq  10^{-24} \times
\left|\frac{\lambda_{11}}{0.1}
\frac{\kappa_{e\mu}}{0.2}
\frac{s_\phi c_\phi}{0.1}
\frac{(500\ {\rm GeV})^2}{m_{h_1^+}^2}\right|^2
\left|\frac{{\cal M}_\epsilon}{100} \right|^2.
\end{equation}
The coupling $\lambda_{11}$ should be restricted by
dijet resonant events at the LHC and pion decays.
Since the rate is a few orders of magnitude smaller than the future experimental goals
\cite{Berryman:2016slh,Sato:2022vny} from the future muon source of NuFact \cite{NeutrinoFactory:2002azy},
we do not consider the constraints here.
We note that
one obtains a roughly $10^5$ times larger value than Eq.~(\ref{Rmue+})
if an estimate given in Ref.~\cite{Sato:2022vny} is used.

\subsection{LHC bounds}

The LHC has the potential to produce pairs of
doubly charged scalars:
\begin{equation}
q \bar q \to Z/\gamma^* \to S^{++} S^{--}.
\end{equation}
The bound on the scalar mass depends on
Br($S^{++} \to l^+ l^{\prime +}$) ($l,l' = e$ or $\mu$).
Due to the difference in the $Z$ boson couplings to the scalars,
the bounds on the triplet and singlet scalars are a little different.
If Br($S^{++} \to l^+ l^{\prime +}$) is 100\%,
the mass bounds are about a TeV \cite{Aaboud:2017qph,ATLAS:2022pbd}.
If the branching ratio is lower,
a lighter scalar can be allowed.
Various works on the doubly charged scalar at colliders can be found \cite{Chiang:2012dk,Chen:2014qda,Dev:2018sel,Crivellin:2018ahj,BhupalDev:2018tox,Padhan:2019jlc,Fuks:2019clu,Enomoto:2021fql,Bai:2021ony,Ashanujjaman:2021txz,Dey:2022whc,Guedes:2025jqu}.
In this process, the doubly charged scalars are produced through electroweak interactions,
and thus the scalar couplings to leptons $y_{ab}$ cannot be directly measured,
although ratios of the couplings such as $y_{ee}/y_{\mu\mu}$ will be able to be measured
from the scalar decays.

Recently, the CMS collaboration reported
the observation of lepton-lepton scattering in $pp$ collisions,
including same-sign $\mu^\pm\mu^\pm$ and $e^\pm\mu^\pm$ final states \cite{CMS:2026bxj}.
The analysis is based on lepton-induced processes in the proton and provides evidence for
the lepton content of the proton.
A useful physical picture is that the proton contains a small lepton component generated
by photon splitting,
$p\to p^{(*)}+\gamma^*$, $\gamma^*\to l^+ l^-$.
One of the leptons participates in the hard scattering, while the other lepton is typically
very forward and may escape the central detector.
The doubly charged scalar can be singly produced resonantly by lepton-lepton scattering at the LHC.
In this case,
the scalar couplings to leptons can also be constrained.
The single production of the doubly charged scalar has been studied in Ref.~\cite{Buonocore:2020nai}.
While it seems that no strong constraint can be obtained at the present luminosity
because of the small lepton distribution in the proton,
much larger luminosities in the future will be able to constrain the model parameters.

The lepton collider has the merits of clean signals and precision measurements of couplings.
LFV and LNV can be directly observed
at lepton colliders, while such violation may not be obvious because of the lack of detection
of the very forward leptons at the LHC.
Cooperation between hadron and lepton colliders may be important to measure the scalar couplings
to leptons in the future.

\subsection{Comment on {\boldmath$e^- p$} collider}

Finally, we discuss the possible LNV processes at $e^- p$ colliders:
\begin{equation}
e^- p \to e^+ X,  \qquad e^- p \to \mu^+ X,
\end{equation}
for $\Delta L_e = -2, \Delta L_\mu=0$ and $\Delta L_\mu = \Delta L_e = -1$ processes,
respectively,
where $X$ does not contain leptons.

At the $\mu^+ e^-$ colliders,
\begin{equation}
\mu^+ e^- \to \bar\nu_\mu e^+ jj, \qquad \mu^+ e^- \to \mu^- \nu_e jj
\end{equation}
processes are also possible for the $(\Delta L_e, \Delta L_\mu) = (-2,0), (0,2)$ LNV processes.
On the other hand,
the $\Delta L_\mu = \Delta L_e = -1$ LNV processes at the lowest order are
\begin{equation}
\mu^+ e^- \to \bar\nu_\mu \mu^+ jj, \qquad \mu^+ e^- \to e^- \nu_e jj.
\end{equation}
However, the SM processes via $WZ$ scattering,
$\mu^+ e^- \to \mu^+ \nu_e W^-, \bar\nu_\mu e^- W^+$,
can mimic the $\Delta (L_e - L_\mu) = 0$ LNV processes.
In the case of $e^- p \to \mu^+ X$,
to mimic this process in the SM,
two extra neutrinos would have to be emitted, which would likely be suppressed by phase space.
Therefore, in order to observe the $\Delta (L_e - L_\mu) = 0$ LNV process,
the $e^- p$ collider has an advantage over the $\mu^+ e^-$ collider.

HERA is the only lepton-proton collider in the world to date.
The center-of-mass energy at HERA is about 320 GeV,
and a stringent bound on the doubly charged scalar mass cannot be obtained.
At the next-generation Electron-Ion Collider \cite{Accardi:2012qut},
the signal with $\mu^+$ emission may be observed in the future.

\section{Setups for colliders and detectors}
\label{sec4}

This section summarizes the collider-related topics relevant to studying physics at a muon collider:
the polarization of the muon beam, cuts for event generation, and
missing-energy vetoes for cLFV processes.

\subsection{Muon polarization at lepton colliders}

We first describe the polarization of the $\mu^+$ beam at the collider.
The positive muons are produced from pion decays at the surface of a target onto which high-energy protons are incident.
The negative pions are captured on target nuclei, and thus the so-called surface muon beam consists of positive muons only.
The surface muons stop in a target, and muonium is emitted from the target.
The muonium is ionized by a laser, after which cold muons are generated.
The cold muons are then reaccelerated to produce a muon beam with a small energy spread.
The surface muons are often referred to as ``left-handed'' because they originate from pion decays via the weak interaction. 
For the positive muons considered here, this corresponds to the charge-conjugate state, 
$(\mu_L)^c=(\mu^c)_R$.
While different conventions for polarization are used, we define the muon polarization in this paper as
\begin{equation}
P = \pm \frac{\langle {\bm \sigma} \cdot {\bm p} \rangle}{| {\bm p} | },
\end{equation}
for $\mu^{\mp}$, with the upper and lower signs taken in the same order,
where ${\bm \sigma}$ is the spin operator and ${\bm p}$ is the muon momentum.
Thus, the polarization of both $\mu^{\pm}$ produced from the decays of $\pi^{\pm}$ via the weak interaction,
$\pi^{\pm} \to \mu^{\pm} \nu_\mu(\bar\nu_\mu)$, is $+1$.

Even if the original surface muon beams are fully polarized,
the polarization is reduced just after muonium is formed:
the $|\!\uparrow_\mu \downarrow_e\rangle$ and
$|\!\downarrow_\mu \uparrow_e\rangle$ states are mixed,
while the $|\!\uparrow_\mu \uparrow_e\rangle$ and
$|\!\downarrow_\mu \downarrow_e\rangle$ states are preserved.
The muon polarization is reduced by a factor of $1/2$ in the simplest case.
The polarization of the cold muons may depend on additional details:
(1) the polarization can be further reduced when muonium is formed;
(2) the muon polarization can be modified during the ionization of muonium; and
(3) the mixing of the $|\!\uparrow_\mu \downarrow_e\rangle$ and
$|\!\downarrow_\mu \uparrow_e\rangle$ states
can depend on the magnetic field strength.
In this paper, we assume that the polarization of the $\mu^+$ beam is $P = 1/2$
to explicitly show the numerical results.
It is straightforward to convert the numerical results to those for a different value of $P$.

\subsection{Cuts for event generation}

Future lepton colliders will be much cleaner than hadron colliders,
and high-precision tracking and energy resolution are expected.
Therefore, the generator-level cuts should be modified relative to the LHC-based default cuts
in MadGraph \cite{Alwall:2014hca}.
The detector designs have not yet been fully determined,
and we should keep the generator-level cuts as loose as possible
as long as cuts are required to regulate infrared singularities.

In this paper, we adopt loose cuts on the transverse momenta of leptons ($e^\pm$ or $\mu^\pm$),
photons, and jets:
\begin{equation}
p_T^\ell > 5\ {\rm GeV}, \quad p_T^\gamma > 5\ {\rm GeV}, \quad p_T^j > 10\ {\rm GeV}.
\end{equation}
For the angular separation between two particles
in the pseudorapidity ($\eta$) and azimuthal angle ($\phi$) plane,
$\Delta R \equiv \sqrt{(\Delta \eta)^2 + (\Delta \phi)^2}$,
we use
\begin{equation}
\Delta R_{ll} > 0.05, \quad
\Delta R_{l\gamma} > 0.05,
\quad
\Delta R_{lj} > 0.05,
\quad
\Delta R_{jj} > 0.1.
\end{equation}

Muon colliders have a unique source of background from
$\mu^+ \to e^+ \nu_e \bar\nu_\mu$.
To reduce this background,
it has been proposed that shielding nozzles be placed in the detector near the collision point along the
beam axes \cite{AlAli:2021let,Accettura:2023ked}.
The geometry of the nozzles determines the angular acceptance near the beam axes,
rather than a pseudorapidity cut.
A smaller value of the minimum polar angle allows particles closer to the beam axis to be detected,
and is therefore advantageous for collider phenomenology.
The optimal nozzle geometry is an important issue for future muon colliders.
In this paper, we adopt a conservative value of the minimum polar angle,
$\theta^{\rm cut} = 10^\circ$
(which corresponds to a pseudorapidity of 2.44).
When it is useful to consider a more aggressive angular acceptance,
we will also present the results for $\theta^{\rm cut} = 5^\circ$
(which corresponds to a pseudorapidity of 3.13).

\subsection{Missing-energy veto for charged lepton flavor violating processes}

We consider processes involving cLFV in this paper,
and search for NP events with fully visible final states.
For such signal events, the sum of the energies of the visible final-state particles
is expected to be close to the center-of-mass energy $\sqrt{s}$.
In contrast, flavor-conserving SM processes that mimic the LFV signal
can involve neutrino emission.
The neutrinos escape detection and carry away part of the collision energy,
resulting in missing energy.
Therefore, requiring the total visible energy to be close to $\sqrt{s}$
can efficiently reject such SM background processes.
One advantage of a lepton collider is that,
except when most particles escape along the beam axis and cannot be detected,
there is no need to apply sophisticated cuts to reduce the SM background.

\section{$\Delta L_\mu = -\Delta L_e = 2$ LFV processes in the lepton colliders}
\label{sec5}
In this section,
we explore the $\mu^+ \mu^+ \to e^+ e^+$ and $\mu^+ \mu^+ \to e^+ e^+ V$ ($V=\gamma, Z$)
processes induced by the doubly charged scalar at the $\mu^+ \mu^+$ collider.
One of the merits of lepton colliders is that the couplings can be measured accurately.
To measure the couplings, it is important to know the scalar mass
and whether the scalar belongs to a triplet or a singlet.
We discuss how the scalar representation can be distinguished from the 
$\Delta L_\mu = -\Delta L_e = 2$ LFV processes,
and the model parameters can be determined
in combination with the Mu-to-$\overline{\rm Mu}$ transition.

\subsection{$\mu^+ \mu^+ \to e^+ e^+$}

The simplest $\Delta L_\mu = -\Delta L_e = 2$ process is
$\mu^+ \mu^+ \to e^+ e^+$ at the $\mu^+\mu^+$ collider.
The helicity cross section for fully polarized muons is given by
\begin{equation}
\sigma(\mu_H \mu_H \to e_H e_H) = \frac{|y_{ee}y_{\mu\mu}|^2}{2\pi} \frac{s}{(s-m_S^2)^2},
\label{sigma-muH}
\end{equation}
where $H = L,R$.
Here, we use the chirality-like symbols $L$ and $R$ to denote the helicity
states, since the sign of the helicity may be confused with the electric charge.
Because we neglect the fermion masses,
we do not need to distinguish between helicity and chirality.
As noted before, we denote $y=\kappa$, $m_S = m_\Delta$ for the triplet scalar,
and $y=g$, $m_S = m_k$ for the singlet scalar.
The cross sections for triplet- and singlet-scalar mediation
are given by
\begin{align}
\sigma_\Delta (\mu^+ \mu^+ \to e^+ e^+) &= \frac{(1 + P)^2}{4} \sigma(\mu_L \mu_L \to e_L e_L), \\
\sigma_k (\mu^+ \mu^+ \to e^+ e^+) &= \frac{(1 - P)^2}{4} \sigma(\mu_R \mu_R \to e_R e_R).
\end{align}

Here, we present an example of a flavor-conserving SM process that mimics the LFV process
described in the previous section.
A possible SM process that mimics $\mu^+ \mu^+ \to e^+ e^+$ is
\begin{equation}
\mu^+ \mu^+ \to e^+ e^+ \nu_e \nu_e \bar{\nu}_\mu \bar{\nu}_\mu,
\end{equation}
since the neutrinos are not detected.
The cross section of this SM process is calculated to be 1.9 fb using MadGraph.
At a lepton collider,
the energies of the final-state positrons can, in principle, be measured accurately.
The sum of the positron energies in the signal process,
$\mu^+ \mu^+ \to e^+ e^+$,
is expected to be $\sqrt{s}$ within the energy resolution.
In contrast, the SM background process has missing energy carried away by the neutrinos.
If we require the energy of each neutrino to be less than 100 GeV in MadGraph,
which is a relatively loose requirement for $\sqrt{s} = 2$ TeV,
the cross section of the SM process is already reduced to the order of a zeptobarn.
Therefore, the missing-energy veto based on the measured sum of the positron energies 
is expected to effectively reject this SM background.

From Eq.~(\ref{G1andG2}),
the product of the couplings, $|y_{ee} y_{\mu\mu}|$, is given by
the coefficient of the Mu-to-$\overline{\rm Mu}$ transition operator
and the scalar mass:
\begin{equation}
|y_{ee} y_{\mu\mu}| = 4\sqrt2 m_S^2 |G_i| ,
\end{equation}
where $i=1$ for the triplet and $i=2$ for the singlet.
We plot $\sigma(\mu^+ \mu^+ \to e^+ e^+)$ as a function of the scalar mass
for fixed $|G_i|/G_F$.
It is easy to see that the cross section is proportional to $(|G_i|/G_F)^2$,
except when the scalar mass is close to $\sqrt{s}$.
At $m_S \sim \sqrt{s}$,
one needs to introduce the total decay width $\Gamma_S$ of the doubly charged scalar,
\begin{equation}
\frac{1}{(s-m_S^2)^2} \to \frac{1}{(s-m_S^2)^2 + m_S^2 \Gamma_S^2}.
\end{equation}
The partial decay width is given by
\begin{equation}
\Gamma (S^{++} \to l_a^+ l_b^+) = \frac{|y_{ab}|^2 m_S}{4\pi (1+\delta_{ab})}.
\end{equation}
In order to plot the figures around $m_S \sim \sqrt{s}$,
we assume
that $S^{++}$ decays only to $e^+ e^+$ and $\mu^+ \mu^+$,
and that $y_{ee} = y_{\mu\mu}$ for simplicity,
which gives
\begin{equation}
\Gamma_S =  \frac{(|y_{ee}|^2 + |y_{\mu\mu}|^2) m_S}{8\pi} =  \frac{\sqrt2 |G_i| m_S^3}{\pi}.
\label{assume-tdw}
\end{equation}
One can see that the cross section at $m_S = \sqrt{s}$
does not depend on $|G_i|$ under this assumption.

\begin{figure}[t]
\centering
\includegraphics[width=0.85\textwidth]{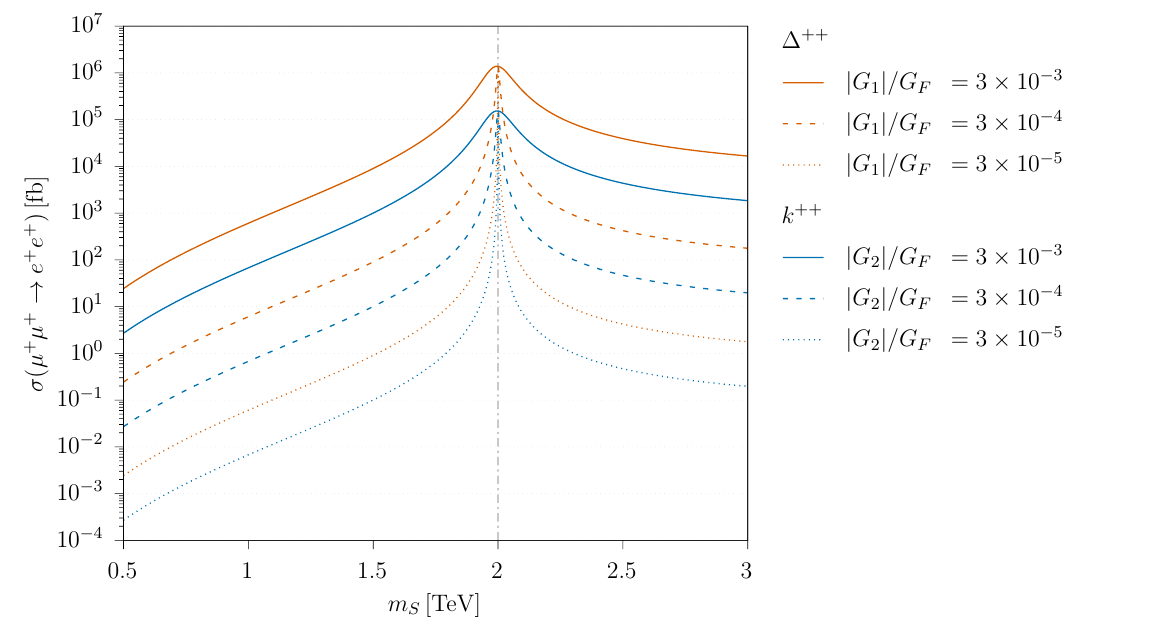}
\caption{%
Cross section of $\mu^+\mu^+\to e^+e^+$ as a function of the
doubly charged scalar mass in the case of $\sqrt{s}=2\ \mathrm{TeV}$ and $P=1/2$.
Orange curves correspond to $\Delta^{++}$ mediation and blue
curves to $k^{++}$ mediation.
The solid, dashed, and dotted
curves represent $|G_i|/G_F=3\times10^{-3}$, $3\times10^{-4}$, and
$3\times10^{-5}$, respectively.
The behavior around $m_S \sim \sqrt{s}$ depends on the decay width of the scalar.
We suppose that the doubly charged scalar decays
only to $e^+e^+$ and $\mu^+\mu^+$
with equal magnitudes as explained in the text.}
\label{fig:xsec-mm2ee-mass}
\end{figure}

Fig.~\ref{fig:xsec-mm2ee-mass} shows the cross sections
of the $\mu^+ \mu^+ \to e^+ e^+$ process
as a function of the scalar mass for $\sqrt{s} = 2$ TeV and
$P = 1/2$.
We display three benchmark values of $|G_1|/G_F$ for the triplet-mediated case, $\sigma_\Delta$,
and the corresponding values of $|G_2|/G_F$ for the singlet-mediated case, $\sigma_k$.
The ratio of $\sigma_\Delta$ to $\sigma_k$
is $(1+P)^2/(1-P)^2$, which is nine for $P=1/2$,
for the same scalar masses and $|G_i|$.

The plot in Fig.~\ref{fig:xsec-mm2ee-mass}
tells us that
we can obtain the allowed values of the scalar mass
from a measurement of the cross section of the
$\mu^+ \mu^+ \to e^+ e^+$ process at the collider
if
the Mu-to-$\overline{\rm Mu}$ transition is observed
and $|G_i|$ is measured.
We note
that it is impossible to determine whether the doubly charged scalar is a singlet or a triplet
from the $\mu^+ \mu^+ \to e^+ e^+$ process alone
(unless the experiment is performed by varying the beam polarization).

Obviously,
for $m_S \gg \sqrt{s} $,
one finds
\begin{equation}
\sigma(\mu_H \mu_H \to e_H e_H) \simeq  \frac{s}{2\pi} \frac{|y_{ee}y_{\mu\mu}|^2}{m_S^4} = \frac{s}{2\pi} (4\sqrt{2} |G_i|)^2,
\end{equation}
from Eq.~(\ref{sigma-muH}),
and the cross section with fixed $|G_i|$ is insensitive to the scalar mass.

Although the cross section of the $\mu^+ e^- \to \mu^- e^+$ process
in a $\mu^+ e^-$ collider
can also be combined with the Mu-$\overline{\rm Mu}$ transition
to probe the scalar mass in a similar way,
the planned center-of-mass energy, $\sqrt{s} = 346$ GeV, of the $\mu^+ e^-$ collider
is not suitable for probing scalar masses of order TeV.
As we will see in the next subsection,
the $\mu^+ \mu^+$ collider has another process that is well suited
for probing the scalar mass.

\subsection{$\mu^+ \mu^+ \to e^+ e^+ \gamma/Z(\to jj)$}

The $\mu^+ \mu^+ \to e^+ e^+ V$ ($V=\gamma, Z$) processes
provide important probes of LFV with
$\Delta L_\mu = -\Delta L_e = 2$.

\begin{figure}[t]
\centering
\includegraphics[width=0.95\textwidth]{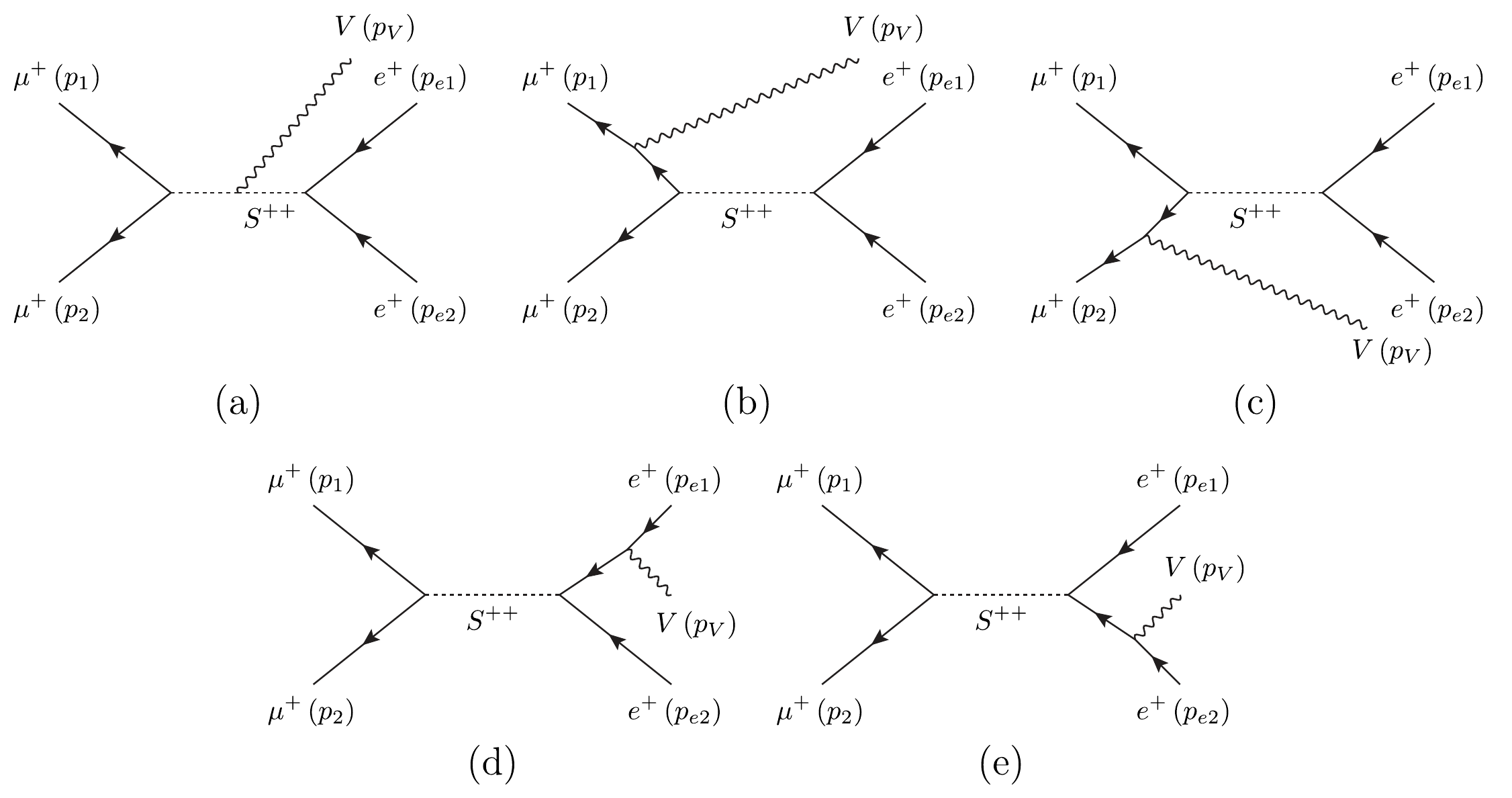}
\caption{Tree-level diagrams for
$\mu^+\mu^+\to e^+e^+V$ with $V=\gamma,Z$, mediated by a doubly
charged scalar $S^{++}=k^{++},\Delta^{++}$.
Diagram (a) represents scalar-line radiation, diagrams (b) and (c) represent
initial-state radiation, and diagrams (d) and (e) represent final-state
radiation.
The incoming-muon momenta are denoted by $p_1$ and $p_2$,
the outgoing-positron momenta by $p_{e1}$ and $p_{e2}$, and the
vector-boson momentum by $p_V$.}
\label{fig:diagrams-mm2eeV}
\end{figure}

The tree-level diagrams of these processes are shown in
Fig.~\ref{fig:diagrams-mm2eeV}, where $S^{++}$ denotes either
$k^{++}$ or $\Delta^{++}$.
Diagram (a) represents vector-boson radiation from the doubly charged scalar line,
diagrams (b) and (c) represent initial-state radiation from either incoming
muon, and diagrams (d) and (e) represent final-state radiation from either
outgoing positron.

For
$m_S < \sqrt{s} - m_V$,
diagrams (a), (b), and (c) are more important than diagrams (d) and (e)
due to resonances in the doubly charged scalar propagators.
In this case, in the narrow-width approximation (NWA), in which the total decay width of the scalar is narrow,
one obtains
\begin{equation}
\sigma (\mu^+ \mu^+ \to e^+ e^+ V) \simeq \sigma (\mu^+ \mu^+ \to S^{++} V)\, {\rm Br} (S^{++} \to e^+ e^+).
\end{equation}
To describe the diagrams in a little more detail,
the contribution of the transverse modes of the vector boson $V$ from diagram (a) vanishes.
For $V=Z$, there is a longitudinal contribution from diagram (a),
while this contribution is suppressed by $m_Z^2/(s-m_S^2)$
for $m_S < \sqrt{s} - m_Z$.

\begin{figure}[t]
\centering
\includegraphics[width=0.85\textwidth]{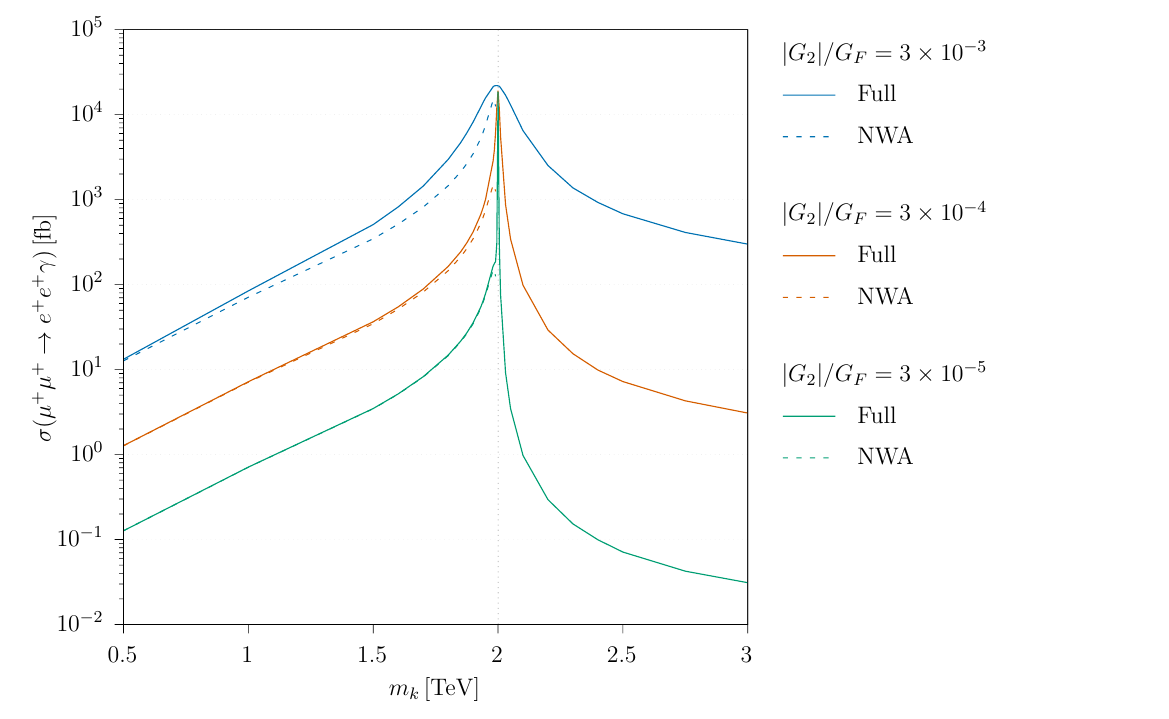}
\caption{%
Cross section of $\mu^+\mu^+\to e^+e^+\gamma$ as a function
of the singlet-scalar mass at $\sqrt{s}=2\ {\rm TeV}$ and $P=1/2$.
The full calculation is shown by solid curves and the
NWA calculation by dashed curves.
Blue, orange, and green denote the benchmark values
$|G_2|/G_F=3\times10^{-3}$, $3\times10^{-4}$, and
$3\times10^{-5}$, respectively.
We impose
$p_T^{e,\gamma}>5\ {\rm GeV}$,
$\Delta R_{ee},\Delta R_{e\gamma}>0.05$,
and a $10^\circ$ angular acceptance.
In order to fix the scalar decay widths, we suppose that the doubly charged scalar decays
only to $e^+e^+$ and $\mu^+\mu^+$
with equal magnitudes as explained in the text.
}
\label{fig:xsec-mm2eeg-mass}
\end{figure}

Figure~\ref{fig:xsec-mm2eeg-mass} shows
the cross section of the
$\mu^+\mu^+\to e^+e^+\gamma$ process in the case of singlet-scalar mediation.
In the case of triplet-scalar mediation,
the cross section is obtained simply by multiplying the singlet-scalar result by
$(1+P)^2/(1-P)^2 = 9$ for $P=1/2$.
The plot is given as a function of the scalar mass for fixed benchmark values of the
effective coupling $|G_2|$ of the Mu-$\overline{\rm Mu}$ transition operator.
We use the same prescription for the decay width as in Eq.~\eqref{assume-tdw},
and then, $\Gamma_k \propto |G_2|$.
For $m_k \ll \sqrt{s}$, the cross section in the NWA is proportional to
$|g_{ee}g_{\mu\mu}|^2/\Gamma_k$ and hence to $|G_2|/G_F$.
The full calculation follows this behavior for the two smaller
benchmarks, $|G_2|/G_F = 3\times10^{-4}$ and $3\times10^{-5}$.
For $|G_2|/G_F=3\times10^{-3}$, the finite scalar width (especially for heavier scalar masses) and the
non-resonant contributions from diagrams (d) and (e) result in a visible departure from the NWA.
Above $m_k=\sqrt{s}$, on-shell associated scalar production is impossible
and the NWA curve terminates, whereas the full calculation retains the
off-shell contribution,
which scales as $(|G_2|/G_F)^2$.

We note in particular the contribution from diagrams (d) and (e).
The contributions from diagrams (d) and (e) contain
large collinear photon radiation,
namely, the emission of a photon at a very small angle relative to a positron.
This can be understood from the positron propagators in the diagrams that emit photons.
We use $\Delta R_{e\gamma}$ cuts to suppress this large logarithmic contribution.
In Fig.~\ref{fig:xsec-mm2eeg-mass}, we use
$\Delta R_{e\gamma}^{\rm cut} = 0.05$.

In the same way as for the $\mu^+ \mu^+ \to e^+ e^+$ process,
one can extract information about the scalar mass from the cross section
of the $\mu^+ \mu^+ \to e^+ e^+\gamma$ process
if the Mu-to-$\overline{\rm Mu}$ transition is observed
and $|G_i|/G_F$ is fixed.
However, while the cross section plotted in Fig.~\ref{fig:xsec-mm2eeg-mass}
is not affected by the total decay width $\Gamma_k$ when $m_k$ is sufficiently greater than $\sqrt{s}$, 
it depends on the total width when $m_k < \sqrt{s}$.
We need additional information for $m_k < \sqrt{s}$,
which can be obtained from
the differential cross section of the $\mu^+ \mu^+ \to e^+ e^+\gamma$ process
as below.

The photon energy distribution of the
$\mu^+ \mu^+ \to e^+ e^+\gamma$ process
has a peak for scalar masses $m_S$ below $\sqrt{s}$.
Let $p_1$ and $p_2$ denote the momenta of the initial-state muons,
and let $p_{e1}$, $p_{e2}$, and $p_\gamma$ denote those of the final-state particles:
\begin{equation}
p_1 + p_2 = p_{e1} + p_{e2} + p_\gamma.
\end{equation}
Then, we obtain
\begin{equation}
m_{ee}^2 \equiv (p_{e1}+p_{e2})^2
= (p_1 + p_2)^2 - 2 p_\gamma \cdot (p_1+ p_2)
= s - 2\sqrt{s} E_\gamma.
\end{equation}
The resonant peak of the photon energy distribution is obtained when
$m_{ee} = m_S$, and therefore,
\begin{equation}
E_\gamma^{\rm peak} = E - \frac{m_S^2}{4E},
\label{eq:egamma-peak}
\end{equation}
where $E$ denotes the initial muon energy in the center-of-mass frame, $E=\sqrt{s}/2$.
As a result, for $m_S < \sqrt{s}$,
the scalar mass can be determined from experimental measurements of the photon energy distribution.

We show the approximate analytic expression for the energy distribution
in Appendix~\ref{appendix:A}.
Around the peak region for $m_k < \sqrt{s}$, the differential cross section of
$\mu^+ \mu^+ \to e^+ e^+\gamma$
mediated by the singlet scalar is given by
\begin{equation}
\frac{d\sigma}{dE_\gamma}
\simeq \frac{(1-P)^2}{4} |g_{ee}g_{\mu\mu}|^2
\frac{e^2}{2\pi^3} \frac{1}{ E E_\gamma}
\frac{(E-E_\gamma) (E^2+(E-E_\gamma)^2)
}{(m_{ee}^2-m_k^2)^2 + m_k^2\Gamma_k^2} I_\theta(\theta_{\rm cut}),
\end{equation}
where $I_\theta(\theta_{\rm cut})$ arises from the polar-angle integration
over $\theta_{\rm cut} < \theta < \pi - \theta_{\rm cut}$,
\begin{equation}
I_\theta(\theta_{\rm cut})
= 2 \left(- \ln \tan\frac{\theta_{\rm cut}}{2}
- \cos\theta_{\rm cut}\right).
  \label{I-theta-cut}
\end{equation}
At the peak, $m_{ee} = m_k$, the differential cross section
is proportional to $(1-P)^2 |g_{ee} g_{\mu\mu}|^2/\Gamma_k^2$ (with fixed $m_k$).
By comparing the peak region with the off-peak region, $\Gamma_k$ can be extracted from the energy distribution.
The width can also be extracted
from the fact that the
differential cross section becomes half of its peak value at
$m_{ee}^2 = m_k^2 \pm m_k \Gamma_k$
(ignoring $O(\Gamma_k/m_k)$ corrections).
Although the quantitative extraction of the width depends on the experimental energy resolution,
it is important that the total decay width can be measured directly
without assuming particular scalar decay modes.

Suppose that the scalar mass is measured 
using the photon energy distribution of the $\mu^+ \mu^+ \to e^+ e^+\gamma$ process
for $m_S < \sqrt{s}$.
If the Mu-to-$\overline{\rm Mu}$ transition is also observed,
the product of the couplings, $|y_{ee}y_{\mu\mu}|$, is determined
and the cross section of $\mu^+ \mu^+ \to e^+ e^+$ can be calculated
depending on whether the scalar is a triplet or a singlet.
Therefore, 
combining these pieces of the experimental information allows us to determine 
the representation of the scalar
(unless the muon beam is unpolarized).
The total decay width of the scalar can be also determined by the cross section
and the peak shape of the photon energy distribution of $\mu^+ \mu^+ \to e^+ e^+ \gamma$.

\begin{figure}[tb]
\centering
\includegraphics[width=0.96\textwidth]{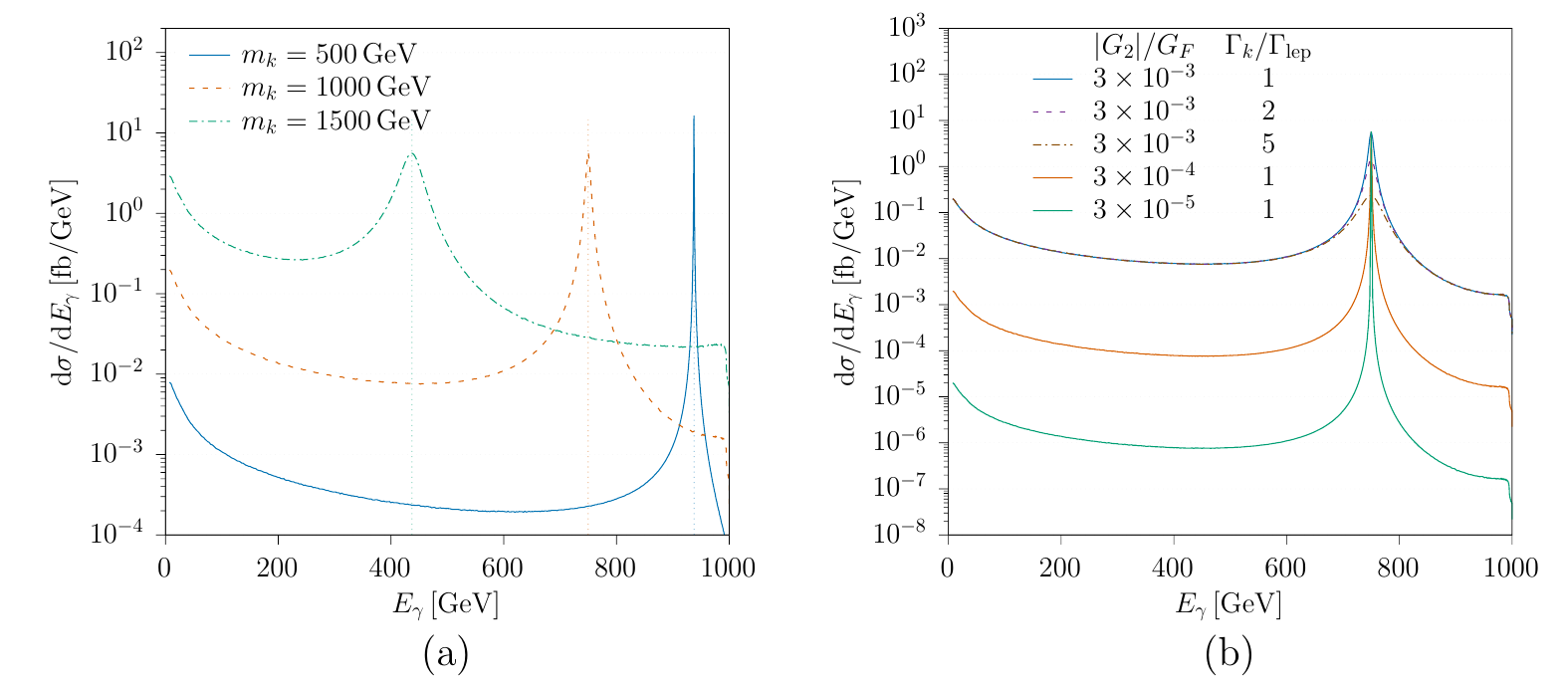}
\caption{%
Photon-energy distribution in
$\mu^+\mu^+\to e^+e^+\gamma$ for a singlet scalar $k^{++}$ at
$\sqrt{s}=2\ {\rm TeV}$ and $P=1/2$.
In panel (a),
$|G_2|/G_F=3\times10^{-3}$ and
${\rm Br}(k^{++}\to e^+e^+)={\rm Br}(k^{++}\to \mu^+\mu^+)=0.50$ are fixed, while
$m_k=500$, $1000$, and $1500$ GeV are compared.
The vertical
dotted lines show Eq.~\eqref{eq:egamma-peak}.
Panel (b) fixes
$m_k=1$ TeV.
The solid blue,
orange, and green curves compare three values of $|G_2|/G_F$
and the total decay width given by Eq.~(\ref{assume-tdw}).
The purple dashed and brown dash-dotted curves keep $|G_2|/G_F=3\times10^{-3}$ fixed and have $\Gamma_k/\Gamma_{\rm lep}=2$ and $5$, respectively, where $\Gamma_{\rm lep}\equiv\Gamma(k^{++}\to e^+e^+)+\Gamma(k^{++}\to\mu^+\mu^+)$.
Both positrons and the
photon satisfy $10^\circ<\theta<170^\circ$; the remaining cuts are those
of Fig.~\ref{fig:xsec-mm2eeg-mass}.}
\label{fig:egamma-spectrum}
\end{figure}

In Fig.~\ref{fig:egamma-spectrum},
we show the photon energy distribution of
$\mu^+ \mu^+ \to e^+ e^+ \gamma$
in the case of singlet-scalar mediation.
In the numerical plots, we present the full tree-level calculations
with all cuts, including $p_T$ and $\Delta R_{ee}$,
while the analytic expression given in Appendix~\ref{appendix:A} ignores some of the cuts,
as described there.
The $p_T$ cut on the positrons affects the distribution just below
$E_\gamma = \sqrt{s}/2$.
The energy distribution depends on the scalar mass $m_k$,
the product of the couplings $|g_{ee}g_{\mu\mu}| = 4\sqrt{2} m_k^2 |G_2|$,
and the total decay width of the scalar $\Gamma_k$.
Panel (a) shows the distribution for three
scalar masses.
Panel (b) illustrates separately how the off-peak
normalization depends on $|G_2|/G_F$, while the peak shape is
sensitive to the total width $\Gamma_k$.

We note the dependence of the cross section of
the $\mu^+ \mu^+ \to e^+ e^+\gamma$ process on the angular acceptance.
Due to the $t$-channel resonance in $\mu^+ \mu^+ \to S^{++} \gamma$
in diagrams (b) and (c) in Fig.~\ref{fig:diagrams-mm2eeV},
the cross section is sensitive to $\theta_{\rm cut}$ for $m_S < \sqrt{s}$.
The angular integration can be completely separated as given in
Eq.~(\ref{I-theta-cut}):
\begin{equation}
I_\theta(\theta_{\rm cut} = 5^\circ) \simeq 4.3, \qquad
I_\theta(\theta_{\rm cut} = 10^\circ) \simeq 2.9.
\end{equation}
The contributions from diagrams (d) and (e) are not sensitive to $\theta_{\rm cut}$,
and they become larger than those from diagrams (b) and (c) for $m_S > \sqrt{s}$.
As a result, the cross section for $m_S > \sqrt{s}$ is insensitive to $\theta_{\rm cut}$.
The near-peak region of the energy distribution shown in
Fig.~\ref{fig:egamma-spectrum}
is also sensitive to $\theta_{\rm cut}$,
while the
off-peak regions, $E_\gamma \ll E_\gamma^{\rm peak}$ and
$E_\gamma \gg E_\gamma^{\rm peak}$,
are insensitive to $\theta_{\rm cut}$ since the contribution from diagrams (d) and (e) dominates there.

We next study the $\mu^+ \mu^+ \to e^+ e^+ Z$ process.
While the helicity cross section, which is a function of the coupling $y$ and mass $m_S$,
is the same for singlet- and triplet-scalar mediation in the photon emission process,
it differs for the $Z$ emission process.
The $Z$ boson couplings to the charged leptons $l$
are given by
\begin{equation}
{\cal L} \supset \frac{g_2}{c_W} Z_\alpha
\overline{l} \gamma^\alpha
\left[\left(-\frac12 + s_W^2\right) P_L + s_W^2 P_R\right] l ,
\label{Lag:Zll}
\end{equation}
where $s_W$ and $c_W$ are the sine and cosine of the weak mixing angle,
respectively,
and $g_2$ is the $SU(2)_L$ gauge coupling.
The $Z$ boson couplings to the scalars are
\begin{equation}
{\cal L} \supset
i \frac{g_2}{c_W} Z_\alpha
\left[(1- 2 s_W^2) \Delta^{--} \partial^\alpha \Delta^{++}
- 2 s_W^2 k^{--} \partial^\alpha k^{++}
  \right]
+ {\rm H.c.}
  \label{Lag:ZSS}
\end{equation}
As a result,
the helicity cross section for triplet-scalar mediation
is that for singlet-scalar mediation multiplied by a factor
$(1/2 - s_W^2)^2/s_W^4$.
Unfortunately, $s_W^2$ is approximately $1/4$,
and the helicity cross sections of $\mu^+ \mu^+ \to e^+ e^+ Z$
for triplet- and singlet-scalar mediation
are not drastically different.
However,
$(1/2 - s_W^2)^2/s_W^4$ is about 1.35
for
$s_W^2 = 0.23122$ \cite{ParticleData:2026uvj}
in the $\overline{\rm MS}$ scheme\footnote{
  We note that
  the default value of the weak mixing angle in MadGraph is defined in the on-shell scheme:
  $s_W^2 = 1-m_W^2/m_Z^2 = 0.222247$.
  }.
Thus, the helicity cross sections differ by more than 30\%.

As explained,
if the Mu-to-$\overline{\rm Mu}$ transition is observed
and $|G_i|/G_F$ is determined,
we can obtain information about the scalar mass,
depending on whether the scalar is a triplet or a singlet
(unless $P=0$).
If more than $O(100)$ events of
$\mu^+ \mu^+ \to e^+ e^+ jj$
are observed,
we can distinguish whether the scalar is a triplet or a singlet
at a statistical significance of more than $3\sigma$,
and hence determine the scalar mass as well.
If $m_S < \sqrt{s}$,
the scalar mass can also be determined from the peak of the photon energy distribution
in $\mu^+\mu^+ \to e^+ e^+\gamma$.
In that case, the $\mu^+ \mu^+ \to e^+ e^+ jj$ process
provides a consistency check
on whether the Mu-to-$\overline{\rm Mu}$ transition receives contributions 
from operators other than $G_1$ and $G_2$
given in Eq.~(\ref{G1-G2-operators}).

\begin{figure}[tb]
\centering
\includegraphics[width=0.85\textwidth]{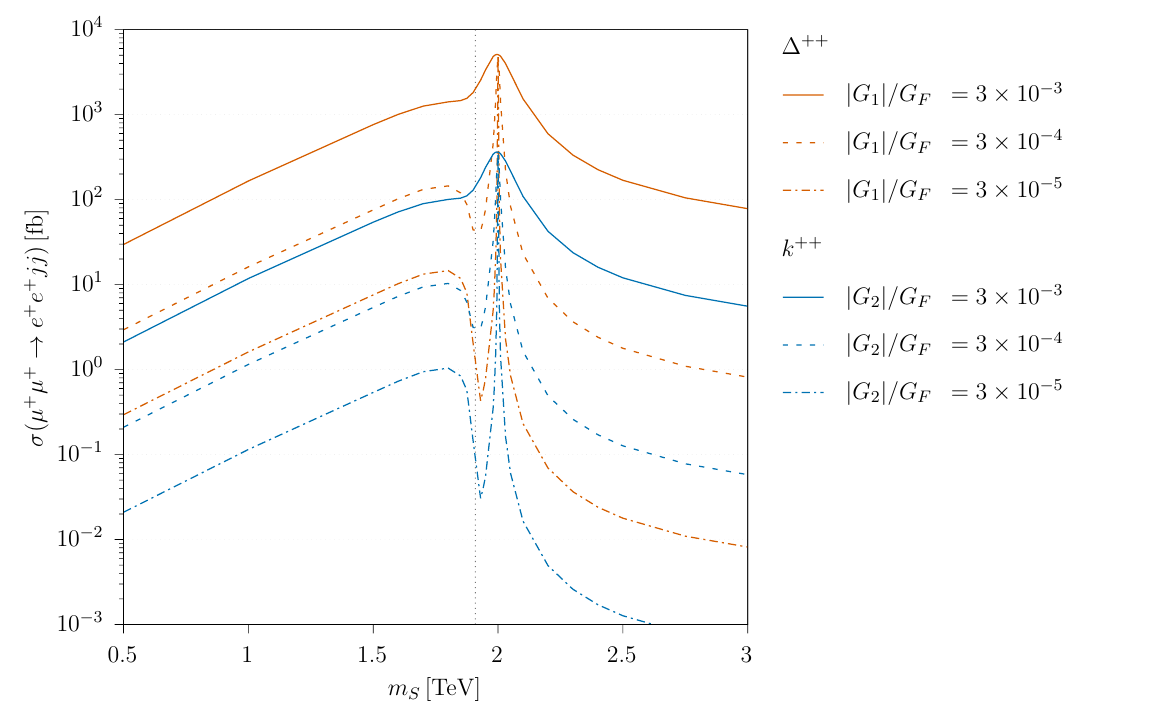}
\caption{%
Cross section of $\mu^+\mu^+\to e^+e^+ jj$ as a
function of the doubly charged scalar mass $m_S$ at
$\sqrt{s}=2\ {\rm TeV}$ and $P=1/2$.
Orange and blue curves denote triplet- and singlet-scalar mediation, respectively;
solid, dashed, and dash-dotted curves correspond to $|G_i|/G_F=3\times10^{-3}$,
$3\times10^{-4}$, and $3\times10^{-5}$.
The triplet curves are
obtained from the singlet calculation using the exact factor in
Eq.~\eqref{eq:triplet-singlet-eeZ-factor}.
We impose
$p_T^e>5\ {\rm GeV}$,
$p_T^j>10\ {\rm GeV}$,
$\Delta R_{ee},\Delta R_{ej}>0.05$, $\Delta R_{jj}>0.1$,
and a $10^\circ$ angular acceptance.
The gray dotted vertical line marks the endpoint of on-shell scalar production,
$m_S=\sqrt{s}-m_Z$.
The scalar decay widths are assumed in the same way as in
Fig.~\ref{fig:xsec-mm2eeg-mass}.}
\label{fig:xsec-mm2eejj-mass}
\end{figure}

In Fig.~\ref{fig:xsec-mm2eejj-mass},
we plot the cross section $\sigma(\mu^+ \mu^+ \to e^+ e^+ jj)$
as a function of the scalar mass $m_S$ for fixed $|G_i|/G_F$.
The assumption for the scalar decay widths used to obtain the cross sections
is the same as before.
The finite $Z$ boson mass
removes the collinear singularity,
and therefore,
NWA holds well below the threshold for on-shell scalar production, $m_S \ll \sqrt{s}-m_Z$,
and the cross section is approximately proportional to $|G_i|/G_F$.
The ratio of the cross sections for $|G_1| = |G_2|$ and the same scalar mass $m_S$ is
\begin{equation}
\frac{\sigma_\Delta (\mu^+ \mu^+ \to e^+ e^+ jj)}
{\sigma_k (\mu^+ \mu^+ \to e^+ e^+ jj)}
= \left( \frac{\frac12 - s_W^2}{s_W^2} \right)^2
\frac{(1+P)^2}{(1-P)^2},
\label{eq:triplet-singlet-eeZ-factor}
\end{equation}
which is about 12.2 for $P=1/2$.

We note that $\mu^+ \mu^+ \to \Delta^+ W^+$, followed by
$\Delta^+ \to e^+ \bar{\nu}_e$, is also possible
in the triplet case.
In this case, however, the final state contains a neutrino,
and the missing-energy veto alone is not sufficient to identify this process
as a cLFV signal.
In fact, $\mu^+ \mu^+ \to e^+ \nu_e \bar{\nu}_\mu \bar{\nu}_\mu W^+$,
which is an SM process,
can mimic this cLFV process.
Since the SM process arises from $WW$ fusion,
$\mu^+ \mu^+ \to \bar{\nu}_\mu \bar{\nu}_\mu W^+ W^+$,
its cross section is not small.
It is important that the representation of the doubly charged scalar
can be determined solely from cLFV processes
that do not involve neutrinos.

Let us summarize this subsection.
There are three NP parameters in the cross section of $\mu^+ \mu^+ \to e^+ e^+ V$:
the product of the couplings $|y_{ee}y_{\mu\mu}|$,
the scalar mass $m_S$, and the scalar decay width $\Gamma_S$.
The cross sections also depend on whether the doubly charged scalar is a component of an
$SU(2)_L$ triplet or a singlet unless the initial muons are unpolarized.
If the scalar mass is sufficiently smaller than $\sqrt{s}$,
the scalar mass and decay width can be measured from the peak
of the photon energy distribution in the $\mu^+ \mu^+ \to e^+ e^+ \gamma$ process.
Suppose that the Mu-to-$\overline{\rm Mu}$ transition is observed
and $|G_i| = |y_{ee} y_{\mu\mu}|/(4\sqrt2 m_S^2)$ is fixed.
Then, we can determine whether the scalar is a triplet or a singlet
from the cross sections of $\mu^+ \mu^+ \to e^+ e^+$
and $\mu^+ \mu^+ \to e^+ e^+ \gamma$.
Using the cross section of $\mu^+ \mu^+ \to e^+ e^+ jj$,
we can verify the consistency of the scenario.
In the case that the scalar mass is larger than $\sqrt{s}$, on the other hand,
the cross sections are insensitive to $\Gamma_S$.
The measurements of the cross sections and $|G_i|$
can determine the scalar mass (unless $m_S \gg \sqrt{s}$)
and whether the scalar is a triplet or a singlet
if a sufficient number of events of $\mu^+ \mu^+ \to e^+ e^+ jj$
can be observed.

\subsection{Determination of the branching ratio of {\boldmath${ S^{++} \to l^+ l^+}$}}

As described in the previous subsection,
the total decay width can be determined from the
photon energy distribution of $\mu^+ \mu^+ \to e^+ e^+ \gamma$
if the scalar mass $m_S$ is sufficiently smaller than $\sqrt{s}$.
Because
$|y_{ee}|^2 + |y_{\mu\mu}|^2 \geq 2 |y_{ee}y_{\mu\mu}| = 8\sqrt{2} |G_i| m_S^2$,
we obtain
\begin{equation}
\Gamma (S^{++} \to e^+ e^+) + \Gamma (S^{++} \to \mu^+ \mu^+)
= \frac{(|y_{ee}|^2 + |y_{\mu\mu}|^2) m_S}{8\pi}
\geq
\frac{\sqrt{2} |G_i| m_S^3}{\pi}.
\end{equation}
The $\Delta L_\mu = - \Delta L_e = \pm 2$ LFV processes depend only on the product
$|y_{ee}y_{\mu\mu}|$,
and therefore only a lower bound on the sum of the two partial widths can be obtained.
Therefore,
in order to determine whether
the two decay modes can account for the total decay width inferred from the distribution,
we need another process to measure $y_{ee}$ or $y_{\mu\mu}$
(unless $\Gamma_S$ is equal to the lower bound).

Elastic $\mu^+ \mu^+$ scattering, $\mu^+ \mu^+ \to \mu^+ \mu^+$,
at the $\mu^+\mu^+$ collider
provides a measurement (or constraint) on the $y_{\mu\mu}$ coupling,
because the scalar contribution depends only on the $y_{\mu\mu}$ coupling.
The differential cross section of $\mu^+ \mu^+ \to \mu^+ \mu^+$ is given by
\begin{align}
\frac{d\sigma}{d\Omega}
=\frac{\alpha^2}{2s}
&\left[
\left(\frac{1+P}{2}\right)^2
s^2 \left(F_{LL}(t) + F_{LL}(u)- \frac{|\kappa_{\mu\mu}|^2}{2\pi \alpha} \frac{1}{s-m_\Delta^2}
\right)^2 \nonumber
\right.\\
& \left. +
\left(\frac{1-P}{2}\right)^2
s^2 \left(F_{RR}(t) + F_{RR}(u) - \frac{|g_{\mu\mu}|^2}{2\pi \alpha} \frac{1}{s-m_k^2}
\right)^2 \nonumber
\right. \\
&\left.
+ 2\frac{1+P}{2}\frac{1-P}{2}
\left(u^2 F_{LR}(t)^2 + t^2 F_{LR}(u)^2 \right) 
\right],
\end{align}
where
\begin{align}
F_{LL}(x) &= \frac{1}{x}
+ \frac{(-\frac12 + s_W^2)^2}{s_W^2 c_W^2}
\frac{1}{x-m_Z^2}, \\
F_{RR}(x) &= \frac{1}{x}
+ \frac{s_W^2}{c_W^2}
\frac{1}{x-m_Z^2}, \\
F_{LR}(x) &= \frac{1}{x}
+ \frac{-\frac12 + s_W^2}{c_W^2}
\frac{1}{x-m_Z^2},
\end{align}
and $s$, $t$, and $u$ are the Mandelstam variables.
The SM process contains $t$- and $u$-channel exchanges of the $\gamma$ and $Z$ gauge bosons,
and
the contribution from the $s$-channel exchange of the doubly charged scalar interferes with the SM amplitudes.
We note that the NP contribution interferes constructively (destructively) with the SM contribution for $\sqrt{s} > m_S$
($\sqrt{s} < m_S$), due to $F_{LL,RR}(t,u) < 0$.
Therefore,
the $\mu^+\mu^+ \to \mu^+\mu^+$ process provides a consistency check
of the results obtained from the cLFV processes described in the previous subsection.
The sensitivity to the $y_{\mu\mu}$ coupling at the $\mu^+\mu^+$ collider
has been studied in Ref.~\cite{Lichtenstein:2023iut}.
The $y_{ee}$ coupling can also be measured at future $e^+ e^-$ colliders,
as it has been constrained by the LEP experiment described in Sec.~\ref{sec3.2}.
Since the product $|y_{ee} y_{\mu\mu}|$ can be determined, as we have discussed,
at least one of $y_{ee}$ and $y_{\mu\mu}$ needs to be measured.
At the $\mu^+ \mu^+$ collider, we can also measure $|y_{\mu\mu}y_{\tau\tau}|$
from the cross section of $\mu^+ \mu^+ \to \tau^+ \tau^+$.
Therefore, we can obtain the branching ratios of the scalar decays to charged-lepton pairs
and investigate
whether the dilepton decays can account for the total decay width
inferred from the measurements at the $\mu^+ \mu^+$ collider.

The $\mu^+ \mu^+ \to S^{++} V$ process violates lepton number
if the scalar has non-leptonic decay modes,
as we will study in the next section.

\section{LNV processes via scalar decays}
\label{sec6}

When the scalars, $k^{++}$, $h^+$, $\Delta^{++}$, and $\Delta^{+}$, decay to quarks,
the processes given in Eqs.~(\ref{mumuhh}), (\ref{mumukZgamma}), (\ref{mumumuek}),
and (\ref{mumu-WD}) are LNV ones, as described in Sec.~\ref{sec2}.
In this section, we describe the decays of these scalars.

\subsection{$\eta$-$h$ scalar mixing}
\label{sec6.1}

We define linear combinations of the Higgs doublets, $\Phi_1$ and $\Phi_2$, as
\begin{align}
\Phi &= \Phi_1 \cos \beta + \Phi_2 \sin\beta, \\
\eta &= -\Phi_1 \sin \beta + \Phi_2 \cos\beta,
\end{align}
where $\tan\beta = \langle \Phi^0_2 \rangle/\langle \Phi^0_1 \rangle$,
so that the vev of $\eta^0$ is zero.
The mass terms of the $SU(2)_L$ singlet $h^+$ and the charged scalar $\eta^+$ in the doublet $\eta$
in Eq.~(\ref{lagrangian-1})
can be written as
\begin{equation}
\left(\begin{array}{cc} \eta^+ & h^+ \end{array}\right)
\left(
\begin{array}{cc}
m_{\eta^+}^2 & v m_{\Phi\Phi h} \\
v m_{\Phi\Phi h} & m_{h^+}^2
\end{array}
\right)
\left(
\begin{array}{c}
\eta^- \\
h^-
\end{array}
\right),
\label{mass-eta-h}
\end{equation}
where $v$ is the vev of $\Phi^0$ responsible for electroweak symmetry breaking.
The mixing angle $\phi$ between $\eta^+$ and $h^+$ is
\begin{equation}
\tan2\phi = \frac{2v m_{\Phi\Phi h}}{m_{h^+}^2 - m_{\eta^+}^2}.
\end{equation}
We denote $m_{h_1^+}$ and $m_{h_2^+}$ ($m_{h_1^+} < m_{h_2^+}$)
as the masses of the eigenstates $h_1^+$ and $h_2^+$.

\subsection{$k^{++}$ decays}

The doubly charged scalar $k^{++}$
can be produced at the $\mu^+\mu^+$ collider
in processes such as
$\mu^+ \mu^+ \to k^{++} \gamma/Z$ and $\mu^+ \mu^+ \to \mu^+ e^- k^{++}$.
When $k^{++}$ is produced at the collider, LNV can be observed via the non-leptonic decays of $k^{++}$.
The width of the $k^{++} \to h_i^+ h_j^+$ decay is
\begin{equation}
\Gamma (k^{++} \to h_i^+ h_j^+)
= \frac{m_{hhk}^2}{8 \pi m_k} \lambda\left(1, \frac{m_{h_i^+}^2}{m_k^2},\frac{m_{h_j^+}^2}{m_k^2}\right)
\times \{ \sin^4\phi, 2\sin^2 \phi\cos^2\phi, \cos^4\phi \},
\end{equation}
where the K\"all\'en function $\lambda$ is
\begin{equation}
\lambda (x,y,z) = \sqrt{x^2 + y^2 + z^2 - 2xy - 2yz - 2zx},
\end{equation}
and the values in the brackets correspond to $(i,j) = (1,1), (1,2), (2,2)$.
The width of the leptonic decays via the $g$ coupling is
\begin{equation}
\Gamma (k^{++} \to l_a^+ l_b^+ ) = \frac{|g_{ab}|^2 m_k}{4\pi (1+\delta_{ab})}.
\end{equation}
If $m_{hhk} \sim m_k$,
the width of $k^{++} \to h_i^+ h_j^+$ can be larger than
those of the leptonic decay modes,
as long as the decays are kinematically allowed ($m_k > m_{h_i^+} + m_{h_j^+}$).

If $m_k < 2 m_{h_1^+}$,
$k^{++}$ cannot decay into a pair of charged scalars.
Although $k^{++}$ is an $SU(2)_L$ singlet,
$k^{++}$ can decay into $W^+ W^+$ via a one-loop diagram
if there is a $(\eta \Phi^\dagger)^2$ term,
and then, LNV processes can be observed.
Such a coupling of a doubly charged scalar to same-sign $W$ bosons
has been discussed in the context of radiative neutrino mass generation and neutrinoless double beta decay~\cite{Gustafsson:2012vj,Fukuyama:2022dhe,King:2014uha}.
Since the decay is generated at the loop level,
the leptonic decays are the dominant modes of $k^{++}$ decay
in the current setup, where the scalar couplings $g_{ab}$ are not small.
Therefore, it is preferable for the charged scalars to be light in order to observe the LNV processes.

In order to allow light scalars without suffering from the $h_{1,2}^+ \to \tau^+ \nu_\tau$ decay,
we consider a type-I model for the two Higgs doublets ($\Phi_1$ does not couple to quarks and leptons).
Even in this case, the light scalars are excluded for small $\tan\beta$
since too many $h_{1,2}^+$ can be produced at the LHC
via $gg \to t\bar b h_{1,2}^-$ and $gb \to t h_{1,2}^-$.

Especially in the case of $\mu^+ \mu^+ \to \gamma k^{++}/\Delta^{++}$,
the angular distribution is mostly concentrated around $\theta \sim 0,\pi$.
If the sum of the $h_{1,2}^+$ masses is much smaller than
the doubly charged scalar mass,
the angular distribution of $h_{1,2}^+$ is also concentrated around $\theta \sim 0,\pi$.
Therefore,
it is preferable that the singly charged scalars are not too light
so that as many of their decay products as possible can be detected.

\subsection{Singly charged scalar decays and direct production}

Both the $h_1^+$ and $h_2^+$ scalars can decay into $t\bar b$.
The heavier scalar $h_2^+$ can also decay into $Z h_1^+$ and $W^+ H/A$,
and there are additional decays into jets.
The decay widths can be obtained as
\begin{align}
\Gamma(h_2^+ \to t\bar b) &=
\frac{3}{16\pi} 
\frac{m_{h_2^+}^2 - m_t^2 - m_b^2}{m_{h_2^+}} \lambda\left(1,\frac{m_t^2}{m_{h_2^+}^2},\frac{m_b^2}{m_{h_2^+}^2} \right)
\frac{m_t^2}{v^2} \cot^2\beta \sin^2\phi, \\
\Gamma(h_2^+ \to Z h_1^+) &=
\frac{g_Z^2 m_{h_2^+}^3}{64\pi m_Z^2 }  
\lambda\left(1,\frac{m_{h_1^+}^2}{m_{h_2^+}^2},\frac{m_Z^2}{m_{h_2^+}^2} \right)^3 \cos^2\phi \sin^2\phi.
\end{align}
For $m_{\eta^+} > m_{h^+}$, the heavier charged scalar $h_2^+$ is $\eta^+$-like,
and $h_2^+$ typically decays into $t\bar b$.
The $k^{++}$ scalar decays into a pair of $h_1^+$,
and thus, most of the $k^{++}$ decays are into $t\bar b t\bar b$.
For $m_{\eta^+} < m_{h^+}$, on the other hand,
the lighter charged scalar $h_1^+$ is $\eta^+$-like.
In that case, one finds $\Gamma(h_2^+ \to Z h_1^+) > \Gamma(h_2^+ \to t\bar b)$
unless $\tan\beta$ is small and $m_{h_2^+}$ is light.
As explained above, small $\tan\beta$ with light charged scalars is excluded by the LHC results,
and most of the $k^{++}$ decays in that case contain multiple jets.

The singly charged scalar $h_i^+$ can also be directly produced at the collider.
The cross section of $\mu^+ \mu^+ \to h_i^+ h_j^+$ is
\begin{align}
\sigma (\mu^+ \mu^+ \to h_i^+ h_j^+)
= \frac{(1-P)^2}4 
\frac{|g_{\mu\mu}|^2}{2\pi}
\frac{m_{hhk}^2}{(s-m_k^2)^2} \lambda\left(1, \frac{m_{h_i^+}^2}{s},\frac{m_{h_j^+}^2}{s}\right)
\\ \times \{ \sin^4\phi, 2\sin^2 \phi\cos^2\phi, \cos^4\phi \}. \nonumber
\end{align}
For $m_k = m_{hhk} = 1$ TeV and $\sqrt{s}= 2$ TeV, we obtain
$\sigma \simeq |g_{\mu\mu}|^2 \times 430$ fb for $m_{h_i^+} \ll \sqrt{s}$,
up to the mixings in the brackets.

\subsection{Case of the triplet scalar model}

In the model with the triplet scalar $\Delta$,
the singly charged scalars $\Delta^+$ and $\eta^+$ can be described similarly.
There is an extra neutral scalar $\Delta^0$ in the triplet,
and
a quartic coupling of $\Delta$ contains $\Delta^{++} \Delta^- \Delta^- \Delta^0$.
However, the vev of $\Delta^0$ should be tiny in the current setup because
we are considering a sizable $\kappa$ coupling that generates neutrino masses.
The tiny vev of $\Delta^0$ also makes $\Delta^{++} \to W^+ W^+$ decays small.
Therefore, $\Delta^{++}$ decays to $\Delta^+ W^+$ via gauge interactions.
Unless $\Delta^+$ and $\eta^+$ are mixed, the decay width is suppressed
because the masses of $\Delta^+$ and $\Delta^{++}$ are nearly degenerate.
With the $\Delta^+$ and $\eta^+$ mixing, $\Delta^{++}$ can decay into a $W^+$ boson and an $\eta^+$-like singly charged scalar.
Due to the mixing of $\Delta^+$ and $\eta^+$, $\Delta^{++}$ can decay into four quarks.

\begin{figure}[t]
\centering
\includegraphics[width=0.47\textwidth]{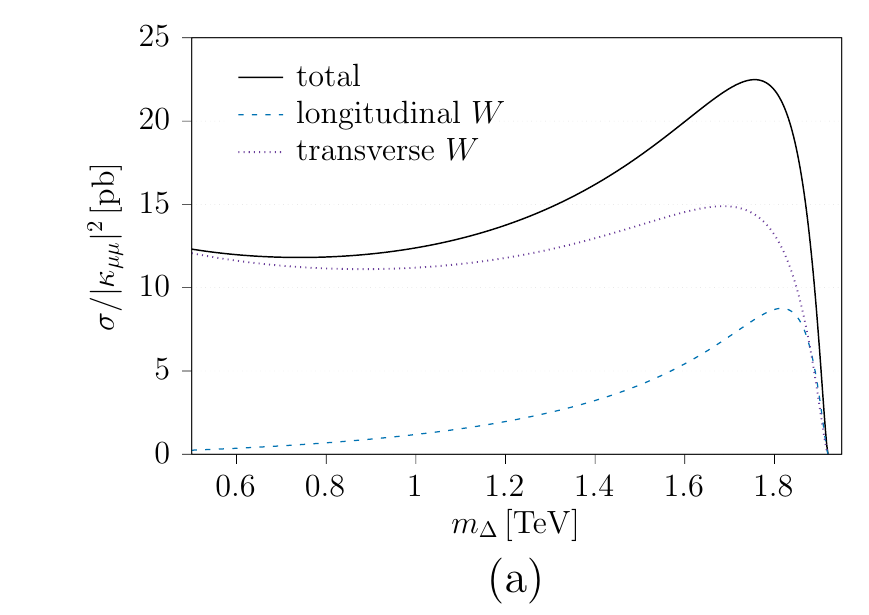}
\includegraphics[width=0.47\textwidth]{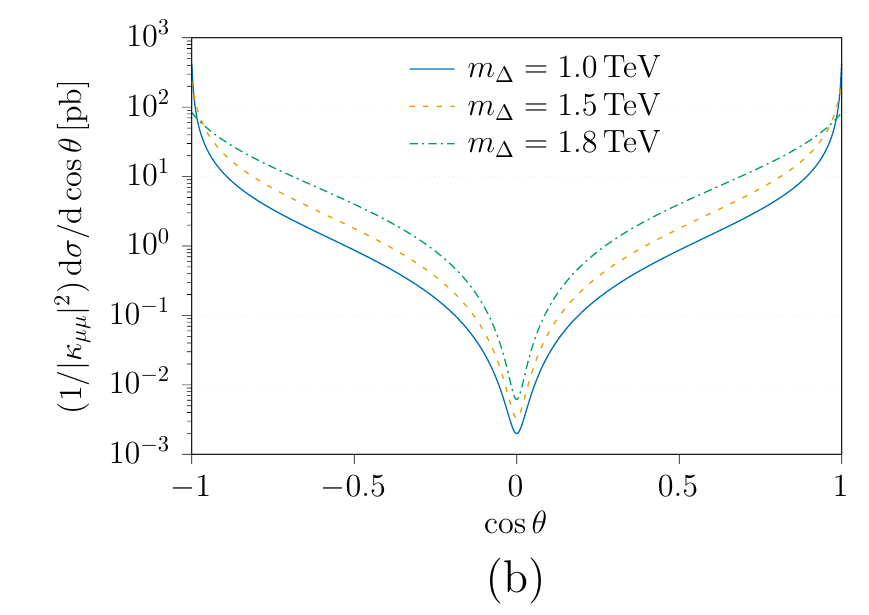}
\caption{%
    Cross section and angular distribution for $\mu^+\mu^+\to W^+\Delta^+$, with both quantities normalized to $|\kappa_{\mu\mu}|^2$.
    In panel (a), the cross section is plotted as a function of $m_\Delta = m_{\Delta^{++}} = m_{\Delta^+}$.
    The longitudinal and transverse $W$ contributions are shown by the dashed and dotted curves, respectively, while the total cross section is shown by the solid curve.
    In panel (b), the angular distribution is shown as a function of $\cos\theta$ for $m_\Delta=1.0,1.5,1.8$\,TeV.
    In both panels, we take $\sqrt{s}=2$\,TeV and $P=1/2$.
}
\label{fig:mm2WDelta_xsec}
\end{figure}

In addition to the $\mu^+ \mu^+ \to \Delta^{++} \gamma/Z$ process,
there is a $\mu^+ \mu^+ \to W^+ \Delta^+$ process
which can be LNV in the triplet scalar case.
The squared amplitude of $\mu^+ \mu^+ \to \Delta^+ W^+$ for the longitudinal $W$ boson is
\begin{align}
|{\cal M}_{\rm L}|^2 =
  \frac{ g_2^2  |\kappa_{\mu\mu}|^2  s}{4m_W^2}
  & \left(
  \frac{4\hat\lambda}{s-m_{\Delta^{++}}^2}
  \right.
  \nonumber\\
& \quad \left.
  + \frac{\hat\lambda - (s- m_W^2 -m_{\Delta^+}^2)\cos\theta}{t}
  +
  \frac{\hat\lambda + (s- m_W^2 -m_{\Delta^+}^2)\cos\theta}{u}
  \right)^2,
\label{longitudinal-mumuDeltaW}
\end{align}
where
$\hat\lambda = \lambda(s,m_W^2,m_{\Delta^+}^2)$, and
the Mandelstam variables, $t$ and $u$, in this process are
\begin{align}
t &= \frac12 \left( m_{\Delta^+}^2 + m_W^2 - s + \hat\lambda \cos\theta\right), \\
u &= \frac12 \left( m_{\Delta^+}^2 + m_W^2 - s - \hat\lambda \cos\theta \right) .
\end{align}
The terms in the brackets in Eq.~(\ref{longitudinal-mumuDeltaW}) are calculated to be
\begin{equation}
\hat\lambda m_W^2 \frac{5m_\Delta^2 +m_W^2-s + (3m_\Delta^2-m_W^2+s)\cos2\theta}{2tu (s-m_\Delta^2)},
\end{equation}
for $m_{\Delta } = m_{\Delta^{++}} = m_{\Delta^+}$.
As a result, the longitudinal contribution is suppressed for $m_W^2 \ll s-m_\Delta^2$.
The squared amplitude for the transverse modes of the gauge boson is
\begin{equation}
|{\cal M}_{\rm T+}|^2 + |{\cal M}_{\rm T-}|^2 =
  \frac12 |\kappa_{\mu\mu}|^2 g_2^2
  (\hat\lambda^2 + 2 s m_\Delta^2) \left(\frac{1}{t} - \frac{1}{u}\right)^2
  \sin^2\theta.
\end{equation}
One can find that
the distribution is concentrated around $\cos\theta \sim \pm 1$
if $m_W^2 \ll s - m_\Delta^2$
due to $t,u$ in the denominators.
If $m_\Delta$ is larger (say, 1.8 TeV in the case of $\sqrt{s} = 2$ TeV),
the angular distribution increases
at points away from $\cos\theta \sim \pm 1$.
We plot the cross section (divided by $|\kappa_{\mu\mu}|^2$) and angular distribution of
$\mu^+ \mu^+ \to W^+ \Delta^+$ for $m_{\Delta } = m_{\Delta^{++}} = m_{\Delta^+}$
in Fig.~\ref{fig:mm2WDelta_xsec}.

As described above,
the longitudinal contribution is suppressed if the masses of
$\Delta^+$ and $\Delta^{++}$ are degenerate.
The vev of the SM Higgs doublet $\Phi$ can split the squared masses of the $\Delta$ multiplet
through a $\Phi \Delta \Delta^\dagger \Phi^\dagger$ term.
However, a squared-mass splitting of $O(v^2)$ (which means that the mass splitting is $O(v^2/m_\Delta)$)
is not enough to increase the contribution of the longitudinal mode.
Similarly to Eq.~(\ref{mass-eta-h}) in Sec.~\ref{sec6.1},
$\Delta^+$ and $\eta^+$ can be mixed by a $\Delta \eta \Phi$ term.
If $m_\Delta \simeq m_\eta$, the mass difference can become $O(v)$
and the longitudinal contribution can become large.
However, a large mass splitting will shift the $\rho$ parameter
due to an oblique correction, and therefore, this is not a favorable scenario.
If $m_\Delta \ll m_\eta$ or $m_\Delta \gg m_\eta$,
the longitudinal contribution can also be sizable when the final state consists of
the $\eta$-like eigenstates
(though a mixing angle should be multiplied by the amplitude).
If the longitudinal contribution can become larger in this way,
the angular distribution around $\theta \sim \pi/2$ can become larger
(while the distribution of the transverse contribution around $\theta \sim \pi/2$ is tiny).

Since the $\Delta^0$ field decays into a pair of neutrinos,
it is preferable that the decay of the singly charged scalar to $\Delta^0$ is kinematically blocked.
Then,
the heavier singly charged scalar can decay into a $W^+$ boson and $\eta^0$,
or into a $Z$ boson and the lighter singly charged scalar $h_1^+$,
leading to the LNV processes
$\mu^+ \mu^+ \to jjjj t\bar b,\ jjjj t\bar t,\ jj t\bar b$.
We note that the main decay modes of the
$\eta^+$-like charged scalar
are not necessarily $t\bar b$
since the inert doublet $\eta$ couplings to quarks are not necessarily
related to the quark masses.

\section{$\Delta L_\mu = \Delta L_e = 1$ LFV \& LNV process in the same-sign lepton collider}
\label{sec7}

As described in Sec.~\ref{sec2}, the $e\mu$ couplings to the doubly charged scalar can induce both LNV and LFV.
As emphasized, such processes are allowed even if
$\Delta L_\mu = - \Delta L_e = \pm 1$ processes, such as
$\mu\to e\gamma$ and $\mu\to 3e$, are eliminated
by a discrete symmetry.
In this section, we study how the $\Delta L_\mu = \Delta L_e = 1$ processes
can be observed in the same-sign lepton collider.
It is more challenging to observe these processes than processes with only LFV.

\begin{figure}[t]
\centering
\includegraphics[width=0.95\textwidth]{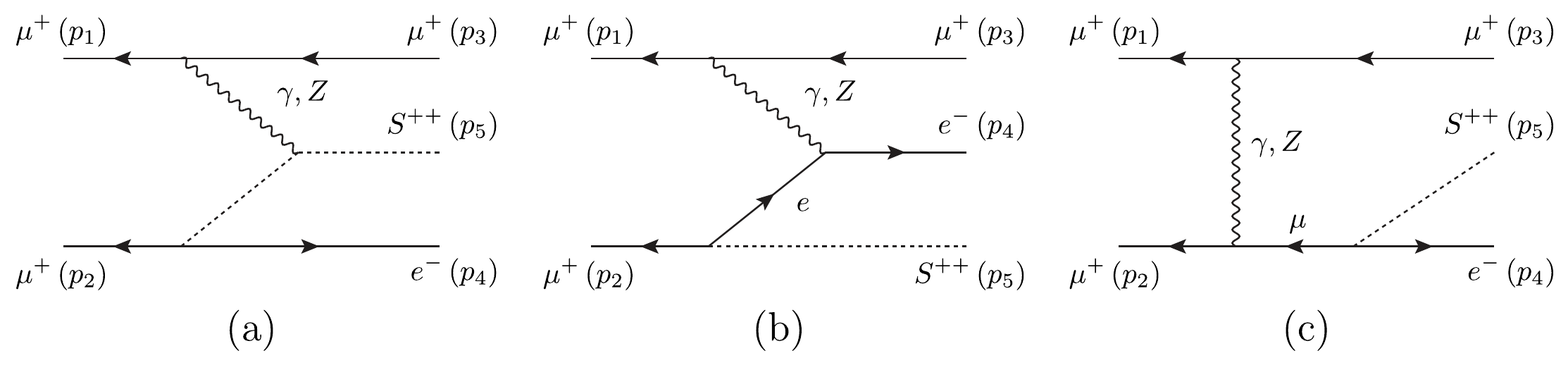}
\caption{%
    Tree-level diagrams for
    $\mu^+\mu^+\to \mu^+e^-S^{++}$, where $S^{++}=k^{++},\Delta^{++}$.
    In diagrams (a), (b), and (c), the exchanged gauge boson emitted from one of the incoming muon lines couples to the scalar line, the electron line, and the other muon line, respectively.
    The momenta of the incoming muons are denoted by $p_1$ and $p_2$, while those of the outgoing muon, electron, and scalar are denoted by $p_3$, $p_4$, and $p_5$, respectively.
    The diagrams obtained by interchanging $p_1$ and $p_2$ are omitted.
}
\label{fig:diagrams_mm2meS}
\end{figure}

We consider
$\mu^+ \mu^+ \to \mu^+ e^- S^{++}$ processes
induced by the ${e\mu}$ couplings in Eqs.~(\ref{lagrangian-1}) and (\ref{lagrangian-2}).
If $S^{++}$ decays to non-leptonic particles as described in the previous section,
this becomes an LNV process with conserved $L_\mu - L_e$.
The Feynman diagrams for the process are given in Fig.~\ref{fig:diagrams_mm2meS}.

\subsection{Cross section}

Let $p_1$ and $p_2$ be the momenta of the initial muons,
and let $p_3$, $p_4$, and $p_5$ be those of the final-state muon, electron, and scalar, respectively,
as given in Fig.~\ref{fig:diagrams_mm2meS}.
We parametrize
\begin{align}
p_1 &= (E,0,0,E), \label{p1} \\
p_2 &= (E,0,0,-E), \\
p_3 & = (1, \sin\theta_3, 0, \cos\theta_3) E_3, \\
p_4 & = (1, \sin\theta_4 \cos\phi, \sin\theta_4 \sin\phi, \cos\theta_4) E_4. \label{p4}
\end{align}
The fifth momentum is determined by momentum conservation, $p_1 + p_2 = p_3 + p_4 + p_5$.
Unless it leads to confusion, we often write $p_5^2 = m_5^2$ for the scalar mass, especially in
Appendix~\ref{appendix:B}, in which the kinematics of the process is given.

The squared amplitude (up to a coupling) for
$\mu_L^c \mu_L^c \to \mu_L^c e_L \Delta^{++}$
is given as
\begin{align}
& |{\cal M}_{LL}^\Delta|^2 \nonumber \\
&= 64 E^2 E_3 E_4 \left|(C_L^\Delta(p_1) + C_L^\Delta(p_2))
2E\left(\cos\frac{\theta_3}2 \cos\frac{\theta_4}2
+ e^{-i\phi} \sin\frac{\theta_3}2 \sin\frac{\theta_4}2 \right) \right.   
 \label{MLLDelta}
\\
&\left.
+ (B_L^\Delta(p_1,p_2)-A_L^\Delta(p_2,p_1))
\sin\frac{\theta_4}2 \left(
e^{-i\phi}(2E-E_4 -E_4 \cos\theta_4)  \sin\frac{\theta_3}2
+  E_4 \cos\frac{\theta_3}2 \sin\theta_4
\right) \right.  \nonumber \\
&\left.
+ (B_L^\Delta(p_2,p_1)-A_L^\Delta(p_1,p_2)) 
\cos\frac{\theta_4}2\left(
(2E-E_4 +E_4 \cos\theta_4)  \cos\frac{\theta_3}2
+ e^{-i\phi} E_4 \sin\frac{\theta_3}2 \sin\theta_4
\right)
 \right|^2 , \nonumber
\end{align}
and the squared amplitude for $\mu_R^c (p_1)\mu_L^c (p_2)\to \mu_R^c e_L \Delta^{++}$ is
\begin{align}
&|{\cal M}_{RL}^\Delta|^2 \nonumber \\
&= 64 E^2 E_3 E_4 \left|C_R^\Delta(p_1)
\cos\frac{\theta_3}2
\left( (2E-E_3 + E_3 \cos\theta_3) \cos\frac{\theta_4}2 + e^{-i\phi} E_3 \sin\theta_3 \sin\frac{\theta_4}2 \right) \right.  \nonumber\\
&\left.
- B_R^\Delta(p_1,p_2)
\left(
 \sin\frac{\theta_3}2 \cos\frac{\theta_4}2 -
 e^{-i\phi} \cos\frac{\theta_3}2 \sin\frac{\theta_4}2
\right)
(E_3 \sin\theta_3 + e^{i\phi} E_4 \sin\theta_4) \right. \nonumber\\
&\left.
- A_R^\Delta(p_1,p_2)
\cos\frac{\theta_4}2
 \left(
   (2E-E_4 + E_4 \cos\theta_4) \cos\frac{\theta_3}2 + e^{i\phi} E_4
\sin\frac{\theta_3}2 \sin\theta_4
 \right)
 \right|^2  . \label{MRLDelta}
\end{align}
The squared amplitude for $\mu_L^c (p_1)\mu_R^c(p_2) \to e_L \mu_R^c \Delta^{++}$ is simply obtained as
\begin{equation}
|{\cal M}_{LR}^\Delta|^2 = |{\cal M}_{RL}^\Delta|^2 ( \theta_3 \to \pi - \theta_3, \theta_4 \to \pi - \theta_4, p_1 \leftrightarrow p_2).
\end{equation}
The coefficients $A_{L,R}^\Delta$, $B_{L,R}^\Delta$ and $C_{L,R}^\Delta$ in the amplitudes are given as
\begin{align}
A_{L,R}^\Delta(p_1,p_2) & = - 4 D_{L,R}^\Delta(p_1) \frac{1}{(p_2-p_4)^2- m_\Delta^2}, \label{AL,AR}\\
B_{L,R}^\Delta(p_1,p_2) & = - 2 D_{L,R}^\Delta(p_1) \frac{1}{(p_2-p_5)^2- m_e^2}, \label{BL,BR}\\
C_{L,R}^\Delta(p_1) & = 2 D_{L,R}^\Delta(p_1) \frac{1}{(p_4+p_5)^2- m_\mu^2}, \label{CL,CR}
\end{align}
and
\begin{align}
D_L^\Delta (p_1) & = \frac{e^2}{(p_1-p_3)^2} +  \frac{g_Z^2 \left(-\frac12 + s_W^2\right)^2}{(p_1-p_3)^2 - m_Z^2}, \\
D_R^\Delta (p_1) & = \frac{e^2}{(p_1-p_3)^2} +  \frac{g_Z^2 s_W^2 \left(-\frac12 + s_W^2\right)}{(p_1-p_3)^2 - m_Z^2},
\end{align}
where $g_Z^2 = g_2^2/c_W^2 = e^2/(s_W^2 c_W^2)$.
Although we neglect the muon and electron masses,
we retain them in $B_{L,R}$ and $C_{L,R}$
to show that these contributions contain the propagators for electrons and muons, respectively.

In the case of the singlet scalar $k^{++}$, the expressions can be basically obtained
by swapping $L$ and $R$, except for the contribution from the $Z$ boson couplings to the charged leptons
and scalars
in Eqs.~(\ref{Lag:Zll}) and (\ref{Lag:ZSS}):
\begin{align}
D_R^k (p_1) & = \frac{e^2}{(p_1-p_3)^2} +  \frac{g_Z^2 \left(s_W^2\right)^2}{(p_1-p_3)^2 - m_Z^2}, \\
D_L^k (p_1) & = \frac{e^2}{(p_1-p_3)^2} +  \frac{g_Z^2 s_W^2 \left(-\frac12 + s_W^2\right)}{(p_1-p_3)^2 - m_Z^2}.
\end{align}
Since $D_R^\Delta = D_L^k$,
we obtain (for $m_k = m_\Delta$)
\begin{equation}
|{\cal M}^k_{LR}|^2 = |{\cal M}^\Delta_{RL}|^2.
\label{MLR-MRL}
\end{equation}
Because of $s_W^2 \sim 1/4$ (and thus, $1/2 - s_W^2 \sim s_W^2$),
one finds
$D_L^\Delta \sim D_R^k$,
and
\begin{equation}
|{\cal M}^k_{RR}|^2 \sim |{\cal M}^\Delta_{LL}|^2.
\label{MRR-MLL}
\end{equation}

The cross sections for the $S^{++} = \Delta^{++}$ case are given as
\begin{equation}
d\sigma_{LL,LR,RL}^\Delta =\frac1{2s} |\kappa_{e\mu}|^2 |{\cal M}_{LL,LR,RL}^\Delta|^2 d{\rm LIPS},
\end{equation}
and
\begin{equation}
d\sigma_\Delta = \frac{(1+P)^2}{4} d\sigma^\Delta_{LL}  + \frac{1-P^2}{4} (d\sigma^\Delta_{LR}+d\sigma^\Delta_{RL}).
\label{dsig-Delta}
\end{equation}
In the case of the singlet $k^{++}$,
the cross section is similarly given as
\begin{equation}
d\sigma_k = \frac{(1-P)^2}{4} d\sigma^k_{RR}  + \frac{1-P^2}{4} (d\sigma^k_{RL}+d\sigma^k_{LR}).
\label{dsig-k}
\end{equation}
The $LR+RL$ contributions of both $\Delta^{++}$ and $k^{++}$ are given by the same expressions
for $\kappa = g$ and $m_\Delta = m_k$
from Eq.~(\ref{MLR-MRL}),
while the $LL$ and $RR$ contributions for $\Delta^{++}$ and $k^{++}$ are approximately the same (for $P=0$)
from Eq.~(\ref{MRR-MLL}).
When the integration variables are chosen to be $E_3$ and the angles (then, $E_4$ is given by Eq.~(\ref{solE4})),
$d$LIPS is given as
\begin{equation}
d{\rm LIPS} = \frac{1}{512\pi^5} E_3 dE_3 d\Omega_3 d\Omega_4 \frac{4E^2-4E E_3 -m_5^2}{(2E- E_3(1-\cos\theta_{34}))^2} .
\end{equation}

\begin{figure}[t]
\centering
\includegraphics[width=0.65\textwidth]{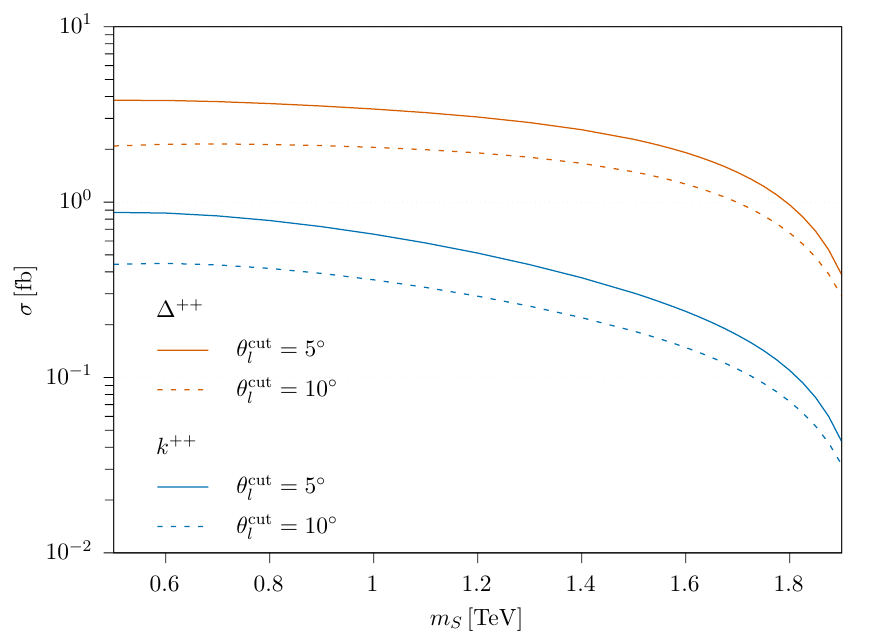}
\caption{%
    Upper bounds on the cross section for $\mu^+\mu^+\to\mu^+e^-S^{++}$.
    The horizontal axis shows the scalar mass $m_S$.
    For each scalar mass, the couplings are set to their LEP upper limits given in Eq.~\eqref{LEPbound:emu}.
    The solid and dashed curves correspond to $\theta_l^{\mathrm{cut}}=5^\circ$ and $10^\circ$, respectively.
    We take $\sqrt{s}=2$\,TeV and $P=1/2$, and impose the following kinematic cuts on the charged leptons: $p_T^l>5$\,GeV and $\Delta R_{ll}>0.05$.
}
\label{fig:mm2meS_cross_section_mass}
\end{figure}

The $e\mu$ couplings are constrained by LEP experiments as given in Eq.~(\ref{LEPbound:emu}).
In addition, the $\kappa_{e\mu}$ coupling can shift the $W$ boson mass prediction in the SM,
and Eq.~(\ref{CDF-W-anomaly-emu}) gives the range that can explain the CDF $W$ boson mass anomaly.
In Fig.~\ref{fig:mm2meS_cross_section_mass},
we show the upper bounds on the cross sections
of the $\mu^+ \mu^+ \to \mu^+ e^- k^{++}/\Delta^{++}$ processes.
Because the distribution of the $\mu^+$ angle (which is denoted as $\theta_3$)
is concentrated around $0$ or $\pi$ (namely, along the beam axis) due to the photon propagator,
as we will describe in more detail later,
we plot the upper bounds for both
$\theta_{l}^{\rm cut} = 5^\circ$ and $10^\circ$
as the angular acceptance of the charged leptons.

\subsection{Distribution}

Let us examine the scalar, electron, and muon propagators in diagrams
(a), (b), and (c), respectively, in Fig.~\ref{fig:diagrams_mm2meS}.
For the discussion below, let $p_1$ be the momentum of the muon that couples to a photon in diagrams (a) and (b).
The denominator of the photon propagator is
\begin{equation}
(p_1 - p_3)^2  = -2EE_3 (1-\cos\theta_3),
\end{equation}
and therefore, the distribution near $\theta_3 \sim 0$ is enhanced.
The contributions from the diagrams obtained by swapping $p_1$ and $p_2$ become resonantly large for $\theta_3 \sim \pi$,
while they do not give a large contribution for $\theta_3 \sim 0$. So we do not consider their contributions for now.

The denominator of the scalar propagator in diagram (a) is
\begin{equation}
(p_2- p_4)^2 - m_5^2 = - 2EE_4 (1+\cos\theta_4) -m_5^2.
\label{scalar-propagator}
\end{equation}
The contribution from diagram (a)
is enhanced for small $E_4$ and/or $\theta_4 \sim \pi$ (namely, when the electron is emitted in the opposite direction to the muon),
especially if the scalar is light.

The denominator of the electron propagator in diagram (b) is
\begin{align}
(p_2 -p_5)^2 &= (p_3 + p_4 - p_1)^2 = (p_3+p_4)^2 - 2 p_1 \cdot (p_3+p_4) \nonumber \\
&= 2E_3 E_4 (1-\cos\theta_{34}) -2EE_3 (1-\cos\theta_3)
-2EE_4 (1-\cos\theta_4),
\end{align}
where $\theta_{34}$ is the angle between the three-dimensional momenta ${\bm p}_3$ and ${\bm p}_4$
in Eq.~(\ref{costheta34}).
One can prove that this can be zero only when $\theta_3 = \theta_4 = 0$ (then, $\theta_{34}=0$).
Therefore, the contribution from diagram (b)
is large when the electron is emitted in the same direction as the muon.

The denominator of the muon propagator in diagram (c) is
\begin{equation}
(p_4+ p_5)^2 = (p_1+p_2-p_3)^2 = (p_1+p_2)^2 - 2 p_3 \cdot (p_1+ p_2) = 4E(E-E_3).
\end{equation}
The contribution from diagram (c) is large when $E_3$ is large.
As described in the kinematics given in Appendix~\ref{appendix:B},
the maximal value of $E_3$ is $E_3^{\rm max} = E- m_5^2/(4E)$.
Therefore, this tendency is pronounced if the scalar is much lighter than $E$.

One often considers helicity conservation to understand the angular distribution.
Actually,
helicity conservation favors the emission of $\mu^+$ and $e^-$ in the same direction
in both cases of $\mu_L^c \mu_L^c \to \mu_L^c e_L \Delta^{++}$
and
$\mu_R^c \mu_L^c \to \mu_R^c e_L \Delta^{++}$.
In the latter case, emission along the beam axis is preferred.
However, the statement based on helicity only refers to the fermion structure,
and does not account for the propagator resonance.
In fact, the argument based on helicity conservation differs from
the enhancement due to the scalar propagator, as shown in
Eq.~(\ref{scalar-propagator}).
Besides, in the limit of $\theta_{34} \to 0$ and $\theta_3 \sim 0$,
one can find that the amplitude ${\cal M}_{LL}^\Delta$ in Eq.~(\ref{MLLDelta}) cancels
among the (a), (b), and (c) contributions
at $E_3 = m_5^2/(4E)$.
Therefore, the distribution around $\cos\theta_{34} \sim 1$ (which is preferred by helicity conservation) is not so
significant, especially if the scalar is not heavy
(the kinematically allowed range of $E_3$ ($0 \leq E_3 \leq E - m_5^2/(4E)$)
does not contain the cancellation for $m_5 \geq \sqrt2 E$).

It is also important to note that
the helicity amplitudes ${\cal M}_{LL}$ and ${\cal M}_{RL}$ given in Eqs.~(\ref{MLLDelta}) and (\ref{MRLDelta})
multiplied by $(p_1 - p_3)^2 = -2EE_3(1-\cos\theta_3)$, which is the denominator of the photon propagator,
vanish in the limit of $\theta_3 \to 0$.
This means that the behavior of the amplitudes for small $\theta_3$ is
$O(1/\theta_3)$, instead of $O(1/\theta_3^2)$ from the photon propagator.
As a result,
none of the three contributions from diagrams (a), (b), and (c)
can become significantly larger than the others
when the distribution is large near $\theta_3 \sim 0$.
For this reason, it is not easy to describe the overall qualitative behavior;
it would likely be easier to understand it by performing specific numerical calculations.
Theoretically,
the above statement follows from
the Ward identity, $q^\mu {\cal A}_\mu = 0$,
where $q= p_1-p_3$ and ${\cal A}_\mu$ is an amplitude
for $\gamma \mu^+ \to e^- \Delta^{++}$.

\begin{figure}[t]
\centering
\includegraphics[width=0.95\textwidth]{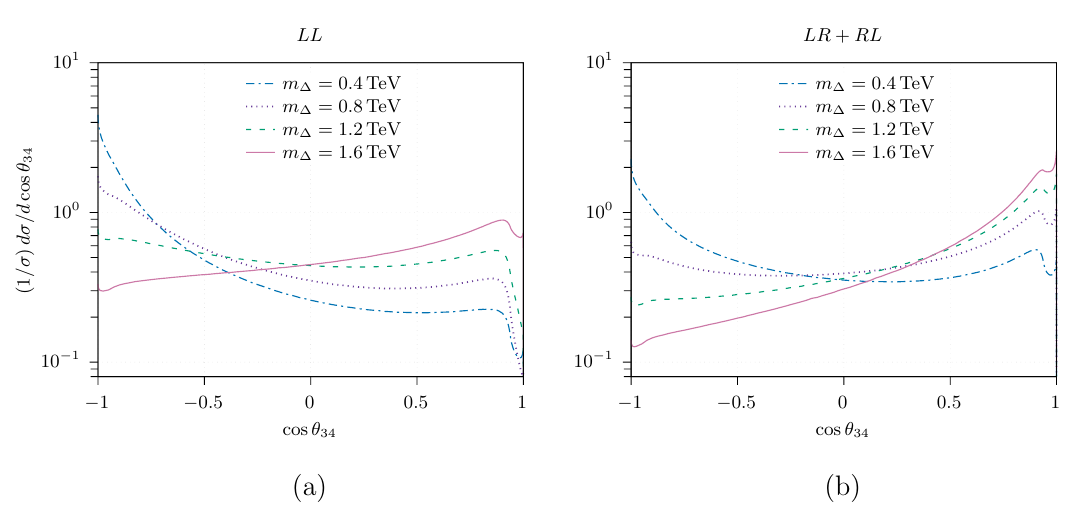}
\caption{%
    Normalized angular distributions for $\mu^+\mu^+\to\mu^+e^-\Delta^{++}$.
    Panels (a) and (b) show the angular distributions
    $d\sigma_{LL}/d\cos\theta_{34}$ and
    $d(\sigma_{LR}+\sigma_{RL})/d\cos\theta_{34}$, respectively,
    for $m_\Delta=0.4,0.8,1.2,1.6$\,TeV.
    Each distribution is normalized to the corresponding integrated cross section.
    We take $\sqrt{s}=2$\,TeV and impose the following kinematic cuts on the charged leptons:
    $\theta_l^{\mathrm{cut}}=10^\circ$, $p_T^l>5$\,GeV, and $\Delta R_{ll}>0.05$.
}
\label{fig:mm2meS_cos34_helicity_mass}
\end{figure}

In Fig.~\ref{fig:mm2meS_cos34_helicity_mass},
we plot the $\cos\theta_{34}$ distributions for $\sigma_{LL}$
and $\sigma_{LR}+\sigma_{RL}$ separately for various scalar masses.
If the scalar is light, the distribution is concentrated around $\cos\theta_{34} \sim -1$.
As the scalar becomes heavier, it shifts toward $\cos\theta_{34} \sim 1$.
In the case of $LL$, the shift occurs after the scalar has become heavier than in the case of $RL+LR$.
This is a consequence
of the fact that the $LL$ amplitude has a cancellation at $E_3 = m_5^2/(4E)$ for $\theta_{34} \to 0$ as described above.

\begin{figure}[t]
\centering
\includegraphics[width=0.65\textwidth]{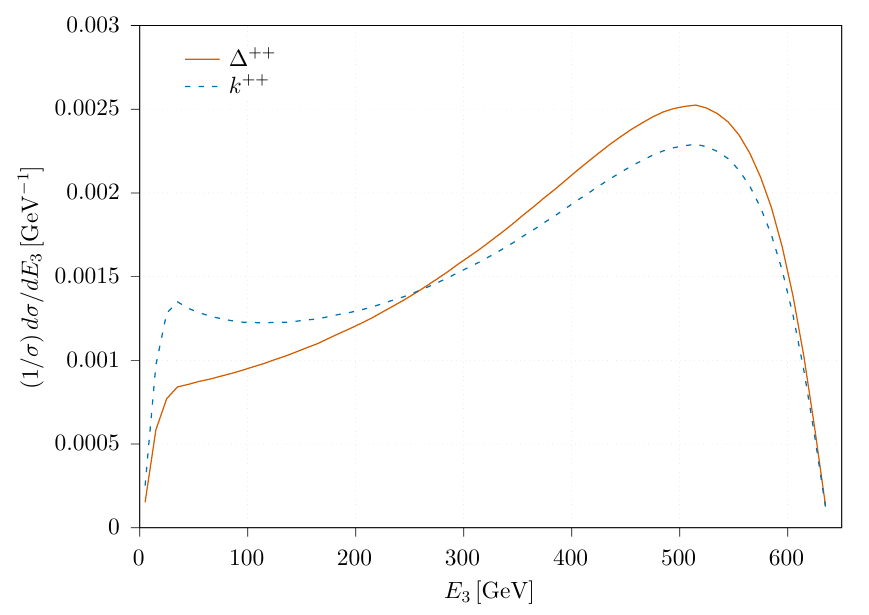}
\caption{%
    Normalized energy distribution of the final-state muon in $\mu^+\mu^+\to\mu^+e^-S^{++}$.
    The solid and dashed curves correspond to $S^{++}=\Delta^{++}$ and $k^{++}$, respectively.
    We take $\sqrt{s}=2$\,TeV, $P=1/2$, and $m_S=1.2$\,TeV.
    We impose the following kinematic cuts on the charged leptons:
    $\theta_l^{\mathrm{cut}}=10^\circ$, $p_T^l>5$\,GeV, and $\Delta R_{ll}>0.05$.
}
\label{fig:mm2meS_triplet_singlet_e3_m1200}
\end{figure}

The differential cross sections for $\Delta^{++}$ and $k^{++}$ are given
as a linear combination of $LL \sim RR$ and $LR+RL$ in Eqs.~(\ref{dsig-Delta}) and (\ref{dsig-k}).
Because the $LL\sim RR$ and $LR+RL$ helicity cross sections are different,
the distributions for $\Delta^{++}$ and $k^{++}$ are different in principle unless the initial muons are unpolarized.
In Fig.~\ref{fig:mm2meS_triplet_singlet_e3_m1200},
we show the energy distributions (normalized by their cross sections)
for $\Delta^{++}$ and $k^{++}$
as a typical case for $m_S = 1.2$ TeV.

\subsection{Remarks on the small-angle distribution of $\mu^+$}

As described in Sec.~\ref{sec4}, the muon beam is accompanied
by a large beam-induced background,
and a shielding nozzle along the beam axis is considered
as one of the measures to mitigate this background.
However, the nozzle reduces the acceptance for final-state particles
emitted at small angles with respect to the beam axis.
The small-angle distribution of the final-state muon is enhanced
due to the photon propagator.
We have presented the cross sections in Fig.~\ref{fig:mm2meS_cross_section_mass}
with a cut on the charged-lepton scattering angle,
$\theta_l^{\rm cut}$.
The cross section in the excluded small-angle region,
$|\cos\theta_l|>|\cos\theta_l^{\rm cut}|$,
can be comparable to, or even larger than, that in the region considered in the plot,
$|\cos\theta_l|<|\cos\theta_l^{\rm cut}|$.

In the expressions shown above, the muon mass has been neglected.
In the physical process, however, the muon mass is nonzero and the final state contains a massive scalar.
As a result, the kinematically allowed region does not reach $q^2=0$,
where $q=p_1-p_3$, and the forward singularity is regulated.
For small $|q^2|$, the exchanged photon can be regarded as a quasi-real photon.
In fact, within the equivalent photon approximation (EPA) \cite{vonWeizsacker:1934nji,Williams:1935dka},
the cross section in the small-angle region can be estimated from the cross section for
$\gamma\mu^+\to e^-S^{++}$ convoluted with the photon distribution function.
We impose the small-angle acceptance such that the final-state muons can be detected,
allowing the process to be clearly identified as an LNV process.
Although the events may still be identifiable as signals of NP
without detecting the final-state muons, we do not consider this possibility in the present paper.

\begin{figure}[t]
\centering
\includegraphics[width=0.47\textwidth]{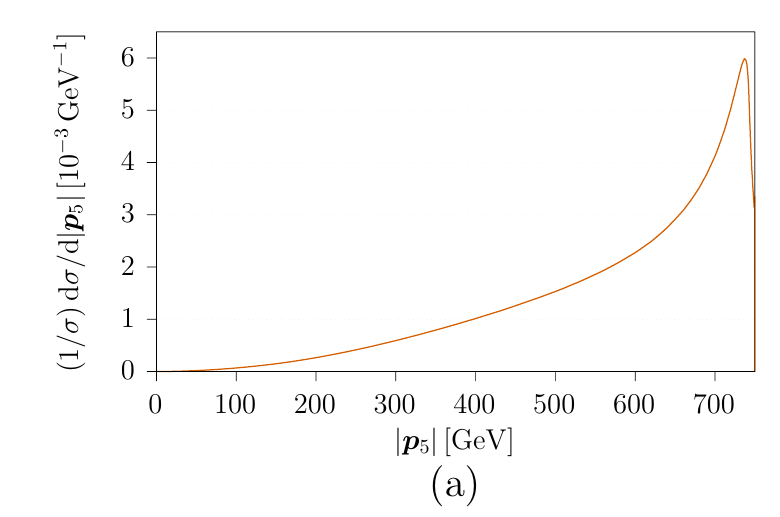}
\includegraphics[width=0.47\textwidth]{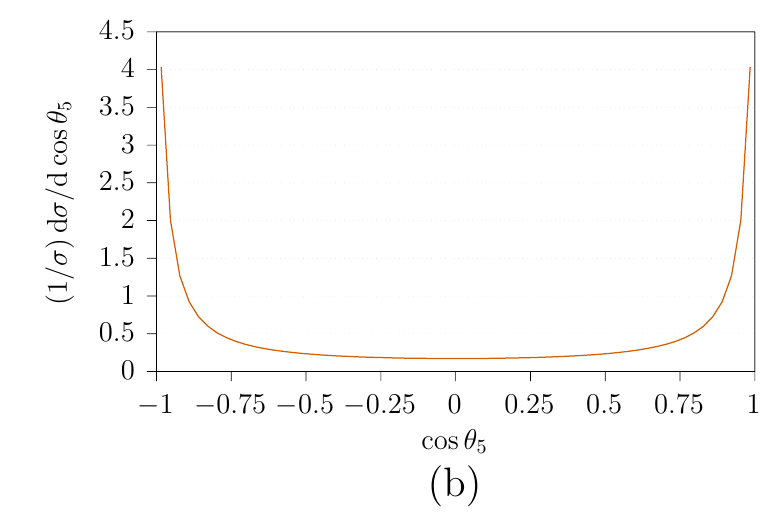}
\caption{%
    Normalized momentum and angular distributions of the scalar in $\mu^+\mu^+\to\mu^+e^-\Delta^{++}$.
    Panels (a) and (b) show the distributions of the scalar momentum magnitude $|\bm{p}_5|$ and $\cos\theta_5$, respectively.
    We take $\sqrt{s}=2$\,TeV, $P=1/2$, and $m_\Delta=1$\,TeV.
    We impose the following kinematic cuts on the charged leptons:
    $\theta_l^{\mathrm{cut}}=10^\circ$, $p_T^l>5$\,GeV, and $\Delta R_{ll}>0.05$.
}
\label{fig:mm2meS_costheta5_distribution}
\end{figure}

We note that even if a very forward muon cannot be detected,
a sizable fraction of the scalars can still be emitted in the backward direction.
This backward enhancement can be partly understood from the propagator structure of the process.
In particular, there are contributions containing the factor $B_{L,R} (p_1,p_2)$ in Eq.~(\ref{BL,BR}).
The $1/(p_1-p_3)^2$ factor enhances forward muon emission,
while the $1/(p_2-p_5)^2$ favors scalar emission in the opposite direction.
As we have noted,
this does not fully explain the qualitative behavior
since the terms corresponding to the diagrams cancel each other.
To state only the results, higher energy of the final-state muon and lower energy of the electron are favored,
and consequently, from a kinematic standpoint,
the scalar is preferentially emitted with larger momentum in the direction opposite to the muon.
In Fig.~\ref{fig:mm2meS_costheta5_distribution},
we show the distributions of the scalar momentum $|{\bm p}_5|$
and the angle $\theta_5$ between the scalar momentum and the beam axis.
The distribution is enhanced at large momentum near the maximal value, $E - m_5^2/(4E)$,
and at small angle with respect to the beam axis ($\cos\theta_5 \sim \pm 1$).
The plots are drawn for $m_5 =1$ TeV in the triplet $\Delta$ case as an example.
The tendency of the distribution does not depend on the scalar mass.

The detectability of the scalar decay products depends on the boost of the scalar.
A highly boosted scalar produces collimated decay products, which may be lost in the shielding nozzle,
whereas a moderately boosted scalar can produce decay products with a wider angular distribution.
The detectability of these decay products depends on the scalar boost
and hence on the scalar mass and the collider energy
because the distribution near the maximal scalar energy is large
as described above,
and the boost factor at the maximal scalar energy is
$\gamma = E/m_5 (1+m_5^2/(4E^2))$.
In the case of $m_5 \agt E$,
even if a scalar is emitted on the opposite side of the muon,
its decay products are unlikely to be collimated
and are more likely to be emitted in diverse directions away from the beam axis.

A muon collider can also be viewed as an effective photon-muon or photon-photon collider.
In particular, photon- and $Z$-boson fusion processes are important physics targets at muon colliders.
For this reason, methods for detecting or monitoring very forward muons may be developed in the future.
According to the EPA calculation, the cross section for events with very forward muons increases
with increasing $\sqrt{s}$.
On the other hand, increasing $\sqrt{s}$ also increases the boost of the produced scalar,
which can make its decay products more collimated
and may reduce the detector acceptance depending on the shielding geometry.
Therefore, when upgrading the collider after a scalar has been discovered
or if the scalar mass has already been determined at a hadron collider,
it is preferable to choose the collider energy in order to balance the production cross section
and the detectability of the decay products.

\section{Conclusions and discussions}
\label{sec8}

We have investigated the LFV and LNV processes via doubly charged scalars $S^{++}$
at the $\mu$TRISTAN $\mu^+\mu^+$ collider.
As conclusions,
we discuss the following three topics:
(1) determination of the couplings and scalar mass by combining Mu-to-$\overline{\rm Mu}$ transitions
and $\mu^+\mu^+ \to e^+ e^+ V$ $(V= \gamma, Z)$
for $\Delta L_\mu = -\Delta L_e = 2$ processes,
(2) the merit of employing an inert $SU(2)_L$ doublet for LNV processes,
(3) $\Delta L_e = \Delta L_\mu = 1$ LFV and LNV processes at muon colliders.

The important parameters
for the $\Delta L_\mu = -\Delta L_e = 2$ processes
are
the product of the scalar couplings to charged leptons, $|y_{ee} y_{\mu\mu}|$,
and the doubly charged scalar mass $m_S$.
If $m_S$ is smaller than the center-of-mass energy $\sqrt{s}$,
the cross sections of $\mu^+\mu^+ \to e^+ e^+ V$ processes
depend on the scalar decay width $\Gamma_S$.
The cross sections also depend on
whether the doubly charged scalar is a triplet or a singlet.
These parameters can be determined by combining measurements of
the probability of the Mu-to-$\overline{\rm Mu}$ transition
and
the photon energy distribution of the $\mu^+\mu^+ \to e^+ e^+\gamma$ process
if $m_S < \sqrt{s}$.
In the case $m_S > \sqrt{s}$,
additional information from the cross section of
$\mu^+\mu^+ \to e^+ e^+ jj$ can determine the parameters except for $\Gamma_S$,
and can also determine whether the doubly charged scalar is a triplet or a singlet.
We expect that the $y_{\mu\mu}$ coupling can be measured
by the $\mu^+ \mu^+ \to \mu^+ \mu^+$ scattering process.
Using these measurements,
the branching ratio of the scalar decay to charged leptons,
$\mathrm{Br}(S^{++} \to \ell^+ \ell^+)$,
can be determined.
If a doubly charged scalar exists at the TeV scale,
it is likely that its mass and the branching ratios of its decays
would have been determined at a high-luminosity hadron collider
before the lepton collider is built.
At the hadron collider, three-charged-lepton final states can arise from
$u \bar d \to \Delta^{++} \Delta^-$
if the scalar belongs to a triplet,
in addition to four-charged-lepton final states
from the pair production of doubly charged scalars,
$q \bar q \to S^{++} S^{--}$.
Therefore, the hadron collider can in principle distinguish
whether the scalar is a triplet or a singlet.
However, it is still meaningful that these parameters can be determined
and cross-checked solely through measurements at the lepton collider.

In the triplet scenario of the doubly charged scalar,
we have introduced an inert $SU(2)_L$ doublet $\eta$
as in Eq.~(\ref{lagrangian-2})
in order to observe LNV while avoiding constraints from neutrino masses.
In the minimal type-II seesaw model with the triplet,
the cross section of the LNV process is small because it is related to the tiny neutrino masses.
A similar situation can be utilized in the framework with right-handed neutrinos.
With an SM singlet fermion $N$ and the inert doublet $\eta$,
we consider the Lagrangian
$-{\cal L} \supset \overline{N_R} \ell \eta + 1/2 M_N \overline{N_R^c} N_R$.
Then, due to the Majorana mass $M_N$,
$\mu^+ \mu^+ \to \eta^+ \eta^+$ is possible.
If the inert doublet couples to quarks and $\eta^+$ decays to $t\bar b$, one can observe the LNV process
at the $\mu^+\mu^+$ collider.
When $\eta$ acquires a vev and the Dirac neutrino mass is generated,
neutrino masses are generated by the seesaw mechanism,
and the LNV process at the collider is suppressed due to the tiny neutrino masses.
If $\eta$ does not acquire a vev,
neutrino masses are not generated at tree level.
Indeed, in that case, one can assign lepton numbers as
$\ell: 1$, $N_R :0$, $\eta : -1$.
The $\eta$ couplings to quarks violate lepton number.
These couplings can induce neutrino masses through a two-loop diagram.
The existence of a tree-level $(\eta \Phi^\dagger)^2$ term can also induce neutrino masses
through a one-loop diagram \cite{Ma:2006km},
and thus, the coefficient of the term should be small.
In this way,
the inert doublet can be utilized to construct a phenomenological model
with a visible LNV process at the lepton collider.

Observation of the $\Delta L_\mu = \Delta L_e = 1$ LFV and LNV process,
$\mu^+ \mu^+ \to \mu^+ e^- S^{++}$, $S^{++} \to t\bar b t\bar b$, is more challenging
than $\mu^+ \mu^+ \to t\bar b t\bar b$.
The $\Delta L_\mu = \Delta L_e = 1$ process can also be generated by
right-handed neutrinos with $Z_3$ flavor symmetry.
In the case of doubly charged scalar mediation,
a hadron collider can produce pairs of the scalars,
and the existence of the $y_{e\mu}$ coupling
can be found from the decays of the scalars.
In the $\mu^+\mu^-$ collider,
the $\Delta L_\mu = \Delta L_e = 1$ LFV and LNV process
can be investigated through $S^{++}S^{--}$ pair production,
followed by $S^{++} \to t\bar b t\bar b$ and
$S^{--} \to \mu^- e^-$.
This leads to the final state
$\mu^- e^- t\bar b t\bar b$.
If pair production is kinematically allowed for $\sqrt{s} > 2m_S$,
its cross section can be larger than that of
$\mu^+\mu^+ \to \mu^+ e^- S^{++}$,
since the former is a two-body production process,
whereas the latter is a three-body production process.
In the case of single production of the doubly charged scalar,
the cross sections at same- and opposite-sign muon colliders are of the same order,
though there are fewer diagrams for the process at the same-sign muon collider.
As mentioned in the text,
it is easier to detect the decay products of the scalar if the scalar is not highly boosted,
and it is preferable that $\sqrt{s}$ not be too large compared to the scalar mass.
If we suppose that the scalar mass is known in advance from a future hadron collider,
this would likely influence the choice of $\sqrt{s}$.

\section*{Acknowledgements}

This work was supported in part by JSPS KAKENHI Grant Numbers JP25H00653 (T.F.), JP23K13106, and JP24K07061 (Y.U.).

\appendix

\section{Photon energy distribution of $\mu^+ \mu^+ \to e^+ e^+ \gamma$}
\label{appendix:A}

In this appendix, we show the photon energy distribution of the
$\mu^+ \mu^+ \to e^+ e^+ \gamma$ process.

It is convenient to parameterize the four-momenta
as follows:
\begin{align}
p_1 &= (1,\sin\theta,0,\cos\theta)E, \\
p_2 &= (1,-\sin\theta, 0, -\cos\theta)E, \\
p_{e1} &= (1, \sin\theta_1 \cos\phi, \sin\theta_1 \sin\phi, \cos\theta_1)E_{e1}, \\
p_{e2} &= (1, -\sin\theta_2 \cos\phi, -\sin\theta_2 \sin\phi, \cos\theta_2)E_{e2}, \\
p_\gamma &= (1,0,0,1) E_\gamma,
\end{align}
where $p_1$ and $p_2$ are the initial muon momenta,
$p_{e1}$ and $p_{e2}$ are the final-state positron momenta,
and $p_\gamma$ is the photon momentum.
The usual coordinate system in which $p_1$ and $p_2$ point along the
$z$-axis can be obtained by rotating the momenta in the $x$-$z$ plane:
\begin{align}
& p_1 \to  (1,0,0,1)E, \\
& p_2 \to  (1,0,0,-1)E, \\
& p_{e1} \to (1, -\sin\theta \cos\theta_1 + \cos\theta \sin\theta_1 \cos\phi,
\sin\theta_1 \sin\phi,
\cos\theta \cos\theta_1 + \sin\theta \sin\theta_1 \cos\phi )E_{e1},
\label{pe1-wu-coordinate}\\
& p_{e2} \to (1, -\sin\theta \cos\theta_2 - \cos\theta \sin\theta_2 \cos\phi,
-\sin\theta_2 \sin\phi,
\cos\theta \cos\theta_2 - \sin\theta \sin\theta_2 \cos\phi )E_{e2}, \\
& p_\gamma \to  (1, -\sin\theta, 0, \cos\theta)E_\gamma,
\label{pgamma-wu-coordinate}
\end{align}
Momentum conservation,
$p_1+p_2=p_{e1}+p_{e2}+p_\gamma$, gives
\begin{align}
& 2E = E_{e1} + E_{e2} + E_\gamma, \\
& E_{e1} \sin\theta_1 = E_{e2} \sin\theta_2, \\
& E_{e1}\cos\theta_1 + E_{e2}\cos\theta_2 + E_\gamma = 0.
\end{align}
Two of $(E_{e1}, E_{e2}, E_\gamma, \theta_1,\theta_2)$ are free.
We choose $E_\gamma$ and $E_{e1}$ as the free variables.
We then obtain
\begin{align}
\cos \theta_1 &=
\frac{-E_{e1}^2 + E_{e2}^2 -E_\gamma^2}{2E_{e1} E_\gamma}
=
\frac{2E^2 + E_{e1} E_\gamma - 2 (E_{e1} + E_\gamma) E}{E_{e1} E_\gamma},
\label{costh1} \\
\cos \theta_2
&= \frac{E_{e1}^2 - E_{e2}^2 -E_\gamma^2}{2E_{e2} E_\gamma}
=
-\frac{2E^2 -(E_{e1}+E_\gamma)(2E-E_\gamma)}
{(2E-E_{\gamma}-E_{e1})E_\gamma},
\label{costh2} \\
E_{e2} &= 2E - E_{e1} - E_\gamma.
\label{Ee2}
\end{align}
From $-1 \leq \cos\theta_1 \leq 1$, one obtains
\begin{equation}
E-E_\gamma \leq E_{e1} \leq E.
\end{equation}
For $E_{e1}=E$, one finds $\cos\theta_1=-1$.
For $E_{e1}=E-E_\gamma$, one finds $\cos\theta_1=1$.
We note that the positron and photon are emitted in the same (opposite)
direction for $\cos\theta_1=1$ ($\cos\theta_1=-1$), by definition.

We show the results in the case of $\Delta^{++}$ mediation.
The $k^{++}$ case is obtained simply by flipping the helicity and
replacing the coupling and mass by those of $k^{++}$ below.
The helicity amplitudes (up to couplings), where $\lambda$ denotes the
photon helicity, are given by
\begin{align}
M_1 (\lambda) &=
4\sqrt2 E^2 \, {\bf M}_1(\lambda)
\frac{1}{(p_{e1}+p_{e2})^2 - m_\Delta^2}
\left(\frac{1}{(p_\gamma-p_2)^2} -
\frac{1}{(p_\gamma-p_1)^2}\right), \\
M_2 (\lambda) &=
8\sqrt2 E^2 \, {\bf M}_2(\lambda)
\frac{1}{s-m_\Delta^2}
\left(\frac{1}{(p_{e1}+p_\gamma)^2} -
\frac{1}{(p_{e2}+p_\gamma)^2}\right),
\end{align}
where $M_1$ corresponds to diagrams (b) and (c), and $M_2$ corresponds
to diagrams (d) and (e) in Fig.~\ref{fig:diagrams-mm2eeV}.
As described in the text, the contribution of diagram (a) vanishes
for a transversely polarized gauge boson.
We obtain ${\bf M}_1$ and ${\bf M}_2$ using
Eqs.~(\ref{costh1})--(\ref{Ee2}),
\begin{align}
{\bf M}_1 (+) &=
- \sqrt{E_{e1} E_{e2}} \sin\frac{\theta_1+\theta_2}{2}
= -\sqrt{E(E-E_\gamma)}, \\
{\bf M}_1 (-) &=
\frac{E-E_\gamma}{E}
\sqrt{E_{e1} E_{e2}}
\sin\frac{\theta_1+\theta_2}{2}
=
\frac{E-E_\gamma}{E}\sqrt{E(E-E_\gamma)}, \\
{\bf M}_2 (+) &=
\sqrt{E_{e1} E_{e2}}
\sin\frac{\theta_1}{2}\sin\frac{\theta_2}{2}e^{-i\phi}
%
=
\frac{E-E_\gamma}{E_\gamma}
\sqrt{(E-E_{e1})(E_{e1}+E_\gamma-E)}e^{-i\phi}, \\
{\bf M}_2 (-) &=
-\sqrt{E_{e1} E_{e2}}
\cos\frac{\theta_1}{2}\cos\frac{\theta_2}{2}e^{i\phi}
=
-\frac{E}{E_\gamma}
\sqrt{(E-E_{e1})(E_{e1}+E_\gamma-E)}e^{i\phi}.
\end{align}

The differential cross section for
$\mu_L\mu_L\to e_L e_L\gamma$ is given by
\begin{equation}
d\sigma_{LL} = \frac{1}{4s}\rho \, d{\rm LIPS},
\end{equation}
where
\begin{equation}
\rho =
16 |\kappa_{ee} \kappa_{\mu\mu}|^2 e^2
\sum_{\lambda=\pm}
(M_1(\lambda)+M_2(\lambda))
(M_1(\lambda)+M_2(\lambda))^*.
\end{equation}
The phase-space integral is
\begin{equation}
d{\rm LIPS} =
\frac{1}{128\pi^4}
dE_\gamma\, dE_{e1}\, d(\cos\theta)\, d\phi,
\end{equation}
where one trivial angular integration has already been performed.
The differential cross section can be written as
\begin{equation}
\frac{d^2\sigma_{LL}}{dE_\gamma d(\cos\theta)}
=
\frac{1}{512\pi^4 s}
\int d\phi
\int_{E-E_\gamma}^E dE_{e1}\, \rho.
\end{equation}
Unless we impose rapidity (or angular acceptance) cuts and $p_T$ cuts
on the positrons, the $\phi$ integration is trivial for
$M_i M_i^*$ ($i=1,2$), while the integration of the
$M_1 M_2^*+M_1^*M_2$ term vanishes.
In order to obtain analytical expressions,
we ignore the rapidity and $p_T$ cuts here
(these cuts are applied in the numerical plots in the figures).

The denominators of the positron propagators in $M_2(\lambda)$,
$(p_{ei}+p_\gamma)^2=2E_{ei}E_\gamma(1-\cos\theta_i)$,
are calculated using Eqs.~(\ref{costh1})--(\ref{Ee2}) as
\begin{align}
(p_{e1}+p_\gamma)^2
&= 2 E_{e1} E_\gamma (1-\cos\theta_1)
= 4 E (E_{e1} +E_\gamma - E), \\
(p_{e2}+p_\gamma)^2
&= 2 E_{e2} E_\gamma (1-\cos\theta_2)
= 4 E (E-E_{e1}).
\end{align}
They become zero at
$E_{e1}=E-E_\gamma$ for $\cos\theta_1=1$
and at $E_{e1}=E$ for $\cos\theta_2=1$.
One therefore needs cuts near $E_{e1}=E-E_\gamma$ and $E$.
In the collider analyses, $\Delta R_{e\gamma}$ cuts play an important
role in avoiding these divergences from the positron propagators.

Supposing that $|\theta_1|\ll 1$, we obtain the angular differences
between $p_\gamma$ and $p_{e1}$ from
Eqs.~(\ref{pe1-wu-coordinate}) and
(\ref{pgamma-wu-coordinate}) as
\begin{equation}
\Delta \theta = \theta_1 \cos\phi,
\qquad
\Delta \phi = \theta_1 \sin\phi \, \csc\theta.
\end{equation}
By definition, we obtain
\begin{equation}
\Delta R_{e\gamma}
\equiv \sqrt{(\Delta\eta)^2 + (\Delta\phi)^2}
= \sqrt{\csc^2\theta\,(\Delta\theta)^2 + (\Delta\phi)^2}
= \csc\theta \,\theta_1.
\end{equation}
Our purpose is to perform the positron energy integration in the
$M_2 M_2^*$ term:
\begin{align}
I &=
\int_{E-E_\gamma+\Delta_1}^{E-\Delta_2} dE_{e1}\,
(E-E_{e1})(E_{e1}+E_\gamma-E)
\left(
\frac{1}{E_{e1}+E_\gamma-E}
-\frac{1}{E-E_{e1}}
\right)^2
\\
&=
4(\Delta_1+\Delta_2-E_\gamma)
+E_\gamma
\ln\frac{(E_\gamma-\Delta_1)(E_\gamma-\Delta_2)}
{\Delta_1\Delta_2}.
\end{align}
Assuming $|\theta_1|,|\theta_2|\ll1$, we obtain from
Eqs.~(\ref{costh1}) and (\ref{costh2}),
\begin{equation}
\Delta_1 =
\frac{E_\gamma(E-E_\gamma)}{4E}\theta_1^2,
\qquad
\Delta_2 =
\frac{E_\gamma(E-E_\gamma)}{4E}\theta_2^2.
\end{equation}
Therefore, the positron energy cuts, $\Delta_1$ and $\Delta_2$, are
related to $\Delta R_{e\gamma}$ by
\begin{equation}
\Delta_1 = \Delta_2 =
\frac{E_\gamma(E-E_\gamma)}{4E}
\sin^2\theta
(\Delta R_{e\gamma}^{\rm cut})^2.
\end{equation}
Then, we obtain
\begin{equation}
I \simeq
2E_\gamma
\left(
\ln\frac{4E}{E-E_\gamma}
-2
-2\ln(\Delta R_{e\gamma}^{\rm cut}\sin\theta)
\right).
\end{equation}

We note that a small $\Delta R_{ee}$ occurs only for
$\theta_1,\theta_2\simeq\pi$ kinematically.
Similarly to $\Delta R_{e\gamma}$, we find
\begin{equation}
\Delta R_{ee}
=
(2\pi-\theta_1-\theta_2)\csc\theta,
\end{equation}
for $|\pi-\theta_i|\ll1$.
Therefore, using Eqs.~(\ref{costh1})--(\ref{Ee2}), we obtain
\begin{equation}
(\Delta R_{ee})^2
=
\frac{4E(E-E_\gamma)}{E_{e1}E_{e2}}
\csc^2\theta,
\end{equation}
for small $\Delta R_{ee}$.
The resulting analytic expression for the integration imposing the $\Delta R_{ee}$ cut becomes
complicated.
The cut is important only when $E_\gamma$ is very close to $E$,
and therefore, we ignore the $\Delta R_{ee}$ cut to show the analytic
expression below
(the $\Delta R_{ee}$ cut is also applied numerically in the plots
showing the numerical results in Sec.~\ref{sec5}).

The differential cross section for the photon energy distribution
of $\mu_L\mu_L\to e_L e_L\gamma$ is
\begin{align}
\frac{d\sigma_{1LL}}{dE_\gamma}
&=
|\kappa_{ee}\kappa_{\mu\mu}|^2
\frac{e^2}{2\pi^3}
\frac{1}{E E_\gamma}
\frac{(E-E_\gamma)(E^2+(E-E_\gamma)^2)}
{(4E(E-E_\gamma)-m_\Delta^2)^2+m_\Delta^2\Gamma_\Delta^2}
I_\theta(\theta_{\rm cut}),\\
\frac{d\sigma_{2LL}}{dE_\gamma}
&\simeq
|\kappa_{ee}\kappa_{\mu\mu}|^2
\frac{e^2}{2\pi^3}
\frac{1}{(s-m_\Delta^2)^2}
\frac{E^2+(E-E_\gamma)^2}{E_\gamma}
\left(
\ln\frac{E}{E-E_\gamma}
-2\ln(\Delta R_{e\gamma}^{\rm cut})
\right),
\end{align}
where
$\sigma_{LL}\simeq\sigma_{1LL}+\sigma_{2LL}$
($\sigma_{iLL}$ comes from $M_iM_i^*$),
ignoring angular and $p_T$ cuts on the positrons and the
$\Delta R_{ee}$ cut, and
\begin{equation}
I_\theta(\theta_{\rm cut})
=
\int_{\theta_{\rm cut}}^{\pi-\theta_{\rm cut}}
\cot^2\theta\sin\theta\,d\theta
=
2\left(
-\ln\tan\frac{\theta_{\rm cut}}{2}
-\cos\theta_{\rm cut}
\right)
\simeq
2(\eta_\gamma^{\rm cut}-1).
\end{equation}
We have used
\begin{equation}
\int_0^\pi
(-\ln\sin\theta)\sin\theta\,d\theta
=
2(1-\ln2).
\end{equation}

It is straightforward to see that
$\sigma_{1LL}$ is larger than $\sigma_{2LL}$ for
$s(=4E^2)>m_\Delta^2$ due to the resonance at
$m_{ee}^2\sim m_\Delta^2$, where
$m_{ee}^2\equiv(p_{e1}+p_{e2})^2=4E(E-E_\gamma)$,
while $\sigma_{2LL}$ is larger than $\sigma_{1LL}$ for
$m_\Delta^2>s$.
It is interesting to note that
$I_\theta(\theta_{\rm cut})$ governs the dependence on the photon
polar-angle cut $\theta_{\rm cut}$ for $s>m_\Delta^2$,
while the dependence on $\theta_{\rm cut}$ is reduced for
$m_\Delta^2>s$.

Finally, we derive the expression in the NWA.
Using the representation of the Dirac delta function,
\begin{equation}
\delta(x)
=
\lim_{\epsilon\to0}
\frac{1}{2\pi i}
\left(
\frac{1}{x-i\epsilon}
-
\frac{1}{x+i\epsilon}
\right)
=
\lim_{\epsilon\to0}
\frac{\epsilon}{\pi}
\frac{1}{x^2+\epsilon^2},
\end{equation}
we obtain
\begin{equation}
\int dE_\gamma
\frac{X}
{(4E(E-E_\gamma)-m_\Delta^2)^2+m_\Delta^2\Gamma_\Delta^2}
\to
\frac{1}{4E}
\frac{\pi}{m_\Delta\Gamma_\Delta}
X
\big|_{E_\gamma=E-\frac{m_\Delta^2}{4E}},
\end{equation}
for a narrow decay width $\Gamma_\Delta$.
The decay width of $\Delta^{++}\to e^+e^+$ is
\begin{equation}
\Gamma_e
\equiv
\Gamma(\Delta^{++}\to e^+e^+)
=
\frac{|\kappa_{ee}|^2m_\Delta}{8\pi}.
\end{equation}
Then, we find in the NWA,
\begin{equation}
\frac{d\sigma_{1LL}}{d(\cos\theta)}
\simeq
|\kappa_{\mu\mu}|^2
\frac{e^2}{\pi}
\frac{\Gamma_e}{\Gamma_\Delta}
\frac{1}{s^2}
\frac{s^2+m_\Delta^4}{s-m_\Delta^2}
\cot^2\theta,
\end{equation}
which corresponds to the differential cross section for
$\mu_L^+\mu_L^+\to\Delta^{++}\gamma$ multiplied by
${\rm Br}(\Delta^{++}\to e^+e^+)=\Gamma_e/\Gamma_\Delta$.

\section{Kinematics of a 2-body $\to$ 3-body process}
\label{appendix:B}

In this appendix, we describe the kinematics of the process $1+2 \to 3 + 4 + 5$,
in which only the fifth particle is massive.

Let $p_1$ and $p_2$ be the momenta of the initial-state particles,
and let $p_3$, $p_4$, and $p_5$ be those of the final-state particles.
We assume that only the particle with momentum $p_5$ is massive, $p_5^2 = m_5^2$,
while the masses of the other particles are neglected, $p_1^2= p_2^2= p_3^2=p_4^2=0$.
We parametrize the momenta as in Eqs.~(\ref{p1})--(\ref{p4}).
The energy of the fifth particle is denoted by $E_5$.

From momentum conservation, $p_1 + p_2 = p_3 + p_4 +p_5$, we obtain
\begin{equation}
4E^2 - 4E E_5 + m_5^2 = 2E_3 E_4 (1-\cos\theta_{34}).
\label{(p1+p2-p5)^2}
\end{equation}
The left-hand side is
\begin{equation}
(p_1 + p_2 - p_5)^2 = (p_1+p_2)^2 - 2 p_5 \cdot (p_1+ p_2) + p_5^2 = 4E^2 - 4E E_5 + m_5^2,
\end{equation}
and the right-hand side is
\begin{equation}
(p_3+ p_4)^2 = 2 p_3 \cdot p_4 = 2 E_3 E_4 (1-\cos\theta_{34}),
\end{equation}
where $\theta_{34}$ is the angle between the three-dimensional momenta, ${\bm p}_3$ and ${\bm p}_4$:
\begin{equation}
\cos\theta_{34} = \cos\theta_3 \cos\theta_4 + \sin\theta_3 \sin\theta_4 \cos\phi.
\label{costheta34}
\end{equation}

Let us derive the kinematically allowed region of the final-state energies.
From Eq.~(\ref{(p1+p2-p5)^2}),
we obtain
\begin{equation}
E_5 \leq E+ \frac{m_5^2}{4E},
\end{equation}
where the equality holds when $\cos\theta_{34} = 1$.
From energy conservation, $2E = E_3 + E_4 + E_5$,
we also obtain
\begin{equation}
E_3 + E_4 \geq E - \frac{m_5^2}{4E}.
\end{equation}
Eliminating $E_5$ using energy conservation, $2E = E_3 + E_4 + E_5$,
we obtain
\begin{equation}
-4E^2 + 4E(E_3+E_4) +m_5^2 = 2E_3 E_4 (1-\cos\theta_{34}).
\label{(p1+p2-p5)^2withEconv}
\end{equation}
Solving for $E_4$, we obtain
\begin{equation}
E_4 = \frac12 \frac{4E^2 -4EE_3 - m_5^2}{2E - E_3(1-\cos\theta_{34})}.
\label{solE4}
\end{equation}
From Eq.~(\ref{(p1+p2-p5)^2withEconv}),
we obtain
\begin{equation}
-4E^2 + 4E(E_3+E_4) +m_5^2 \leq 4 E_3 E_4,
\end{equation}
where the equality holds when $\cos\theta_{34} = -1$.
Then, we obtain
\begin{equation}
(E-E_3) (E-E_4) \geq \frac{m_5^2}{4},
\label{(E-E3)(E-E4)}
\end{equation}
which gives the maximal value of $E_3$ for a given $E_4$:
\begin{equation}
E_3 \leq E - \frac{m_5^2}{4(E-E_4)}.
\end{equation}
The independent ranges of $E_3$ and $E_4$ are
\begin{equation}
0 \leq E_{3,4} \leq E - \frac{m_5^2}{4E}.
\end{equation}
Eq.~(\ref{(E-E3)(E-E4)}) also gives a lower bound on $E_5$ for a given $E_4$:
\begin{equation}
E_5 \geq E-E_4 + \frac{m_5^2}{4(E-E_4)}.
\end{equation}
The minimum value of $E_5$ is attained at $E_3 = E_4 = E- m_5/2$,
where particle 5 is at rest and $E_5 = m_5$.


\begin{thebibliography}{99}

\bibitem{MEGII:2025gzr}
K.~Afanaciev \textit{et al.} [MEG II],
``New limit on the ${\mu ^+ \rightarrow e^+ \gamma }$ decay with the MEG II experiment,''
Eur. Phys. J. C \textbf{85}, no.10, 1177 (2025)
[erratum: Eur. Phys. J. C \textbf{85}, no.11, 1317 (2025)]
\doi{10.1140/epjc/s10052-025-14906-3}
[arXiv:2504.15711 [hep-ex]].

\bibitem{Bellgardt:1987du}
U.~Bellgardt \textit{et al.} [SINDRUM],
``Search for the Decay $\mu^+ \to e^+ e^+ e^-$,''
Nucl. Phys. B \textbf{299}, 1-6 (1988)
\doi{10.1016/0550-3213(88)90462-2}

\bibitem{SINDRUMII:2006dvw}
W.~H.~Bertl \textit{et al.} [SINDRUM II],
``A Search for muon to electron conversion in muonic gold,''
Eur. Phys. J. C \textbf{47}, 337-346 (2006)
\doi{10.1140/epjc/s2006-02582-x}

\bibitem{Babu:1988ki}
K.~S.~Babu,
``Model of `Calculable' Majorana Neutrino Masses,''
Phys. Lett. B \textbf{203}, 132-136 (1988)
\doi{10.1016/0370-2693(88)91584-5};
%
K.~S.~Babu and C.~Macesanu,
``Two loop neutrino mass generation and its experimental consequences,''
Phys. Rev. D \textbf{67}, 073010 (2003)
\doi{10.1103/PhysRevD.67.073010}
[arXiv:hep-ph/0212058 [hep-ph]].

\bibitem{Gustafsson:2012vj}
M.~Gustafsson, J.~M.~No and M.~A.~Rivera,
``Predictive Model for Radiatively Induced Neutrino Masses and Mixings with Dark Matter,''
Phys. Rev. Lett. \textbf{110}, no.21, 211802 (2013)
[erratum: Phys. Rev. Lett. \textbf{112}, no.25, 259902 (2014)]
\doi{10.1103/PhysRevLett.110.211802}
[arXiv:1212.4806 [hep-ph]];
%
``Radiative neutrino mass generation linked to neutrino mixing
and $0\nu\beta\beta$-decay predictions,''
Phys. Rev. D \textbf{90}, no.1, 013012 (2014)
\doi{10.1103/PhysRevD.90.013012}
[arXiv:1402.0515 [hep-ph]].

\bibitem{Liu:2016mpf}
Z.~Liu and P.~H.~Gu,
``Extending two Higgs doublet models for two-loop neutrino mass generation and one-loop neutrinoless double beta decay,''
Nucl. Phys. B \textbf{915}, 206-223 (2017)
\doi{10.1016/j.nuclphysb.2016.12.001}
[arXiv:1611.02094 [hep-ph]].

\bibitem{Hamada:2022mua}
Y.~Hamada, R.~Kitano, R.~Matsudo, H.~Takaura and M.~Yoshida,
``$\mu$TRISTAN,''
PTEP \textbf{2022}, no.5, 053B02 (2022)
\doi{10.1093/ptep/ptac059}
[arXiv:2201.06664 [hep-ph]].

\bibitem{Hamada:2022uyn}
Y.~Hamada, R.~Kitano, R.~Matsudo and H.~Takaura,
``Precision $\mu^+\mu^+$ and $\mu^+e^-$ elastic scatterings,''
PTEP \textbf{2023}, no.1, 013B07 (2023)
\doi{10.1093/ptep/ptac174}
[arXiv:2210.11083 [hep-ph]].

\bibitem{Mukherjee:2026kmp}
L.~Mukherjee and A.~Sarkar,
``Probing anomalous quartic gauge couplings via vector boson scattering at the same-sign muon collider,''
Phys. Rev. D \textbf{114}, no.5, 055042 (2026)
\doi{10.1103/zw44-5zh8}
[arXiv:2605.03116 [hep-ph]].

\bibitem{Hamada:2024ojj}
Y.~Hamada, R.~Kitano, R.~Matsudo, S.~Okawa, R.~Takai, H.~Takaura and L.~Treuer,
``Higgs boson production at $\mu^+\mu^+$ colliders,''
Phys. Rev. D \textbf{110}, no.11, 113011 (2024)
\doi{10.1103/PhysRevD.110.113011}
[arXiv:2408.01068 [hep-ph]].

\bibitem{Lichtenstein:2023iut}
G.~Lichtenstein, M.~A.~Schmidt, G.~Valencia and R.~R.~Volkas,
``Complementarity of $\mu$TRISTAN and Belle II in searches for charged-lepton flavour violation,''
Phys. Lett. B \textbf{845}, 138144 (2023)
\doi{10.1016/j.physletb.2023.138144}
[arXiv:2307.11369 [hep-ph]].

\bibitem{Yang:2023ojm}
J.~L.~Yang, C.~H.~Chang and T.~F.~Feng,
``Leptonic di-flavor and di-number violation processes at high energy $\mu^\pm\mu^\pm$ colliders,''
Chin. Phys. C \textbf{48}, no.4, 043101 (2024)
\doi{10.1088/1674-1137/ad17b0}
[arXiv:2302.13247 [hep-ph]];
%
K.~Fridell, R.~Kitano and R.~Takai,
``Lepton flavor physics at $\mu^+\mu^+$ colliders,''
JHEP \textbf{06}, 086 (2023)
\doi{10.1007/JHEP06(2023)086}
[arXiv:2304.14020 [hep-ph]];
%
P.~S.~B.~Dev, J.~Heeck and A.~Thapa,
``Neutrino mass models at $\mu$TRISTAN,''
Eur. Phys. J. C \textbf{84}, no.2, 148 (2024)
\doi{10.1140/epjc/s10052-024-12496-0}
[arXiv:2309.06463 [hep-ph]];
%
A.~Das, J.~Li, S.~Mandal, T.~Nomura and R.~Zhang,
``Testing tree level TeV scale seesaw scenarios in $\mu$TRISTAN,''
Phys. Rev. D \textbf{112}, no.3, 035008 (2025)
\doi{10.1103/df3g-32t9}
[arXiv:2410.21956 [hep-ph]];
%
C.~W.~Chiang, K.~Enomoto and M.~Y.~Liao,
``Doubly charged Higgs boson at same-sign lepton colliders,''
Phys. Rev. D \textbf{112}, no.11, 115003 (2025)
\doi{10.1103/yh2v-qhgm}
[arXiv:2506.17541 [hep-ph]];
%
F.~Huang and J.~Sun,
``Probing maximal flavor changing $Z'$ in $U(1)_{L_\mu-L_\tau}$ at $\mu$TRISTAN,''
Phys. Lett. B \textbf{872}, 140047 (2026)
\doi{10.1016/j.physletb.2025.140047}
[arXiv:2507.04614 [hep-ph]];
%
G.~Li and J.~Sun,
``Probing flavor-diagonal couplings of doubly charged scalar at low and high energies,''
Phys. Rev. D \textbf{114}, no.1, 015027 (2026)
\doi{10.1103/khts-4dcn}
[arXiv:2512.07255 [hep-ph]];
%
J.~George, N.~Okada, D.~Sengupta and S.~K.~Vempati,
``Discovery prospects of a singly charged scalar at {\ensuremath{\mu}}TRISTAN,''
Phys. Rev. D \textbf{114}, no.3, 035045 (2026)
\doi{10.1103/3d28-s1rs}
[arXiv:2601.22000 [hep-ph]];
%
M.~Frank, B.~Fuks, C.~Majumdar and S.~Senapati,
``Multiboson Signatures of Doubly Charged Scalars at a Same-Sign Muon Collider,''
[arXiv:2607.15356 [hep-ph]].

\bibitem{Jiang:2023mte}
R.~Jiang, T.~Yang, S.~Qian, Y.~Ban, J.~Li, Z.~You and Q.~Li,
``Searching for Majorana neutrinos at a same-sign muon collider,''
Phys. Rev. D \textbf{109}, no.3, 035020 (2024)
\doi{10.1103/PhysRevD.109.035020}
[arXiv:2304.04483 [hep-ph]];
%
A.~Goudelis, J.~Kriewald, E.~Pinsard and A.~M.~Teixeira,
``cLFV leptophilic $Z^\prime $ as a dark matter portal: prospects for colliders,''
Eur. Phys. J. C \textbf{84}, no.8, 804 (2024)
\doi{10.1140/epjc/s10052-024-13158-x}
[arXiv:2312.14103 [hep-ph]];
%
J.~Heeck and M.~Sokhashvili,
``Lepton flavor violation by two units,''
Phys. Lett. B \textbf{852}, 138621 (2024)
\doi{10.1016/j.physletb.2024.138621}
[arXiv:2401.09580 [hep-ph]];
%
J.~L.~G.~Santiago, D.~Portillo-S\'anchez, G.~Hern\'andez-Tom\'e and J.~Rend\'on,
``Authentic Majorana versus singlet Dirac neutrino contributions to $\mu^+\mu^+\to\ell^+\ell^+$ ($\ell=e,\tau$) transitions,''
Phys. Rev. D \textbf{110}, no.5, 053006 (2024)
\doi{10.1103/PhysRevD.110.053006}
[arXiv:2405.02819 [hep-ph]];
%
L.~Calibbi, T.~Li, L.~Mukherjee and Y.~Yang,
``Probing ALP lepton flavor violation at $\mu$TRISTAN,''
Phys. Rev. D \textbf{110}, no.11, 115009 (2024)
\doi{10.1103/PhysRevD.110.115009}
[arXiv:2406.13234 [hep-ph]];
%
C.~H.~de Lima, D.~McKeen, J.~N.~Ng, M.~Shamma and D.~Tuckler,
``Probing lepton number violation at same-sign lepton colliders,''
Phys. Rev. D \textbf{111}, no.7, 075002 (2025)
\doi{10.1103/PhysRevD.111.075002}
[arXiv:2411.15303 [hep-ph]];
%
S.~Bhattacharya, S.~Datta and A.~Sarkar,
``Probing $\Delta L=2$ lepton number violating SMEFT operators at the same-sign muon collider,''
Phys. Rev. D \textbf{113}, no.5, 055043 (2026)
\doi{10.1103/pmj9-8g25}
[arXiv:2505.20936 [hep-ph]];
%
L.~Calibbi, C.~Hagedorn, M.~A.~Schmidt and J.~Vandeleur,
``Selection rules for charged lepton flavor violating processes from residual flavor groups,''
Phys. Rev. D \textbf{112}, no.7, 075031 (2025)
\doi{10.1103/b6gz-c1c9}
[arXiv:2505.24350 [hep-ph]];
%
K.~Harigaya, R.~Kitano and R.~Takai,
``Lepton number violating signals of a parity symmetric model at $\mu$TRISTAN,''
JHEP \textbf{04}, 085 (2026)
\doi{10.1007/JHEP04(2026)085}
[arXiv:2509.24680 [hep-ph]];
%
R.~Kitano, I.~Low, R.~Matsudo, S.~Okawa and S.~Roy,
``Heavy neutral lepton at same-sign muon collider,''
Phys. Rev. D \textbf{114}, no.5, 055006 (2026)
\doi{10.1103/prx6-wnwr}
[arXiv:2510.18390 [hep-ph]];
%
A.~Das, T.~Nomura and K.~Yagyu,
``Lepton number violation from Higgs/$Z$ decays into a pair of right-handed neutrinos,''
Phys. Rev. D \textbf{114}, no.1, 015001 (2026)
\doi{10.1103/ynd4-dh1b}
[arXiv:2512.03302 [hep-ph]];
%
L.~K.~Yan, Z.~L.~Han, F.~L.~Shao and F.~X.~Yang,
``Same-Sign Tetralepton Signature at $\mu$TRISTAN,''
[arXiv:2604.24383 [hep-ph]].

\bibitem{Pontecorvo:1957cp}
B.~Pontecorvo,
``Mesonium and anti-mesonium,''
Sov. Phys. JETP \textbf{6}, 429 (1957).

\bibitem{Feinberg:1961zza}
G.~Feinberg and S.~Weinberg,
``Conversion of Muonium into Antimuonium,''
Phys. Rev. \textbf{123}, 1439-1443 (1961)
\doi{10.1103/PhysRev.123.1439}

\bibitem{Lee:1977tib}
B.~W.~Lee and R.~E.~Shrock,
``Natural suppression of symmetry violation in gauge theories:
Muon- and electron-lepton-number nonconservation,''
Phys. Rev. D \textbf{16}, 1444 (1977)
\doi{10.1103/PhysRevD.16.1444};
%
B.~W.~Lee, S.~Pakvasa, R.~E.~Shrock and H.~Sugawara,
``Muon and Electron Number Nonconservation in a V-A Gauge Model,''
Phys. Rev. Lett. \textbf{38}, 937 (1977)
[erratum: Phys. Rev. Lett. \textbf{38}, 1230 (1977)]
\doi{10.1103/PhysRevLett.38.937}

\bibitem{Halprin:1982wm}
A.~Halprin,
``Neutrinoless Double-Beta Decay and Muonium-Antimuonium Transitions,''
Phys. Rev. Lett. \textbf{48}, 1313-1316 (1982)
\doi{10.1103/PhysRevLett.48.1313}

\bibitem{Fukuyama:2021iyw}
T.~Fukuyama, Y.~Mimura and Y.~Uesaka,
``Models of the muonium to antimuonium transition,''
Phys. Rev. D \textbf{105}, no.1, 015026 (2022)
\doi{10.1103/PhysRevD.105.015026}
[arXiv:2108.10736 [hep-ph]].

\bibitem{Chang:1989uk}
D.~Chang and W.~Y.~Keung,
``Constraints on Muonium-antiMuonium Conversion,''
Phys. Rev. Lett. \textbf{62}, 2583 (1989)
\doi{10.1103/PhysRevLett.62.2583}

\bibitem{Swartz:1989qz}
M.~L.~Swartz,
``Limits on Doubly Charged Higgs Bosons and Lepton Flavor Violation,''
Phys. Rev. D \textbf{40}, 1521 (1989)
\doi{10.1103/PhysRevD.40.1521}

\bibitem{Willmann:1998gd}
L.~Willmann, P.~V.~Schmidt, H.~P.~Wirtz, R.~Abela, V.~Baranov, J.~Bagaturia, W.~H.~Bertl, R.~Engfer, A.~Grossmann and V.~W.~Hughes, \textit{et al.}
``New bounds from searching for muonium to anti-muonium conversion,''
Phys. Rev. Lett. \textbf{82}, 49-52 (1999)
\doi{10.1103/PhysRevLett.82.49}
[arXiv:hep-ex/9807011 [hep-ex]].

\bibitem{Han:2021nod}
C.~Han, D.~Huang, J.~Tang and Y.~Zhang,
``Probing the doubly-charged Higgs with Muonium to Antimuonium Conversion Experiment,''
Phys. Rev. D \textbf{103}, no.5, 055023 (2021)
\doi{10.1103/PhysRevD.103.055023}
[arXiv:2102.00758 [hep-ph]].

\bibitem{Bai:2022sxq}
A.~Y.~Bai, Y.~ Y.~Chen, R.~R.~Fan, Z.~Hou, H.~T.~Jing, H.~B.~Li, Y.~Li, H.~Miao and H.~Peng, \textit{et al.}
``Snowmass2021 Whitepaper: Muonium to antimuonium conversion,''
[arXiv:2203.11406 [hep-ph]].

\bibitem{Bai:2024skk}
A.~Y.~Bai, H.~J.~Cai, C.~L.~Chen, S.~Y.~Chen, X.~R.~Chen, Y.~Chen, W.~B.~Cheng, L.~Y.~Dai, R.~R.~Fan and L.~Gong, \textit{et al.}
``Conceptual design of the muonium-to-antimuonium conversion experiment (MACE),''
Nucl. Sci. Tech. \textbf{37} (2026) no.4, 57
\doi{10.1007/s41365-025-01876-0}
[arXiv:2410.18817 [hep-ex]].

\bibitem{Kawamura:2021lqk}
N.~Kawamura, R.~Kitamura, H.~Yasuda, M.~Otani, Y.~Nakazawa, H.~Iinuma and T.~Mibe,
``A New Approach for Mu - \(\overline{\text{Mu}}\) Conversion Search,''
JPS Conf. Proc. \textbf{33}, 011120 (2021)
\doi{10.7566/JPSCP.33.011120}

\bibitem{Kawamura:2025lhk}
N.~Kawamura, R.~Kitamura, S.~Kanda, S.~Matoba, T.~Mibe, T.~Fukuyama, Y.~Mimura and Y.~Uesaka,
``A new concept of $\textrm{Mu}${\ensuremath{-}}$\overline{\textrm{Mu}}$ conversion search,''
PoS \textbf{EXA-LEAP2024} (2025), 074
\doi{10.22323/1.480.0074}

\bibitem{Schechter:1980gr}
J.~Schechter and J.~W.~F.~Valle,
``Neutrino Masses in $\mathrm{SU}(2)\otimes\mathrm{U}(1)$ Theories,''
Phys. Rev. D \textbf{22}, 2227 (1980)
\doi{10.1103/PhysRevD.22.2227}

\bibitem{Cheng:1980qt}
T.~P.~Cheng and L.~F.~Li,
``Neutrino Masses, Mixings and Oscillations in
$\mathrm{SU}(2)\times\mathrm{U}(1)$ Models of Electroweak Interactions,''
Phys. Rev. D \textbf{22}, 2860 (1980)
\doi{10.1103/PhysRevD.22.2860}

\bibitem{Lazarides:1980nt}
G.~Lazarides, Q.~Shafi and C.~Wetterich,
``Proton Lifetime and Fermion Masses in an SO(10) Model,''
Nucl. Phys. B \textbf{181}, 287-300 (1981)
\doi{10.1016/0550-3213(81)90354-0}

\bibitem{Mohapatra:1980yp}
R.~N.~Mohapatra and G.~Senjanovic,
``Neutrino Masses and Mixings in Gauge Models with Spontaneous Parity Violation,''
Phys. Rev. D \textbf{23}, 165 (1981)
\doi{10.1103/PhysRevD.23.165}

\bibitem{Fukuyama:2022dhe}
T.~Fukuyama, Y.~Mimura and Y.~Uesaka,
``Neutrinoless double beta decay and the muonium-to-antimuonium transition in models with a doubly charged scalar,''
Phys. Rev. D \textbf{106}, no.5, 055041 (2022)
[erratum: Phys. Rev. D \textbf{107}, no.7, 079903 (2023)]
\doi{10.1103/PhysRevD.106.055041}
[arXiv:2206.09691 [hep-ph]].

\bibitem{Zee:1980ai}
A.~Zee,
``A Theory of Lepton Number Violation, Neutrino Majorana Mass, and Oscillation,''
Phys. Lett. B \textbf{93}, 389 (1980)
[erratum: Phys. Lett. B \textbf{95}, 461 (1980)]
\doi{10.1016/0370-2693(80)90349-4}

\bibitem{Zee:1985id}
A.~Zee,
``Quantum Numbers of Majorana Neutrino Masses,''
Nucl. Phys. B \textbf{264}, 99-110 (1986)
\doi{10.1016/0550-3213(86)90475-X}

\bibitem{ALEPH:2013dgf}
S.~Schael \textit{et al.} [ALEPH, DELPHI, L3, OPAL and LEP Electroweak],
``Electroweak Measurements in Electron-Positron Collisions at W-Boson-Pair Energies at LEP,''
Phys. Rept. \textbf{532}, 119-244 (2013)
\doi{10.1016/j.physrep.2013.07.004}
[arXiv:1302.3415 [hep-ex]].

\bibitem{Nomura:2017abh}
T.~Nomura, H.~Okada and H.~Yokoya,
``Discriminating leptonic Yukawa interactions with doubly charged scalar at the ILC,''
Nucl. Phys. B \textbf{929}, 193-206 (2018)
\doi{10.1016/j.nuclphysb.2018.02.011}
[arXiv:1702.03396 [hep-ph]].

\bibitem{Endo:2022kiw}
M.~Endo and S.~Mishima,
``New physics interpretation of W-boson mass anomaly,''
Phys. Rev. D \textbf{106}, no.11, 115005 (2022)
\doi{10.1103/PhysRevD.106.115005}
[arXiv:2204.05965 [hep-ph]].

\bibitem{Deppisch:2012nb}
F.~F.~Deppisch, M.~Hirsch and H.~Pas,
``Neutrinoless Double Beta Decay and Physics Beyond the Standard Model,''
J. Phys. G \textbf{39}, 124007 (2012)
\doi{10.1088/0954-3899/39/12/124007}
[arXiv:1208.0727 [hep-ph]].

\bibitem{KamLAND-Zen:2024eml}
S.~Abe \textit{et al.} [KamLAND-Zen],
``Search for Majorana Neutrinos with the Complete KamLAND-Zen Dataset,''
Phys. Rev. Lett. \textbf{135}, no.26, 262501 (2025)
\doi{10.1103/jkf6-48j8}
[arXiv:2406.11438 [hep-ex]].

\bibitem{Domin:2004tk}
P.~Domin, S.~Kovalenko, A.~Faessler and F.~Simkovic,
``Nuclear ($\mu^-, e^+$) conversion mediated by Majorana neutrinos,''
Phys. Rev. C \textbf{70}, 065501 (2004)
\doi{10.1103/PhysRevC.70.065501}
[arXiv:nucl-th/0409033 [nucl-th]].

\bibitem{Lee:2021hnx}
M.~Lee and M.~MacKenzie,
``Muon to Positron Conversion,''
Universe \textbf{8}, no.4, 227 (2022)
\doi{10.3390/universe8040227}
[arXiv:2110.07093 [hep-ex]].

\bibitem{Berryman:2016slh}
J.~M.~Berryman, A.~de Gouv{\^e}a, K.~J.~Kelly and A.~Kobach,
``Lepton-number-violating searches for muon to positron conversion,''
Phys. Rev. D \textbf{95}, no.11, 115010 (2017)
\doi{10.1103/PhysRevD.95.115010}
[arXiv:1611.00032 [hep-ph]].

\bibitem{Sato:2022vny}
J.~Sato, K.~Sugawara, Y.~Uesaka and M.~Yamanaka,
``Determination of coupling patterns by parallel searches for 
$\mu^-\rightarrow e^+$ and $\mu^-\rightarrow e^{-}$ in muonic atoms,''
Phys. Lett. B \textbf{836}, 137617 (2023)
\doi{10.1016/j.physletb.2022.137617}
[arXiv:2203.01314 [hep-ph]].

\bibitem{NeutrinoFactory:2002azy}
M.~M.~Alsharoa \textit{et al.} [Neutrino Factory and Muon Collider],
``Recent Progress in Neutrino Factory and Muon Collider Research within the Muon Collaboration,''
Phys. Rev. ST Accel. Beams \textbf{6}, 081001 (2003)
\doi{10.1103/PhysRevSTAB.6.081001}
[arXiv:hep-ex/0207031 [hep-ex]].

\bibitem{Aaboud:2017qph}
M.~Aaboud \textit{et al.} [ATLAS],
``Search for doubly charged Higgs boson production in multi-lepton final states with the ATLAS detector using proton\textendash{}proton collisions at $\sqrt{s}=13\,\text {TeV}$,''
Eur. Phys. J. C \textbf{78}, no.3, 199 (2018)
\doi{10.1140/epjc/s10052-018-5661-z}
[arXiv:1710.09748 [hep-ex]].

\bibitem{ATLAS:2022pbd}
G.~Aad \textit{et al.} [ATLAS],
``Search for doubly charged Higgs boson production in multi-lepton final states using 139~fb$^{-1}$ of proton{\textendash}proton collisions at $\sqrt{s}$ = 13~TeV with the ATLAS detector,''
Eur. Phys. J. C \textbf{83}, no.7, 605 (2023)
\doi{10.1140/epjc/s10052-023-11578-9}
[arXiv:2211.07505 [hep-ex]].

\bibitem{Chiang:2012dk}
C.~W.~Chiang, T.~Nomura and K.~Tsumura,
``Search for doubly charged Higgs bosons using the same-sign diboson mode at the LHC,''
Phys. Rev. D \textbf{85}, 095023 (2012)
\doi{10.1103/PhysRevD.85.095023}
[arXiv:1202.2014 [hep-ph]].

\bibitem{Chen:2014qda}
C.~H.~Chen and T.~Nomura,
``Search for $\delta^{\pm\pm}$ with new decay patterns at the LHC,''
Phys. Rev. D \textbf{91}, 035023 (2015)
\doi{10.1103/PhysRevD.91.035023}
[arXiv:1411.6412 [hep-ph]].

\bibitem{Dev:2018sel}
P.~S.~B.~Dev, M.~J.~Ramsey-Musolf and Y.~Zhang,
``Doubly-Charged Scalars in the Type-II Seesaw Mechanism: Fundamental Symmetry Tests and High-Energy Searches,''
Phys. Rev. D \textbf{98}, no.5, 055013 (2018)
\doi{10.1103/PhysRevD.98.055013}
[arXiv:1806.08499 [hep-ph]].

\bibitem{Crivellin:2018ahj}
A.~Crivellin, M.~Ghezzi, L.~Panizzi, G.~M.~Pruna and A.~Signer,
``Low- and high-energy phenomenology of a doubly charged scalar,''
Phys. Rev. D \textbf{99}, no.3, 035004 (2019)
\doi{10.1103/PhysRevD.99.035004}
[arXiv:1807.10224 [hep-ph]].

\bibitem{BhupalDev:2018tox}
P.~S.~Bhupal Dev and Y.~Zhang,
``Displaced vertex signatures of doubly charged scalars in the type-II seesaw and its left-right extensions,''
JHEP \textbf{10}, 199 (2018)
\doi{10.1007/JHEP10(2018)199}
[arXiv:1808.00943 [hep-ph]].

\bibitem{Padhan:2019jlc}
R.~Padhan, D.~Das, M.~Mitra and A.~Kumar Nayak,
``Probing doubly and singly charged Higgs bosons at the $pp$ collider HE-LHC,''
Phys. Rev. D \textbf{101}, no.7, 075050 (2020)
\doi{10.1103/PhysRevD.101.075050}
[arXiv:1909.10495 [hep-ph]].

\bibitem{Fuks:2019clu}
B.~Fuks, M.~Nemev\v{s}ek and R.~Ruiz,
``Doubly Charged Higgs Boson Production at Hadron Colliders,''
Phys. Rev. D \textbf{101}, no.7, 075022 (2020)
\doi{10.1103/PhysRevD.101.075022}
[arXiv:1912.08975 [hep-ph]].

\bibitem{Enomoto:2021fql}
K.~Enomoto, S.~Kanemura and K.~Katayama,
``Probing doubly charged scalar bosons from the doublet at future high-energy colliders,''
Phys. Rev. D \textbf{104}, no.3, 035040 (2021)
\doi{10.1103/PhysRevD.104.035040}
[arXiv:2102.12950 [hep-ph]].

\bibitem{Bai:2021ony}
X.~H.~Bai, Z.~L.~Han, Y.~Jin, H.~L.~Li and Z.~X.~Meng,
``Same-sign tetralepton signature in type-II seesaw at lepton colliders,''
Chin. Phys. C \textbf{46}, no.12, 012001 (2022)
\doi{10.1088/1674-1137/ac2ed1}
[arXiv:2105.02474 [hep-ph]].

\bibitem{Ashanujjaman:2021txz}
S.~Ashanujjaman and K.~Ghosh,
``Revisiting type-II see-saw: present limits and future prospects at LHC,''
JHEP \textbf{03}, 195 (2022)
\doi{10.1007/JHEP03(2022)195}
[arXiv:2108.10952 [hep-ph]].

\bibitem{Dey:2022whc}
S.~Dey, P.~Ghosh and S.~K.~Rai,
``Confronting dark fermion with a doubly charged Higgs in the left{\textendash}right symmetric model,''
Eur. Phys. J. C \textbf{82}, no.10, 876 (2022)
\doi{10.1140/epjc/s10052-022-10778-z}
[arXiv:2202.11638 [hep-ph]].

\bibitem{Guedes:2025jqu}
L.~Guedes, G.~Hoff, F.~S.~Queiroz, Y.~M.~Oviedo-Torres and Y.~Villamizar,
``Search for a doubly charged scalar at the LHC and FCC-hh,''
Eur. Phys. J. C \textbf{86}, no.5, 561 (2026)
\doi{10.1140/epjc/s10052-026-15774-1}
[arXiv:2601.00083 [hep-ph]].

\bibitem{CMS:2026bxj}
CMS Collaboration,
``Observation of lepton-lepton scattering in proton-proton collisions at $\sqrt{s} = 13$ and 13.6 TeV,''
CMS-PAS-SMP-26-008.

\bibitem{Buonocore:2020nai}
L.~Buonocore, P.~Nason, F.~Tramontano and G.~Zanderighi,
``Leptons in the proton,''
JHEP \textbf{08}, no.08, 019 (2020)
\doi{10.1007/JHEP08(2020)019}
[arXiv:2005.06477 [hep-ph]].

\bibitem{Accardi:2012qut}
A.~Accardi, J.~L.~Albacete, M.~Anselmino, N.~Armesto, E.~C.~Aschenauer, A.~Bacchetta, D.~Boer, W.~K.~Brooks, T.~Burton and N.~B.~Chang, \textit{et al.}
``Electron Ion Collider: The Next QCD Frontier: Understanding the glue that binds us all,''
Eur. Phys. J. A \textbf{52}, no.9, 268 (2016)
\doi{10.1140/epja/i2016-16268-9}
[arXiv:1212.1701 [nucl-ex]].

\bibitem{Alwall:2014hca}
J.~Alwall, R.~Frederix, S.~Frixione, V.~Hirschi, F.~Maltoni, O.~Mattelaer, H.~S.~Shao, T.~Stelzer, P.~Torrielli and M.~Zaro,
``The automated computation of tree-level and next-to-leading order differential cross sections, and their matching to parton shower simulations,''
JHEP \textbf{07}, 079 (2014)
\doi{10.1007/JHEP07(2014)079}
[arXiv:1405.0301 [hep-ph]].

\bibitem{AlAli:2021let}
H.~Al Ali, N.~Arkani-Hamed, I.~Banta, S.~Benevedes, D.~Buttazzo, T.~Cai, J.~Cheng, T.~Cohen, N.~Craig and M.~Ekhterachian, \textit{et al.}
``The muon Smasher{\textquoteright}s guide,''
Rept. Prog. Phys. \textbf{85}, no.8, 084201 (2022)
\doi{10.1088/1361-6633/ac6678}
[arXiv:2103.14043 [hep-ph]].

\bibitem{Accettura:2023ked}
C.~Accettura, D.~Adams, R.~Agarwal, C.~Ahdida, C.~Aim{\`e}, N.~Amapane, D.~Amorim, P.~Andreetto, F.~Anulli and R.~Appleby, \textit{et al.}
``Towards a muon collider,''
Eur. Phys. J. C \textbf{83}, no.9, 864 (2023)
[erratum: Eur. Phys. J. C \textbf{84}, no.1, 36 (2024)]
\doi{10.1140/epjc/s10052-023-11889-x}
[arXiv:2303.08533 [physics.acc-ph]].

\bibitem{ParticleData:2026uvj}
F.~Takahashi \textit{et al.} [Particle Data Group],
``Review of Particle Physics,''
Int. J. Mod. Phys. A \textbf{41}, no.22, 2630011 (2026)
\doi{10.1142/s0217751x26300115}

\bibitem{King:2014uha}
S.~F.~King, A.~Merle and L.~Panizzi,
``Effective theory of a doubly charged singlet scalar: complementarity of neutrino physics and the LHC,''
JHEP \textbf{11}, 124 (2014)
\doi{10.1007/JHEP11(2014)124}
[arXiv:1406.4137 [hep-ph]].

\bibitem{vonWeizsacker:1934nji}
C.~F.~von Weizsacker,
``Radiation emitted in collisions of very fast electrons,''
Z. Phys. \textbf{88} (1934), 612-625
\doi{10.1007/BF01333110}

\bibitem{Williams:1935dka}
E.~J.~Williams,
``Correlation of certain collision problems with radiation theory,''
Kong. Dan. Vid. Sel. Mat. Fys. Med. \textbf{13N4} (1935) no.4, 1-50

\bibitem{Ma:2006km}
E.~Ma,
``Verifiable radiative seesaw mechanism of neutrino mass and dark matter,''
Phys. Rev. D \textbf{73}, 077301 (2006)
\doi{10.1103/PhysRevD.73.077301}
[arXiv:hep-ph/0601225 [hep-ph]].

\end{thebibliography}
\end{document}